\documentclass[letterpaper,twocolumn,11pt,preprintnumbers,preprint,aps]{quantumarticle}
\pdfoutput=1

\usepackage[utf8]{inputenc}
\usepackage{amsfonts}
\usepackage{amssymb} 
\usepackage{amsmath} 
\usepackage{bbold}
\usepackage{hyperref}
\usepackage[utf8]{inputenc}
\usepackage[qm,braket]{qcircuit}
\usepackage[dvipsnames,usenames]{xcolor}
\usepackage[ruled,vlined]{algorithm2e}
\usepackage{graphicx}
\usepackage{setspace}
\usepackage{tikz}
\usepackage{subfigure}
\usepackage[normalem]{ulem}
\usepackage{multirow}

\usepackage[numbers,sort&compress]{natbib}
\usepackage[colorlinks = true]{hyperref}

\begin{document}

\title{Fault-tolerant circuit synthesis for universal fault-tolerant quantum computing}

\author{Yongsoo Hwang}
\affiliation{Electronics and Telecommunications Research Institute, 34129 Daejeon, Republic of Korea}
\orcid{0000-0002-1366-4952}
\email{yhwang@etri.re.kr}
\maketitle

\begin{abstract}
We present a quantum circuit synthesis algorithm for implementing universal fault-tolerant quantum computing based on concatenated codes.
To realize fault-tolerant quantum computing, the fault-tolerant quantum protocols should be transformed into executable quantum circuits based on the nearest-neighbor interaction.
Unlike topological codes that are defined based on local operations fundamentally, for the concatenated codes, it is possible to obtain the circuits composed of the local operations by applying the quantum circuit synthesis.
However, by the existing quantum circuit synthesis developed for ordinary quantum computational algorithms, the fault-tolerant of the protocol may not be preserved in the resulting circuit.
Besides, we have to consider something more to implement the quantum circuit of universal fault-tolerant quantum computing.
First, we have not to propagate quantum errors on data qubits when selecting a qubit move path (a sequence of \emph{SWAP} gates) to satisfy the geometric locality constraint.
Second, the circuit should be self-contained so that it is possible to act independently regardless of the situation.
Third, for universal fault-tolerant quantum computing, we require multiple fault-tolerant quantum circuits of multiple fault-tolerant quantum protocols acting on the same input, a logical data qubit.
Last, we need to recall fault-tolerant protocols such as syndrome measure and encoder implicitly include classical control processing conditioned on the measurement outcomes, and therefore have to partition the quantum circuits in time flow to execute the classical control as the architect intended.
We propose the circuit synthesis method resolving the requirements and show how to synthesize the set of universal fault-tolerant protocols for $[[7,1,3]]$ Steane code and the syndrome measurement protocol of $[[23, 1, 7]]$ Golay code.
\end{abstract}

\section{Introduction}\label{sec:introduction}
An $[[n, k, d]]$ quantum error-correcting code encodes a $k$-qubit logical quantum information into $n$ physical (or lower concatenation level) qubits, and protects the encoded information from at most $\lfloor (d-1)/2 \rfloor$-qubit arbitrary quantum error, where $d$ is the code distance of the code.
If the number of arbitrary quantum errors is not more than $\lfloor (d-1)/2 \rfloor$, it is possible to detect and correct the quantum errors.
Otherwise, the logical information encoded into the code is broken, and cannot be recovered.

A quantum computing protocol acting on an error-corrected logical qubit is called \emph{fault tolerant} if the probability that the logical qubit is corrupted by quantum noise during running the protocol is suppressed than that in the non-error-corrected quantum computing.
For that, there should be no chance that arbitrary quantum errors to propagate to more than $\lfloor (d-1)/2 \rfloor$ qubits within the logical qubit in the protocol. 
By the way, we need to say that some physical qubits composing the logical qubit are more important than others.
The qubits those must not be affected by quantum error to keep the fault tolerance are the qubits, called as \emph{data qubit}, holding quantum data.
Therefore, a fault-tolerant quantum computing is not broken even if some qubits that do not belong to data qubits are corrupted by quantum errors.
By having a noisy interaction with data qubits, the quantum errors on the ancilla qubits are propagated to the data qubits, and the encoded qubit will become corrupted if the number of the data qubits where error happened is greater than $\lfloor (d-1)/2 \rfloor$.
Therefore, to retain the fault tolerance, we have to block the error propagation over the data qubits within the encoded logical qubit.

\begin{figure}[t]
\centering
\includegraphics[scale=0.3]{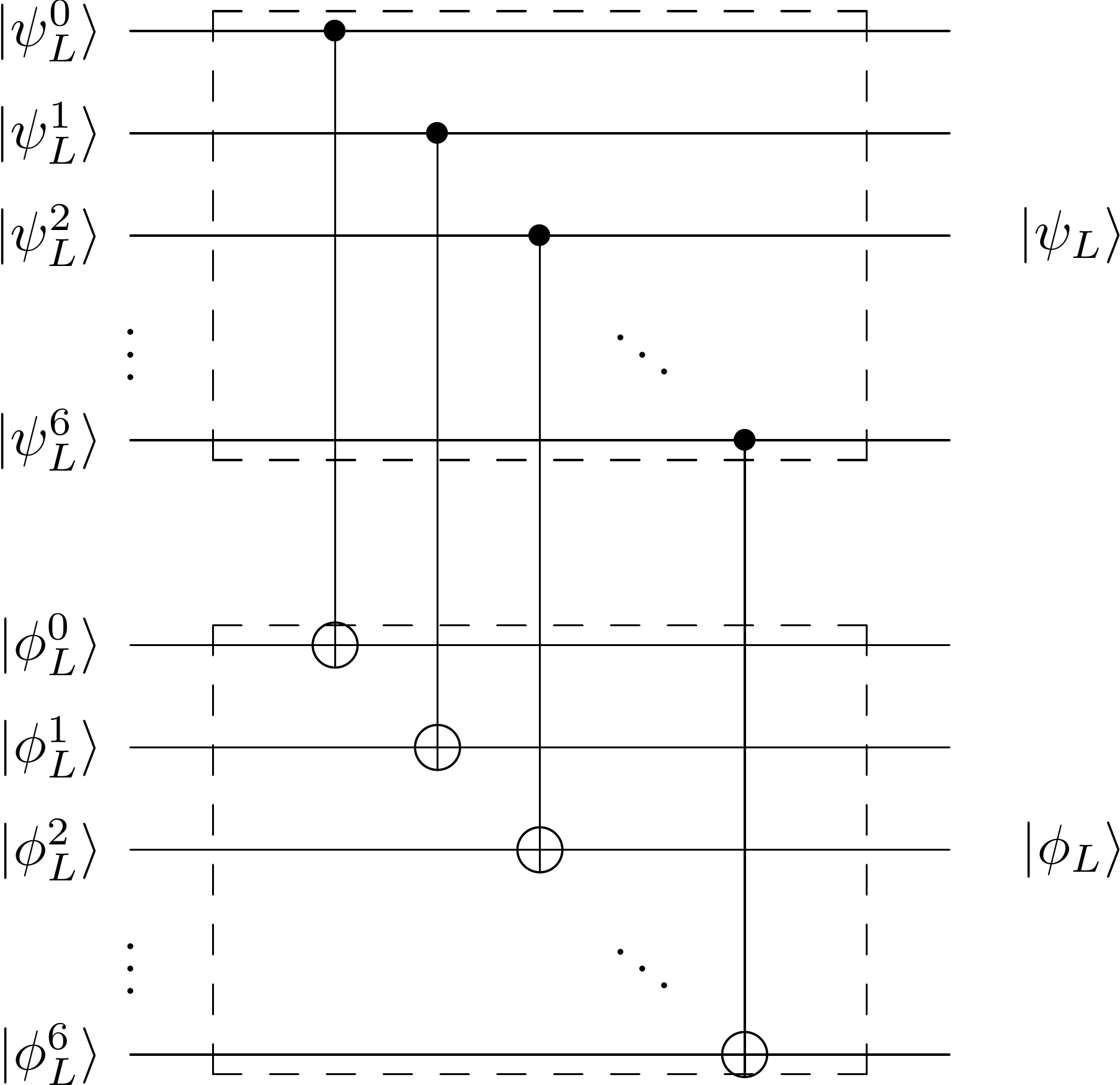}
\caption{
Transversal gate implementation of a logical CNOT gate
}
\label{fig:transversal_cnot}
\end{figure}

Needless to say, a fault-tolerant quantum computing protocol is designed by keeping in mind the rule that does not allow interactions that are likely to be noisy between the data qubits.
A transversal implementation of a logical gate is a typical approach for the fault tolerance (see Figure~\ref{fig:transversal_cnot}).
However, the fact that a protocol has fault tolerance itself does not guarantee that the protocol can be executed in a fault-tolerant way in practice.
A fault-tolerant protocol should be transformed into a fault-tolerant quantum circuit that is executable in the practical situation governed by the geometric locality constraint.
In the transformation, to resolve the constraint that a quantum computing device has, all the multi-qubit quantum gates in the protocol have to be decomposed into a sequence of the local 2-qubit gates acting on the nearest neighbor qubits.
The decomposition can be achieved by introducing a sequence of quantum state move (\emph{SWAP} gates) between distant qubits.

The transformation from the protocol (algorithm) to an executable circuit is called a \emph{circuit synthesis} or \emph{circuit mapping}, and so far a lot of quantum circuit mapping algorithms have been proposed.
Most quantum circuit synthesis algorithms makes an executable quantum circuit by including as least qubit move as possible by exploiting the qubit connectivity of the quantum device, but some advanced methods~\cite{Murali.2019jg,Tannu.2019} additionally consider the real performance of the target quantum device such as the fidelity or the execution time of physical gates.

Back to the fault-tolerant quantum protocol, to implement the circuit of the protocol, we have to consider more than the requirements in the circuit synthesis for an ordinary quantum algorithm.
As mentioned above, a normal circuit synthesis is designed to resolve the locality constraint by introducing the qubit move via \emph{SWAP} gates as little as possible.
For that, the type of quantum information a physical qubit holds is out of concern.
If such a circuit synthesis method is applied to a fault-tolerant quantum protocol, obviously it generates a quantum circuit including as few \emph{SWAP} gates as possible.
By the way, the serious problem here is that the circuit may include \emph{SWAP} gates acting on both data qubits.
As mentioned before, a noisy \emph{SWAP} gate may make both the data qubits corrupted.
Therefore, in the circuit synthesis on a fault-tolerant quantum protocol, a \emph{SWAP} gate between the data qubits should be selected very limitedly.

In addition, there are additional issues to take into account for the circuit synthesis for universal fault-tolerant quantum computing.
We need the fault-tolerant circuits for multiple fault-tolerant quantum protocols, and each circuit has to work correctly regardless of the situation.
Besides, some fault-tolerant quantum protocols such as syndrome measurement involve classical operations conditioned on the result of quantum operations.
Therefore, to implement their circuits exactly as designed, we have to clearly specify the turn for the digital operations in their circuits.

As far as the authors know, nothing has been proposed for the circuit synthesis algorithm specialized for universal fault-tolerant quantum computing, in particular, based on the concatenated codes.
Only for a few well-known codes, some results obtained manually for studying thresholds have been reported~\cite{Svore.2007,Lai.2013,Spedalieri.2008}.
Those results may be optimal for the designated target environment, but for another setting or quantum code, it is not easy to find fault-tolerant circuits by applying their methodology.
For this reason, we believe a circuit synthesis method specialized in fault-tolerant quantum protocols is required.

In the present work, we propose a fault-tolerant circuit synthesis algorithm for universal fault-tolerant quantum computing.
As mentioned above, the algorithm selectively picks \emph{SWAP} gates according to the type of quantum information physical qubits store when selecting the qubit move path.
We call the qubit move not including (or including very limitedly) the \emph{SWAP} gates acting on the data qubits as \textit{fault-tolerant qubit move}.
By taking the fault-tolerant qubit move path, it is possible to retain the fault-tolerance originating from the protocol in the resulting circuit.
However, the fault-tolerant qubit move path itself is not enough for realizing universal fault-tolerant quantum computing.

As is well known, to realize universal fault-tolerant quantum computing, we need multiple fault-tolerant logical gates (\emph{H}, \emph{T}, and \emph{CNOT}), syndrome measurement, and encoding protocol.
According to a quantum code, a fault-tolerant protocol to prepare or distill a special logical ancilla state is also necessary. 
That is, we have to obtain the fault-tolerant circuits of all the above-mentioned protocols.
The most important things for the circuit mapping over the multiple fault-tolerant protocols are first most protocols act on the same configuration of a logical qubit, and second, during the fault-tolerant quantum computing the protocols are applied repeatedly.
Note that the configuration of a logical qubit indicates the arrangement of physical data qubits in a logical qubit (please see Figure~\ref{fig:steane_initial_mapping}).
For the above considerations, the proposed algorithm shares the configuration of a logical qubit for the circuit synthesis over all the protocols and makes the resulting circuit \emph{self-contained} so that it can act independently regardless of the situation.
The details will be discussed in the main body.

Lastly, as mentioned above, to run a fault-tolerant quantum protocol as designed, we have to include the execution of classical processing implicitly contained in the protocol.
For that, we need implement a single fault-tolerant quantum protocol as multiple small fault-tolerant circuits and run the classical operation between the circuits.
In the present work, we achieve it in two directions.
First, one inserts \emph{barrier} statements in the protocol and the circuit synthesis deals with them.
The position where the statements are inserted is, on considering an ordinary fault-tolerant protocol structure, when all the ancilla qubits that were prepared within the protocol are measured.
Second, we first partition the protocol into several sub-protocols and conduct the circuit synthesis on each sub-protocol individually, but with sharing information common to them.

The proposed circuit synthesis algorithm is based on the heuristic circuit mapping algorithm, SABRE (SWAP-based Bi-directional heuristic search algorithm)~\cite{Li.2019}.
The most important reasons we hire the algorithm for our purpose are first it is possible to easily take the type of quantum information each qubit holds into account to find the qubit move path, and second, as discussed in the literature, it works more efficiently than other global optimization methods.

The remainder of this paper first briefly describes SABRE.
We then discuss the four requirements of the circuits for universal fault-tolerant quantum computing, propose our ideas for them, and describe how to implement the ideas with SABRE.
As an example, we show the full snapshots of the syndrome measurement circuit of $[[7, 1, 3]]$ Steane code in sequence.
We then go into the full set of universal fault-tolerant quantum circuits based on Steane code obtained by applying our algorithm.
As the last example, we show the circuit of the syndrome measurement of $[[23, 1, 7]]$ Golay code that is well known as having a high error correction threshold among the concatenated codes.

\section{Review of SABRE}\label{sec:SABRE}

\begin{figure*}[htbp]
\centering
\includegraphics[scale=0.4]{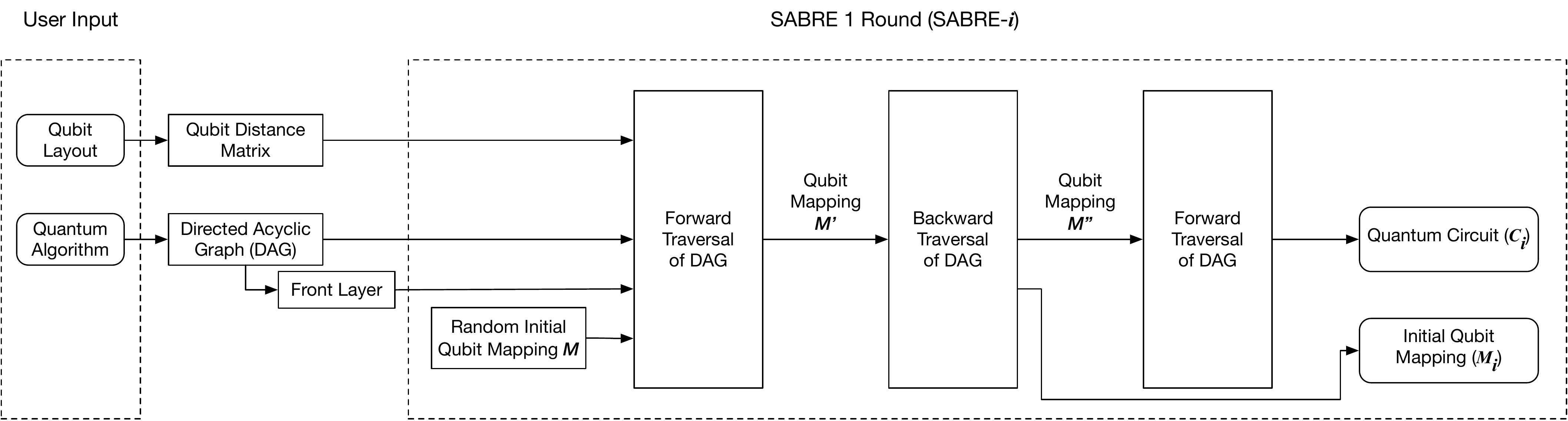}
\caption{
The architecture of SABRE~\cite{Li.2019}. 
The user inputs, \emph{Qubit Layout} and \emph{Quantum Algorithm}, are common to all the iterations.
Usually, \emph{Qubit Layout} is provided as a physical qubit coupling graph of a quantum chip and Quantum Algorithm is given in the QASM format~(\cite{Cross.2017}).
Provided Qubit Layout and Quantum Algorithm, how to make \emph{Qubit Distance Matrix} and \emph{Directed Acyclic Graph} are well described in the literature.
The algorithm is composed of multiple iteration of SABRE graph traversals.
Each iteration begins by picking an initial qubit layout randomly.
After conducting the $i$-th iteration of SABRE (SABRE-$i$), we will have a quantum circuit ($C_i$) and the associated initial qubit mapping table ($M_i$).
The optimal quantum circuit and its associated initial qubit mapping will be obtained by comparing all the results.
}
\label{fig:sabre_architecture}
\end{figure*}

We review the heuristic quantum circuit mapping algorithm SABRE in the present section.
The algorithm generates a quantum circuit believed to be optimal by iterating the procedure composed of picking a random qubit initial mapping and finding the gate sequence based on that mapping as many times as possible.
The standard of the optimality can be flexibly adopted, the number of quantum gates, the circuit depth, or anything.

The main procedure of SABRE which is to find the gate sequence is composed of iterating graph traversals three times in the alternating directions (forward, backward and forward again), where the graph is formed as a directed acyclic graph (\emph{DAG}) from a quantum algorithm. 
For the graph traversals in the forward and backward directions, the graph also should be made for both directions respectively.
Note that a node in DAG includes a quantum instruction of a quantum gate and qubits where the gate acts.
Figure~\ref{fig:sabre_architecture} shows the running flow of SABRE.

In the graph traversal for the forward (backward) direction, the algorithm checks whether each node of the DAG that is generated for the corresponding traversal is executable with respect to the qubit layout of a given quantum chip, and finds required a local optimal \emph{SWAP} quantum gate to make un-executable node executable.
Please note that a quantum gate is called \emph{executable} if it acts locally on the given qubit layout (quantum chip).
An optimal \emph{SWAP} gate is selected from the evaluation of the cost of all the \emph{SWAP} candidate gates which are chosen based on the qubits where the above-mentioned un-executable nodes (gates) act.

The graph traversal is composed of the following two steps.
\begin{enumerate}
\item Find any executable gates in \emph{Front Layer}
\item Append all the executable gates to the circuit or Find a SWAP gate if there does not exist any executable gates
\end{enumerate}
Note that \emph{Front Layer} (FL) is defined as the set of root nodes in the DAG.

If a node in FL, $node_a$, is executable, then it is removed from FL (and DAG) and is appended to the resulting quantum circuit.
In general, the node has succeeding nodes in DAG, and suppose that a node $node_b$ is such one.
Note that $node_b$ belongs to DAG but not in FL.
$node_b$ then has preceding nodes, some of them are already exported to the circuit and the others are not.
If, after exporting $node_a$ to the circuit, none of the preceding nodes of $node_b$ remain in FL, we pull $node_b$ from non-FL to FL.

Otherwise, if none in FL is executable, the algorithm collects several candidate \emph{SWAP} gates and selects an optimal one from the cost evaluation.
Then, by moving qubits via the selected \emph{SWAP} gates, some gates in FL may become executable or close to executable one.
The qubit mapping status is also updated by the \emph{SWAP} operation.
In the literature, two kinds of the cost function are defined, but the heart of both is the same as they evaluate the total qubit move distance to make all the nodes in FL executable.
The graph traversal is terminated if FL becomes empty, that is, when all the nodes of DAG are appended to the quantum circuit.

After each graph traversal is done, the changed qubit mapping table is passed to the next graph traversal in general (see Figure~\ref{fig:sabre_architecture}).
Therefore, the initial qubit mapping and the resulting quantum circuit of the last graph traversal become the qubit mapping and the quantum circuit of a current SABRE round.
The performance of a quantum circuit obtained in each round is affected by the initial qubit mapping chosen randomly.
Therefore, the more iterations, the better performance of a quantum circuit.
\section{Circuit Synthesis for Universal Fault-Tolerant Quantum Computing}\label{sec:requirement}

This section raises four requirements for implementing the quantum circuits for universal fault-tolerant quantum computing.
The first is about how to block the error propagation over the data qubits, the second is about how to make self-contained quantum circuits so that they can be executed independently regardless of situations, and the third is about how to prepare a whole set of fault-tolerant quantum circuits for universal quantum computing.
The last one is about how to execute the fault-tolerant quantum circuits including classical processings in practice.

\subsection{Fault-tolerant qubit move}\label{subsec:fault_tolerance}\label{subsec:condition_1}

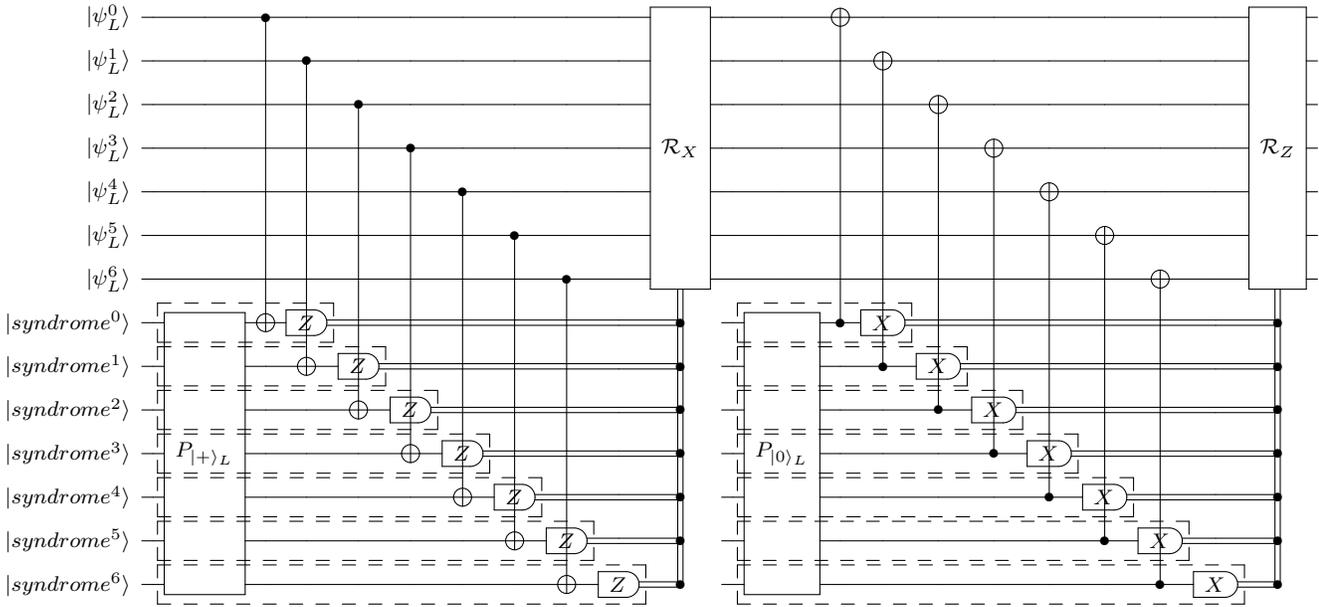
\begin{figure*}[t]
\centerline{ \makebox[0.1\linewidth]{ 
\scriptsize
\Qcircuit @C=0.5 em @R=0.8 em @ !R {
\lstick{|\psi_L^{0}\rangle} 	&\qw &\qw					& \ctrl{7}	& \qw 	&\qw  &\qw &\qw & \qw & \qw & \qw  &\multigate{6}{\mathcal{R}_X} 	&\qw &\qw &\qw & \targ & \qw &  \qw &  \qw &  \qw &  \qw &  \qw &  \qw&  \multigate{6}{\mathcal{R}_Z} &  \qw\\
\lstick{|\psi_L^{1}\rangle} 	&\qw &\qw					& \qw	& \ctrl{7} 	&\qw  &\qw &\qw & \qw & \qw &  \qw &\ghost{\mathcal{R}_X} 		&\qw &\qw & \qw &\qw & \targ &  \qw &  \qw &  \qw &  \qw &  \qw &  \qw &  \ghost{\mathcal{R}_Z} &  \qw\\
\lstick{|\psi_L^{2}\rangle} 	&\qw &\qw					& \qw	& \qw 	&\ctrl{7} &\qw &\qw &\qw & \qw &  \qw &\ghost{\mathcal{R}_X} 		&\qw &\qw & \qw &\qw & \qw &\targ &  \qw &  \qw &  \qw &  \qw &  \qw &  \ghost{\mathcal{R}_Z} &  \qw\\
\lstick{|\psi_L^{3}\rangle} 	&\qw &\qw					&  \qw 	& \qw 	&\qw &\ctrl{7} &\qw & \qw & \qw &  \qw &\ghost{\mathcal{R}_X}	 	&\qw &\qw & \qw &\qw & \qw & \qw &  \targ &  \qw &  \qw &  \qw &  \qw & \ghost{\mathcal{R}_Z} &  \qw\\
\lstick{|\psi_L^{4}\rangle} 	&\qw &\qw					&  \qw	& \qw 	&\qw &\qw &\ctrl{7} & \qw &\qw  &  \qw  &\ghost{\mathcal{R}_X} 	&\qw &\qw & \qw &\qw & \qw & \qw &   \qw & \targ &  \qw &  \qw &  \qw & \ghost{\mathcal{R}_Z} &  \qw \\
\lstick{|\psi_L^{5}\rangle} 	&\qw &\qw					&  \qw	& \qw 	&\qw &\qw &\qw  &\ctrl{7} &\qw &  \qw &\ghost{\mathcal{R}_X} 		&\qw &\qw & \qw &\qw & \qw & \qw &  \qw &  \qw & \targ&  \qw&  \qw  &  \ghost{\mathcal{R}_Z} &  \qw \\
\lstick{|\psi_L^{6}\rangle} 	&\qw &\qw					&  \qw	& \qw 	&\qw &\qw &\qw &\qw &\ctrl{7} & \qw  &\ghost{\mathcal{R}_X}	 	&\qw &\qw & \qw &\qw & \qw & \qw &  \qw &  \qw &  \qw & \targ&  \qw  & \ghost{\mathcal{R}_Z}&  \qw\\
\lstick{|syndrome^0\rangle}  	&\qw & \multigate{6}{P_{|+\rangle_L}}	& \targ 	&\measureD{Z}  &\cw & \cw & \cw & \cw &\cw & \cw &\control \cw \cwx		& &\qw &\multigate{6}{P_{|0\rangle_L}} & \ctrl{-7} & \measureD{X} &  \cw &  \cw &  \cw &  \cw &  \cw &\cw & \control\cw\cwx\\
\lstick{|syndrome^1\rangle}  	&\qw & \ghost{P_{|+\rangle_L}}		&  \qw	&\targ &\measureD{Z} &\cw & \cw & \cw &\cw & \cw &\control \cw \cwx		& &\qw & \ghost{P_{|0\rangle_L}} &\qw &  \ctrl{-7} &  \measureD{X} &  \cw &  \cw &  \cw &  \cw &  \cw & \control\cw\cwx\\
\lstick{|syndrome^2\rangle}  	&\qw & \ghost{P_{|+\rangle_L}}		&  \qw	&\qw &\targ &\measureD{Z} & \cw & \cw & \cw & \cw &\control \cw \cwx		& &\qw & \ghost{P_{|0\rangle_L}} &\qw &  \qw  & \ctrl{-7} &  \measureD{X} &  \cw &  \cw &  \cw &  \cw & \control\cw\cwx\\
\lstick{|syndrome^3\rangle}  	&\qw & \ghost{P_{|+\rangle_L}}		&  \qw	&\qw &\qw & \targ & \measureD{Z} & \cw & \cw & \cw &\control \cw \cwx		& &\qw & \ghost{P_{|0\rangle_L}} &\qw &  \qw &  \qw & \ctrl{-7} &  \measureD{X} &  \cw &  \cw &  \cw & \control\cw\cwx\\
\lstick{|syndrome^4\rangle}  	&\qw & \ghost{P_{|+\rangle_L}}		&  \qw	&\qw &\qw & \qw & \targ & \measureD{Z} & \cw & \cw &\control \cw \cwx	 	& &\qw & \ghost{P_{|0\rangle_L}} &\qw &  \qw &  \qw &\qw & \ctrl{-7}&  \measureD{X} &  \cw &  \cw  & \control\cw\cwx\\
\lstick{|syndrome^5\rangle}  	&\qw & \ghost{P_{|+\rangle_L}}		&  \qw	&\qw &\qw & \qw & \qw & \targ & \measureD{Z} & \cw &\control \cw \cwx	 	& &\qw & \ghost{P_{|0\rangle_L}} &\qw &  \qw &  \qw & \qw &  \qw & \ctrl{-7}&  \measureD{X} &  \cw & \control\cw\cwx\\
\lstick{|syndrome^6\rangle}  	&\qw & \ghost{P_{|+\rangle_L}}		&  \qw	&\qw &\qw & \qw & \qw & \qw & \targ &  \measureD{Z} &\control \cw \cwx	 	& &\qw & \ghost{P_{|0\rangle_L}} &\qw &  \qw &  \qw & \qw & \qw &  \qw & \ctrl{-7} & \measureD{X}  & \control\cw\cwx
\gategroup{8}{3}{8}{5}{0.6em}{--} \gategroup{8}{15}{8}{17}{0.6em}{--}
\gategroup{9}{3}{9}{6}{0.6em}{--} \gategroup{9}{15}{9}{18}{0.6em}{--}
\gategroup{10}{3}{10}{7}{0.6em}{--} \gategroup{10}{15}{10}{19}{0.6em}{--}
\gategroup{11}{3}{11}{8}{0.6em}{--} \gategroup{11}{15}{11}{20}{0.6em}{--}
\gategroup{12}{3}{12}{9}{0.6em}{--} \gategroup{12}{15}{12}{21}{0.6em}{--}
\gategroup{13}{3}{13}{10}{0.6em}{--} \gategroup{13}{15}{13}{22}{0.6em}{--}
\gategroup{14}{3}{14}{11}{0.6em}{--} \gategroup{14}{15}{14}{23}{0.6em}{--}\\
}}}
\caption{Protocol for the syndrome measurement based on steane QEC method~\cite{Steane.1997}.
The dotted boxes for the ancilla qubit $|anc^i\rangle$ are its activation periods.
The outsides of the boxes are the inactivation periods.
In the forward direction graph traversal, the usage status of each qubit becomes activated by preparation and inactivated by measurement. 
On the other hand, in the backward direction, the measurement (preparation) operation makes the status of each qubit activated (inactivated).
In ideal situation, all the ancilla qubits have the same activation period, but in practice due to the locality, some ancilla qubits have longer activation period than others.
Please note that the data qubits $|\psi^{i}_{L}\rangle$ are all activated.
} 
\label{fig:steane_syndrome_measure_full}
\end{figure*}

Previously, we have mentioned that a protocol has the fault-tolerance if the number of quantum errors happened on data qubits is bounded by the capacity of quantum error correction.
In this section, we first propose an idea to block the error propagation over the data qubits and then discuss how to implement the method into SABRE.
A logical qubit encoded by an $[[n, k, d]]$ quantum error-correcting code contains $n$ data qubits.
Besides, in practice, several ancilla qubits for storing error syndrome and verifying the specially prepared ancilla state are additionally needed.
Please see the syndrome measurement protocol in Figure~\ref{fig:steane_syndrome_measure_full}.
There are 7 data qubits $|\psi\rangle_L^{i}$ and 7 syndrome qubits $|syndrome^{i}\rangle$.
Even though it is not explicitly described there, additional qubits are required to verify the prepared logical states $|+\rangle_L$ and $|0\rangle_L$ (see Ref.~\cite{Goto.2016}).
Therefore, by assuming the reuse of the ancilla qubits, to run the protocol we need at least 15 qubits.

Suppose that the qubits $|\psi\rangle^{0}_L$ and $|syndrome^{0}\rangle$ are geometrically adjacent.
Then, the first \emph{CNOT} gate in Figure~\ref{fig:steane_syndrome_measure_full} is executable.
Otherwise, to resolve the locality constraint imposed on the quantum device, we need to move qubits to make them placed in the neighbor.
What here we need to be careful of is that all the quantum gates exploited in quantum computing are noisy!
If a quantum error happens during running a noisy \emph{SWAP} gate acting on two data qubits, both the qubits will be corrupted as
\begin{eqnarray} \nonumber
\widetilde{U_{SWAP}} |\psi\rangle|\phi\rangle &=& (I+\epsilon_2) \cdot U_{SWAP}|\psi\rangle |\phi\rangle \\ \nonumber
&=& |\widetilde{\phi}\rangle|\widetilde{\psi}\rangle,
\end{eqnarray}
where $\epsilon_2$ is an arbitrary 2-qubit error and $U_{SWAP}$ is an ideal noiseless \emph{SWAP} gate.
Therefore, in the circuit synthesis for a fault-tolerant quantum protocol, we have to avoid or limit introducing \emph{SWAP} gates between data qubits for moving the qubits.
Note that for computing algorithms, this is out of concern.

We now discuss how to choose \emph{SWAP} gates to implement an executable circuit and not to spread quantum errors over data qubits at the same time.
The physical qubits constituting a logical qubit have their own lifetime within fault-tolerant quantum computing.
For example, the data qubits hold the quantum data for quantum computing, and therefore their lifetimes are equal to that of the logical qubit.
On the other hand, the syndrome qubits only work in the non-trivial way for the duration of the ancilla preparation and the syndrome measurement.
Therefore, their lifetimes are the duration of those operations.
Then, for the rest, the quantum state they hold is not critical to quantum computing, and therefore they can be played as the communication channel.

We trace the usage status of each qubit within the fault-tolerant quantum protocol.
If a qubit is prepared as a certain quantum state $|0\rangle$ (or $|+\rangle$), the usage status of the qubit turns into \emph{activated}.
On the other hand, its status becomes \emph{inactivated} if a qubit is measured.
During running a fault-tolerant protocol, the usage status of a qubit is changed along the working flow of the protocol.
When activated status, a qubit holds significant quantum state for the protocol, but when inactivated it just stores a kind of garbage information.
Please see Figure~\ref{fig:steane_syndrome_measure_full}.

Regarding the usage status, we classify qubits into data-type qubits and non-data-type qubits.
A \emph{data-type} qubit is defined as a qubit in the activated status, and a \emph{non-data-type} qubit is one in the inactivated.
As mentioned above, a data-type qubit carries a non-trivial quantum information, but a non-data-type qubit holds a kind of garbage information.
Therefore, we have to block or limit \emph{SWAP} gate between the data-type qubits in the circuit synthesis.

As mentioned above, it is not that the \emph{SWAP} gates between the data-type qubits are never allowed.
In general, according to the code distance $d$, it is possible to allow the \emph{SWAP} gates as much as $\lfloor (d-1)/4 \rfloor$ times because arbitrary $\lfloor (d-1)/2 \rfloor$ errors can be recovered.
For example, since a quantum code of $d=5$ can correct arbitrary 2-qubit errors, one-time noisy \emph{SWAP} gate between data qubits does not break a logical qubit.

So far, we have discussed the requirement of how to find the fault-tolerant qubit move by limiting the \emph{SWAP} gates between the data-type qubits.
Then, it is time to discuss how to implement the method into SABRE.
First of all, we need to recall of SABRE is composed of three graph traversals in alternating directions.
Previously, we mentioned that the usage status of a qubit becomes activated (inactivated) if it is prepared (measured).
Such a rule is based on the normal execution flow of a quantum circuit, that is the forward traversal of DAG.
If a quantum circuit is executed in the reverse direction then the status change is triggered by the opposite operation.
That is, in the backward traversal of DAG, the status is turned into activated (inactivated) by measurement (preparation).

We have mentioned that the original SABRE collects all the possible \emph{SWAP} gates that act on the qubits of the nodes in FL, and then picks up the optimal one from them.
But, in the proposed algorithm, we selectively collect the \emph{SWAP} candidates to limit the \emph{SWAP} gates on the data-type qubits.
For that, we first examine the type of the qubits.
There are three kinds of qubit pairs: 1) both non-data-type qubits, 2) one data-type qubit and one non-data-type qubit, and 3) both data-type qubits.
The first and second cases, needless to say, can be collected as the \emph{SWAP} candidates, but for the third one, we need to do something described in what follows.

\begin{figure}[t]
\centering
\includegraphics[scale=0.3]{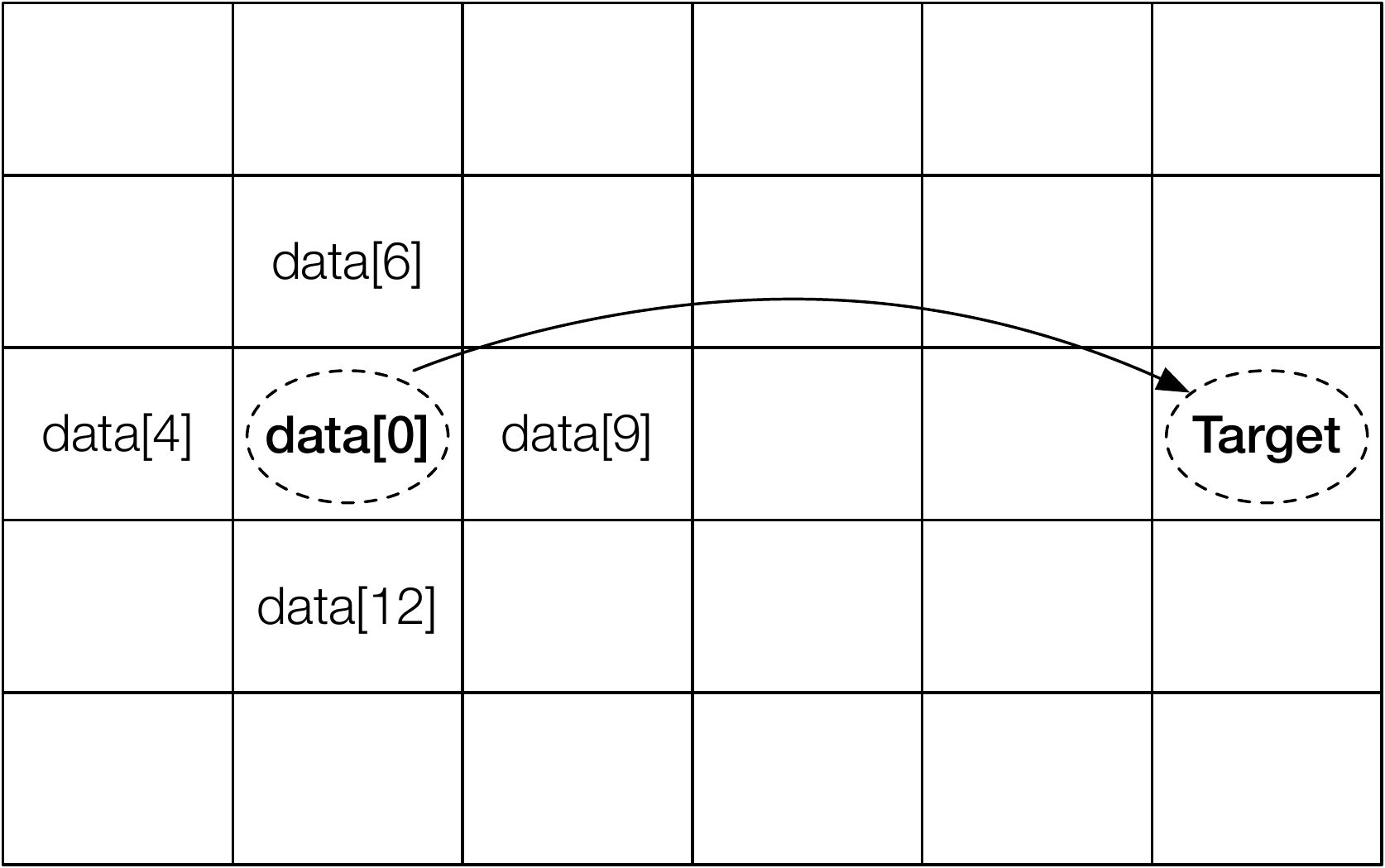}
\caption{
Since the data qubit $data$[0] is surrounded by data qubits, it can not be moved to another physical qubit in a fault-tolerant way.
}
\label{fig:traffic_jam_blocked}
\end{figure}

\begin{figure*}[t]
\centering
\subfigure[]{
	\includegraphics[scale=0.35]{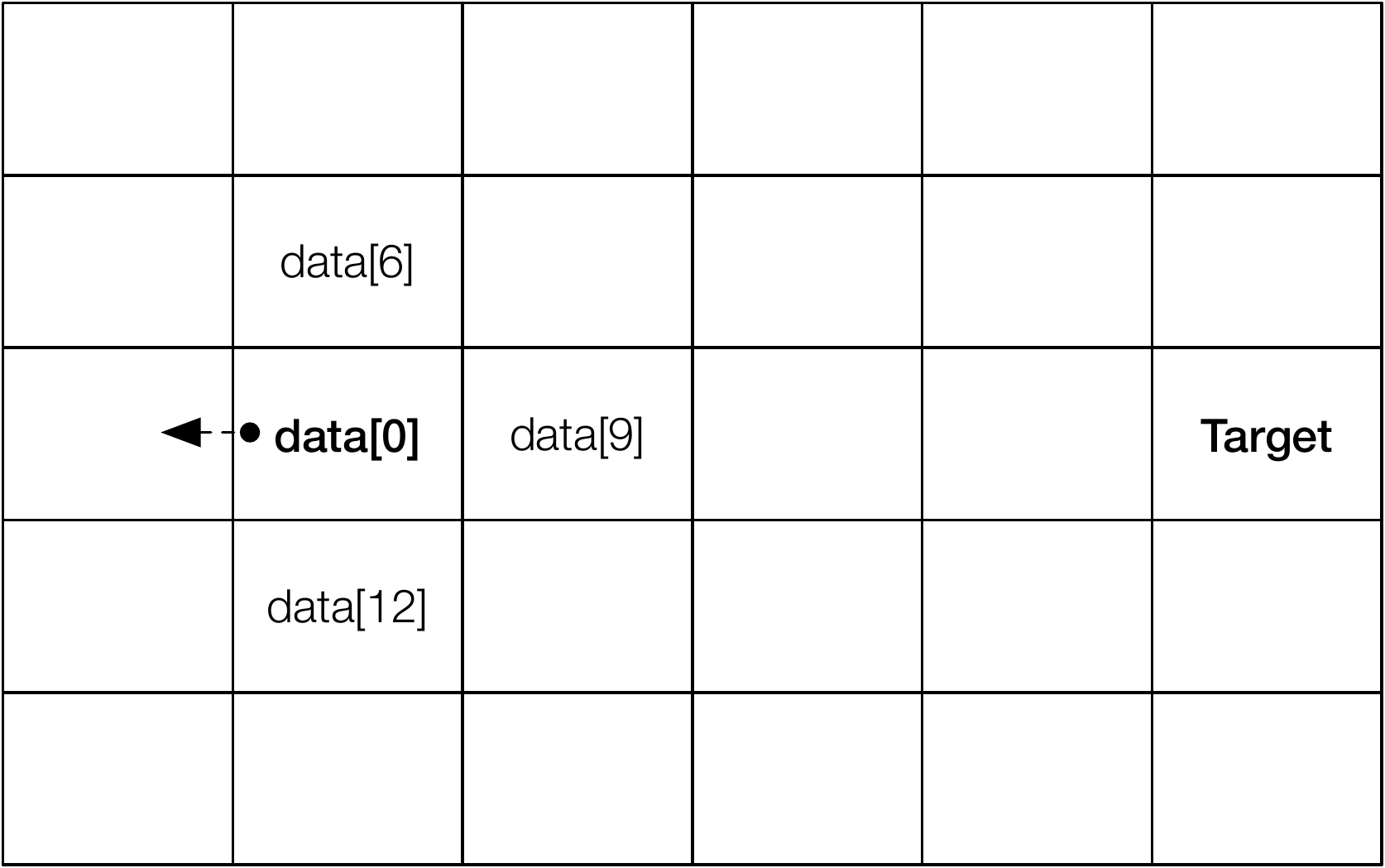}
}
\subfigure[]{
	\includegraphics[scale=0.35]{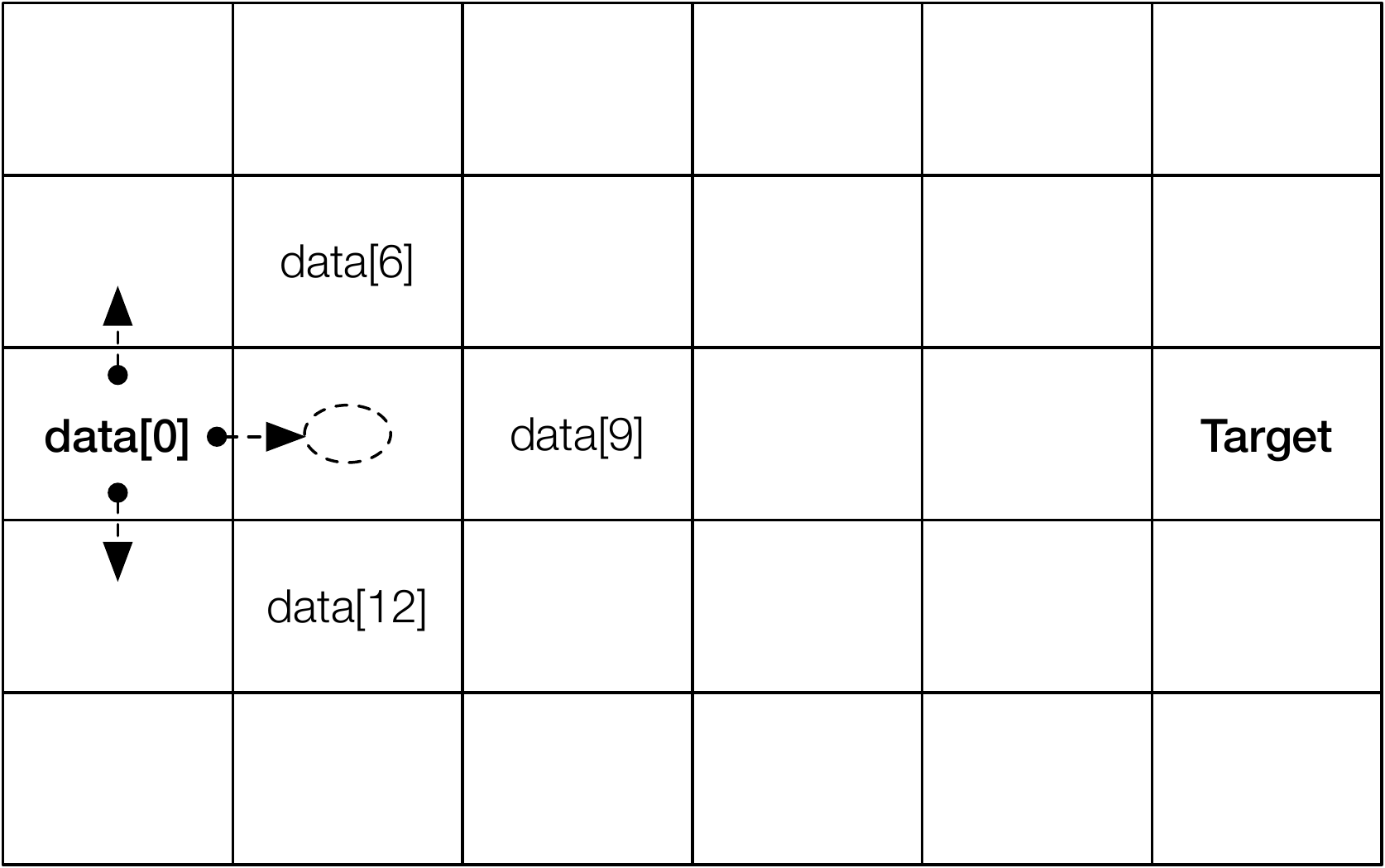}
}
\caption
{
A circuit synthesis falls in an infinite loop when a data qubit surrounded by other data qubits is moved to another place.
All the cells indicate physical qubits.
(a) $data$[0] has only one option for the next move, which is the left cell of which.
(b) The cost function based on the distance between qubits selects the $b$ cell for the next move.
}
\label{fig:traffic_jam_infinite_loop}
\end{figure*}

\begin{figure*}[t]
\centering
\subfigure[]{
	\includegraphics[scale=0.35]{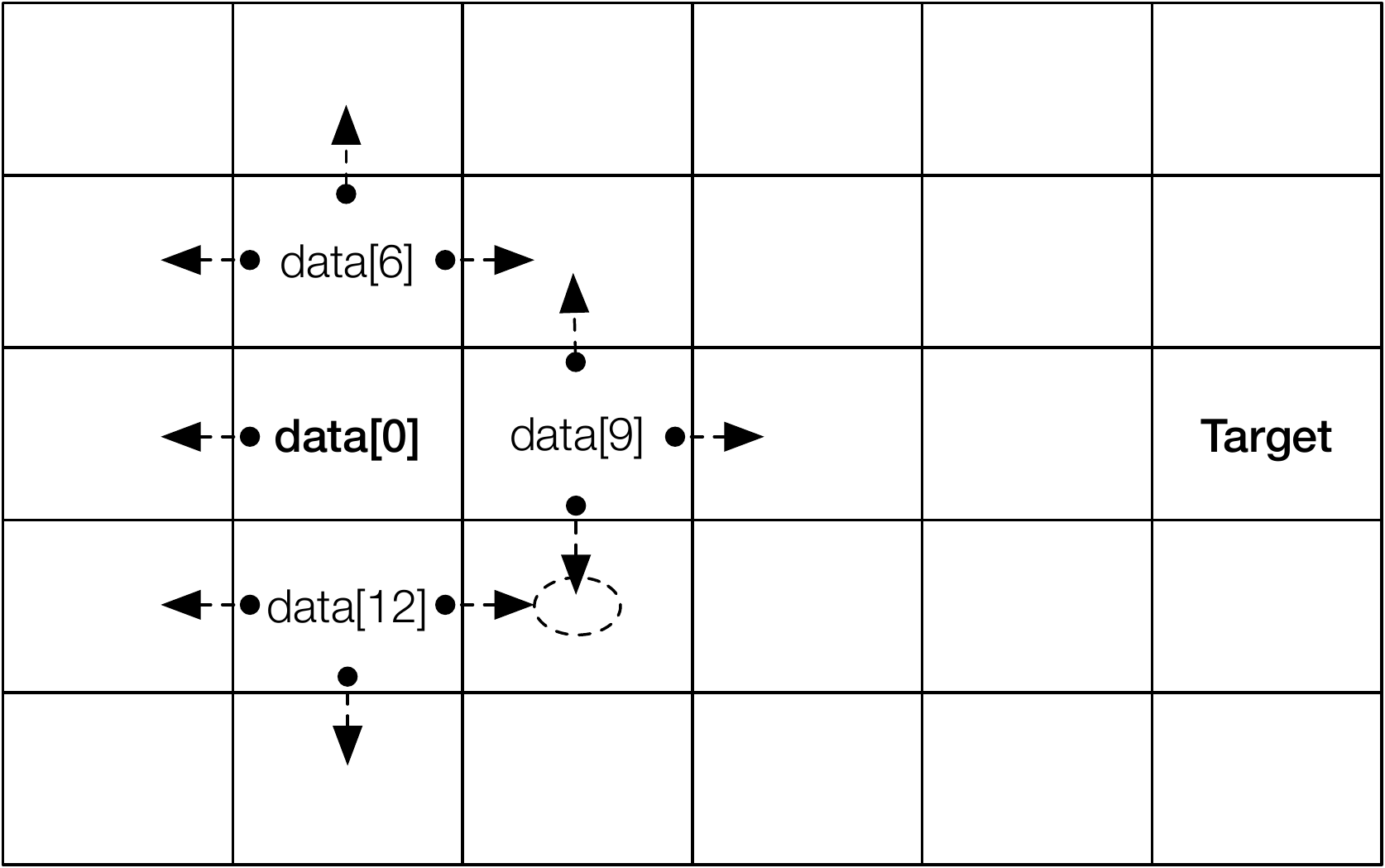}	
}
\subfigure[]{
	\includegraphics[scale=0.35]{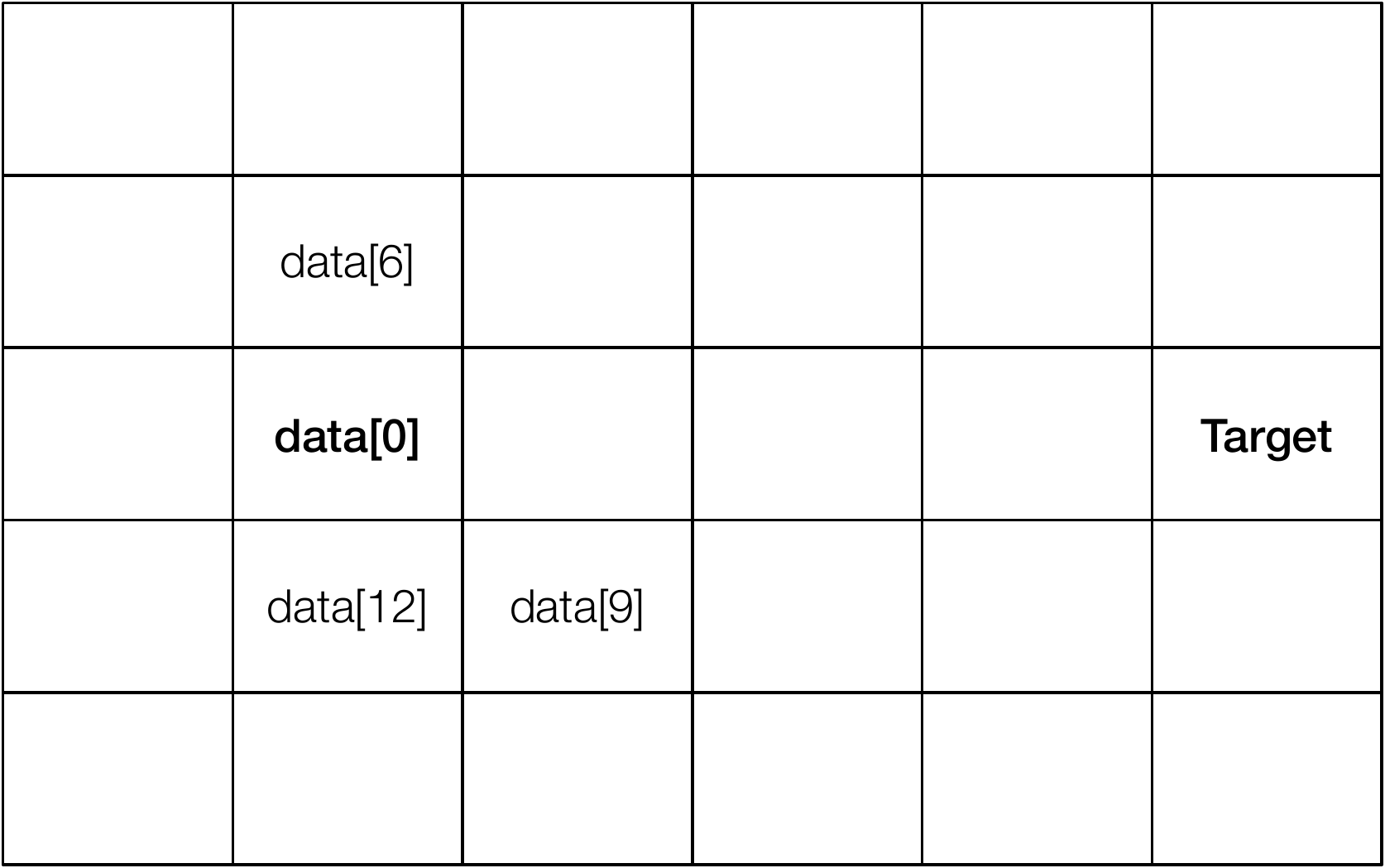}	
}
\subfigure[]{
	\includegraphics[scale=0.35]{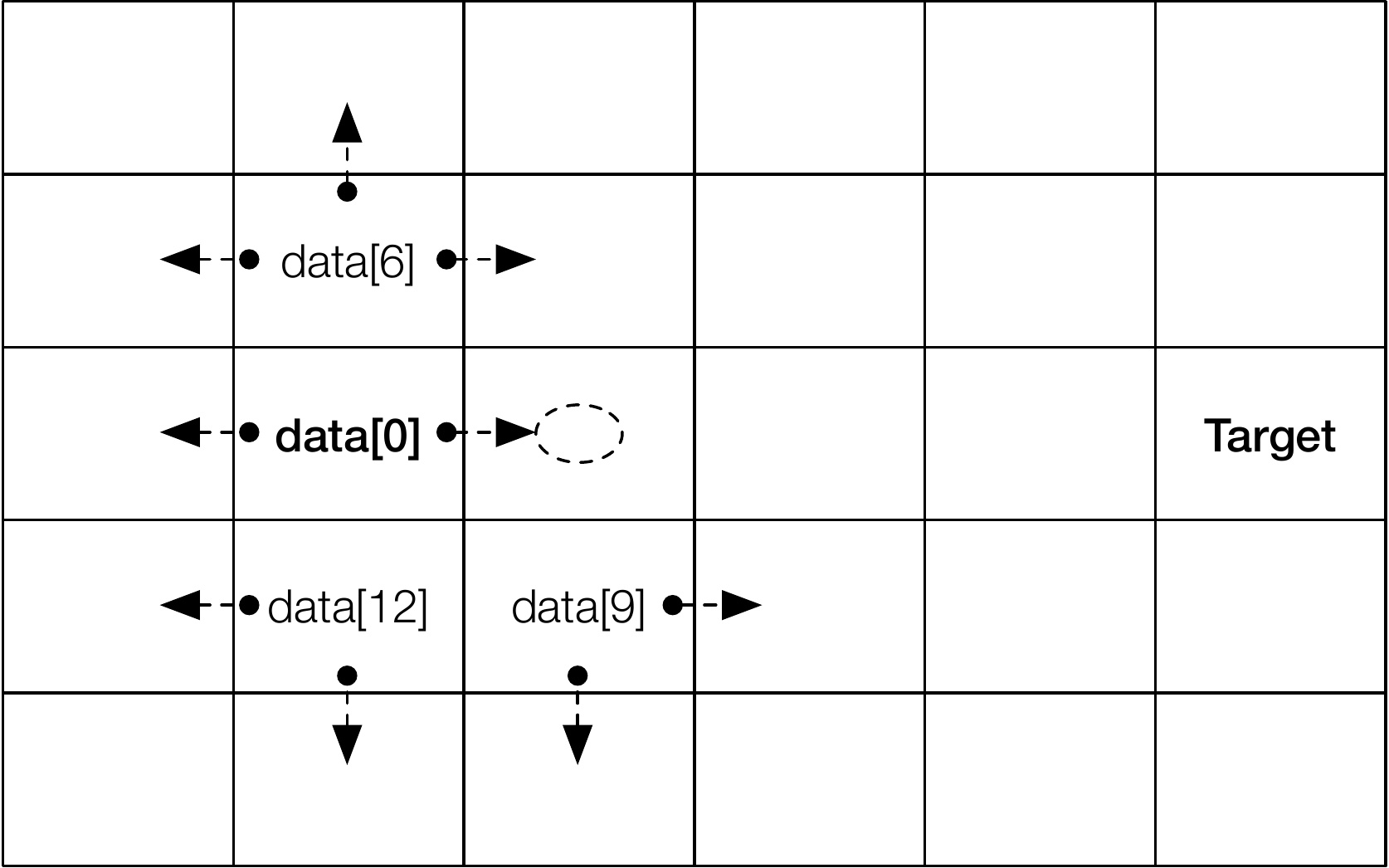}	
}
\subfigure[]{
	\includegraphics[scale=0.35]{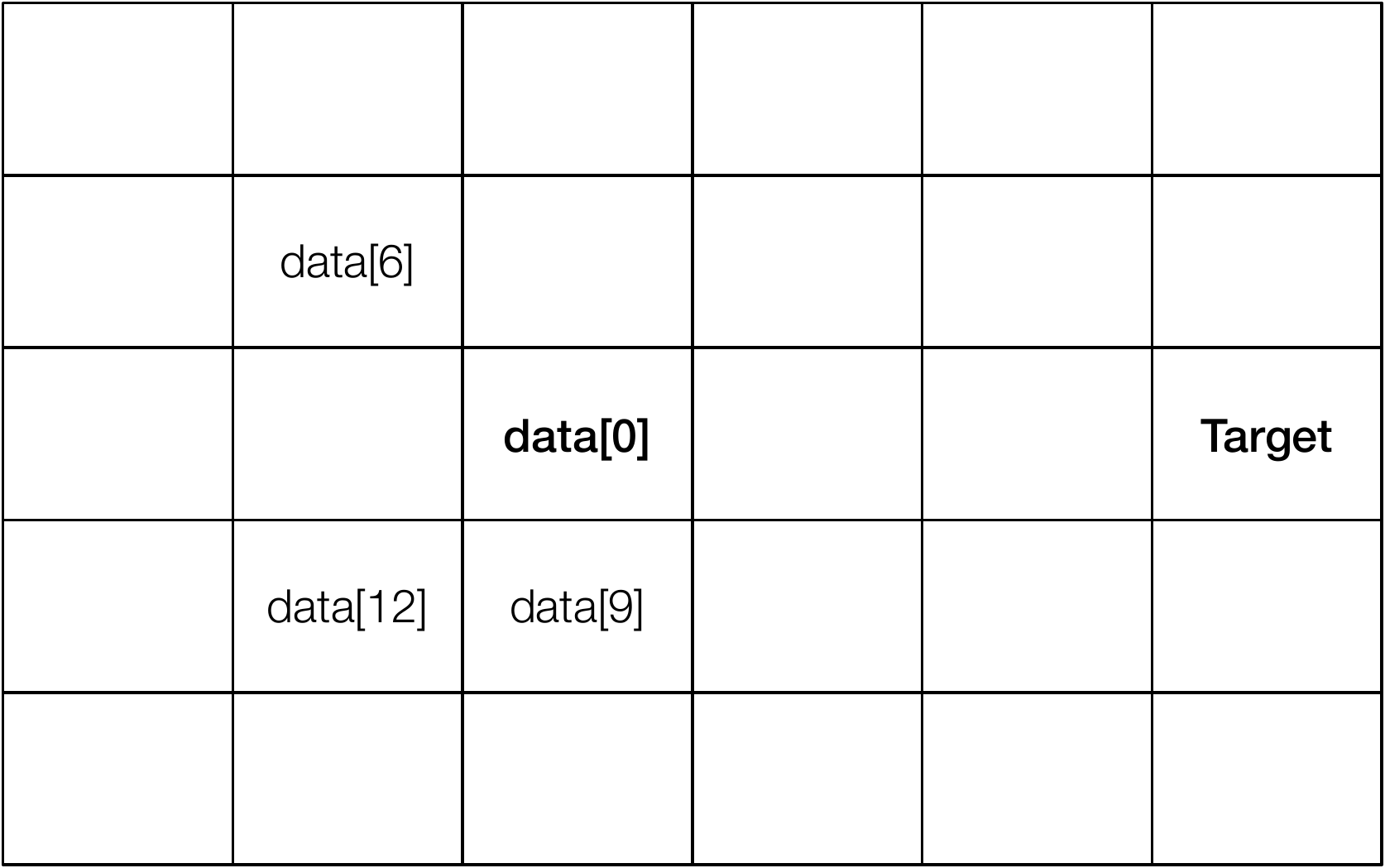}	
}
\caption
{
By first moving the data-type qubits located around the qubit that has to move to the destination, the traffic jams shown in Figures~\ref{fig:traffic_jam_blocked} and~\ref{fig:traffic_jam_infinite_loop} can be relieved and it becomes possible to find a fault-tolerant movement path.
(a) Collect \emph{SWAP} gates from both the data-type qubit and its neighbor data-type qubits.
(b) Based on the cost evaluation, \emph{data[9]} is moved to its neighbor cell of the dotted circle.
(c) Collect \emph{SWAP} gates from both the data-type qubit and its neighbor data-type qubits.
(d) Based on the cost evaluation, \emph{data[0]} moves toward the destination \emph{Target} without interaction with any data-type qubits.
}
\label{fig:traffic_jam_relieved}
\end{figure*}

Suppose that for some reason (which will be discussed later in detail) we need to move the quantum state a data-type qubit holds to another qubit. 
Figure~\ref{fig:traffic_jam_blocked} shows that since $data$[0] is currently surrounded by data-type qubits, the quantum state in $data[0]$ cannot be moved to $Target$ without interaction with one data-type qubit.
Figure~\ref{fig:traffic_jam_infinite_loop} shows another situation where finding a fault-tolerant movement path may fall in an infinite loop.
To relieve such traffic jams, we need to move other data-type qubits around $data$[0] to somewhere beforehand.
We then are able to move $data$[0] to $Target$ without interaction with any data-type qubits.
In this regard, we collect the \emph{SWAP} gates acting on the qubits that are positioned next to the data-type qubit $data[0]$ as \emph{SWAP} candidates.
Figure~\ref{fig:traffic_jam_relieved} shows the effect of this rule.

Previously, we mentioned that the \emph{SWAP} gate between both the data-type qubits can be allowed as much as $\lfloor (d-1)/4 \rfloor$ times.
For that, in the beginning, we designate the number of maximally allowed interactions between data-type qubits as $allowable\_max\_interaction=\lfloor (d-1)/4 \rfloor$, where $d$ is the code distance.
Then, when collecting the \emph{SWAP} candidates we can add a \emph{SWAP} gate acting on a pair of both data-type qubits if the number of the allowed interaction so far is less than the predetermined limit.
Later, after cost evaluation on the \emph{SWAP} candidates, if the \emph{SWAP} gate acting on both the data-type qubits is chosen as an optimal one, we then increment the number of the allowed interaction by 1.

The quantum circuit generated by the rule we have mentioned so far includes very limited interactions between data-type qubits.
Therefore, a quantum error that happened randomly does not propagate to data qubits beyond the capacity of the quantum error correction.
Then, we can say that the circuit works in the fault-tolerant manner.
However, it is not enough for universal fault-tolerant quantum computing.
In the following subsections, we raise additional requirements and propose our solutions for them.

\subsection{Self-contained quantum circuit}\label{subsec:condition_2}

Suppose that you have a fault-tolerant quantum circuit and its initial qubit mapping table of the syndrome measurement protocol by applying the method described in the previous section.
Then, it is possible to measure the error syndrome by running the circuit to a logical qubit if the logical qubit is configured like the initial qubit mapping table.
Otherwise, if the logical qubit is not formed as the mapping table, the circuit does not work as intended.
Needless to say that the quantum circuit should be consistent with the initial qubit mapping.

\begin{figure*}[t]
\centering
\includegraphics[scale=0.5]{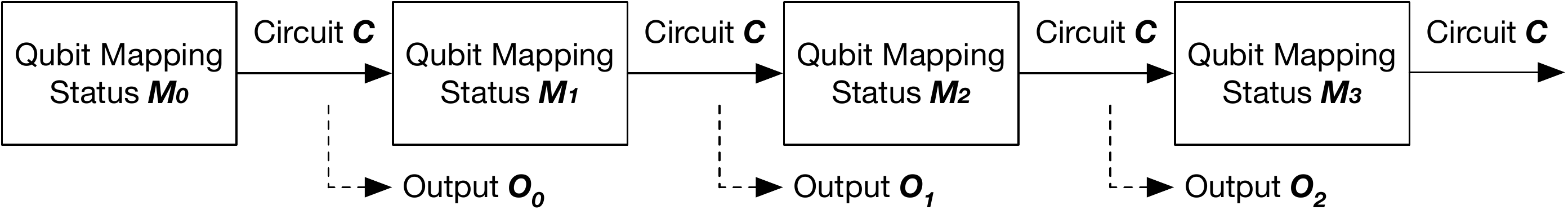}	
\caption{
The repetitive execution of a quantum circuit including SWAP gates introduced by circuit mapping does not work as expected because of the qubit mapping problem, ($O_{0}\neq O_{1}\neq O_{2}=\cdots$).
}
\label{fig:non_self_contained_circuit}
\end{figure*}

Fault-tolerant quantum computing performs error correction periodically.
Then, does the fault-tolerant circuit obtained above always work correctly for the fault-tolerant quantum computing?
Unfortunately, it does not.
The circuit works correctly for one time only, the first time.
The reason is as follows.
If a quantum circuit includes \emph{SWAP} gates introduced by the circuit synthesis, by executing the circuit the qubit mapping will become different from the initial mapping.
Therefore, running the circuit again on the logical qubit will make the output different than expected because the initial mapping for the circuit is not fulfilled.
Please see Figure~\ref{fig:non_self_contained_circuit}.
Please note that \emph{SWAP} gates originally included in the protocol itself do not raise such problems.

Therefore, to make the syndrome measurement circuit work always correctly throughout the fault-tolerant quantum computing, after obtaining the syndrome values we have to move the physical qubits in the logical qubit back to their initial positions.
In the present work, we call the circuit that always works correctly regardless of the situation a \emph{self-contained} circuit.
For that, our algorithm automatically moves the data qubits to the designated positions after the main body of a fault-tolerant protocol.
Please note that since the lifetimes of the ancilla qubits are bounded by a protocol, we only focus on the data qubit (not data-type qubit) for the self-contained property.

To implement the move operation into SABRE, we automatically add the instruction ``\emph{Move} $data[i]$ $destination[i]$" for all the data qubits before generating DAG (see Figure~\ref{fig:sabre_architecture}).
The first argument $data[i]$ is the name of a data qubit used in the protocol, and the second argument $destination[i]$ is the destination of the move.
$destination[i]$ can be a specific physical qubit index or a symbolic value.
However, even if it is provided as a symbolic value in the beginning it is translated to a specific index as the circuit mapping proceeds.

We now describe how to deal with the operation \emph{Move} in the proposed algorithm.
SABRE basically examines whether a quantum instruction in FL is executable or not based on the geometric locality.
In the case of a \emph{CNOT} (or \emph{SWAP}) gate, it determines as executable if a control qubit is positioned next to a target qubit.
On the other hand, for the $Move$, we check whether the argument qubit is located at the designated position or not.
If the \emph{Move} is executable, then the algorithm does not do anything after removing it in FL.
Note that if a \emph{CNOT} gate is executable, then it is exported to the circuit.

In this work, the operation that moves all the data qubits to the designated position is called \emph{Move-Back} because the target position is usually its initial location.
The necessity of the \emph{Move-Back} depends on a fault-tolerant protocol.
Here we describe the details of the \emph{Move-Back} for the syndrome measurement protocol, and other protocols will be discussed in the following section.

\begin{figure*}[t]
\centering
\includegraphics[scale=0.4]{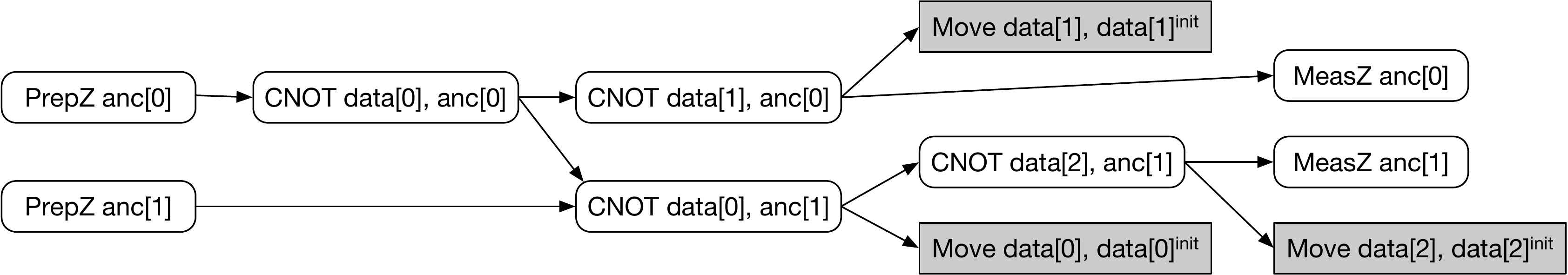}
\caption{
DAG of 3-qubit code syndrome measurement circuit. 
The \emph{Move} instructions for all the data qubits are included at the last.
}
\label{fig:DAG_move}
\end{figure*}

\begin{figure*}[t]
\centering
\subfigure[]{
	\includegraphics[scale=0.45]{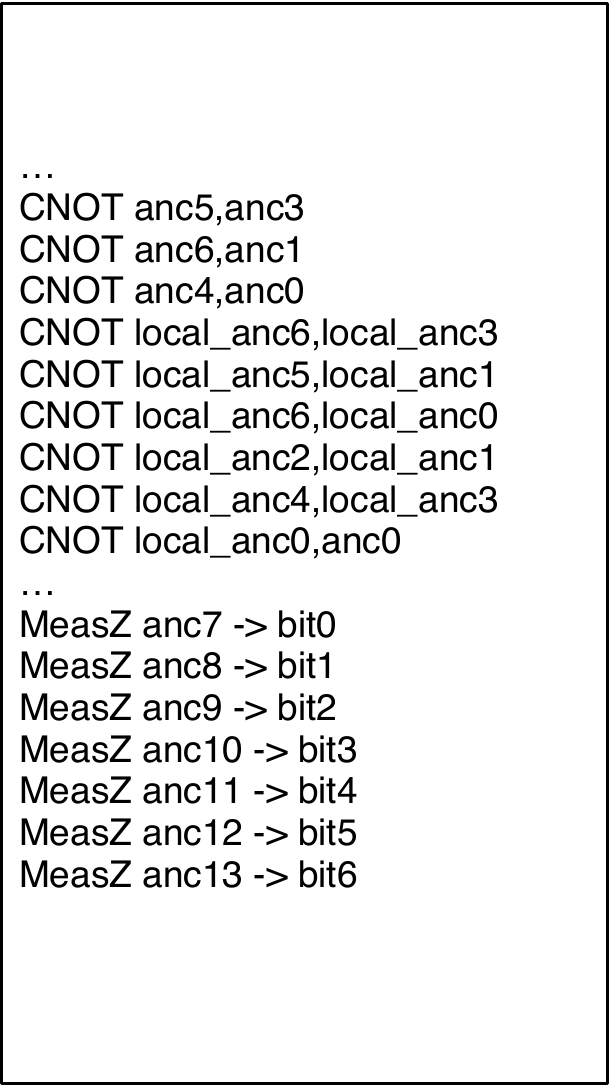}
}
\subfigure[]{
	\includegraphics[scale=0.45]{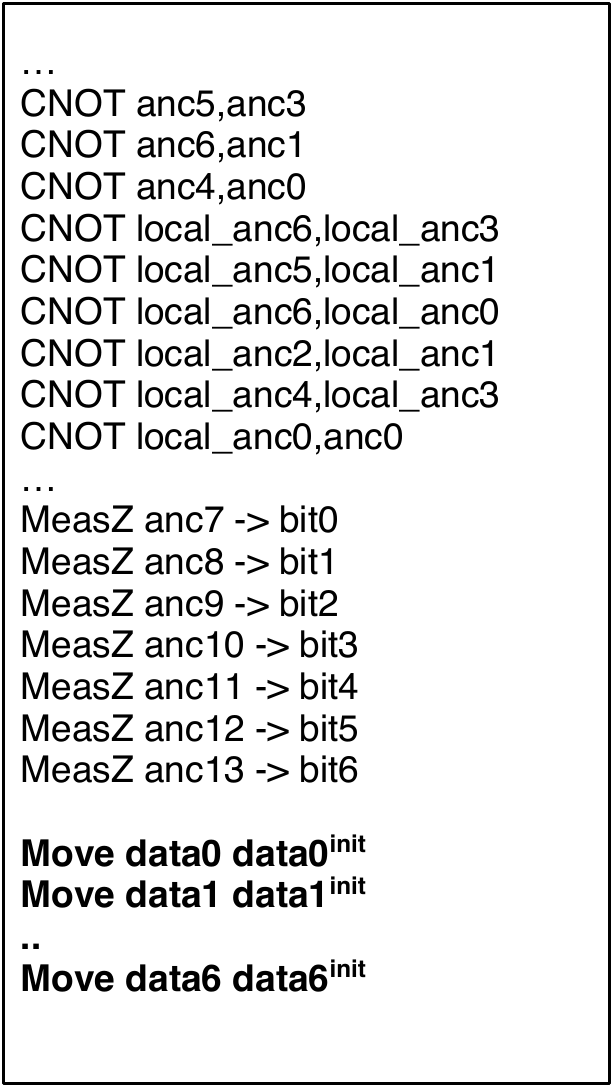}	
}
\subfigure[]{
	\includegraphics[scale=0.45]{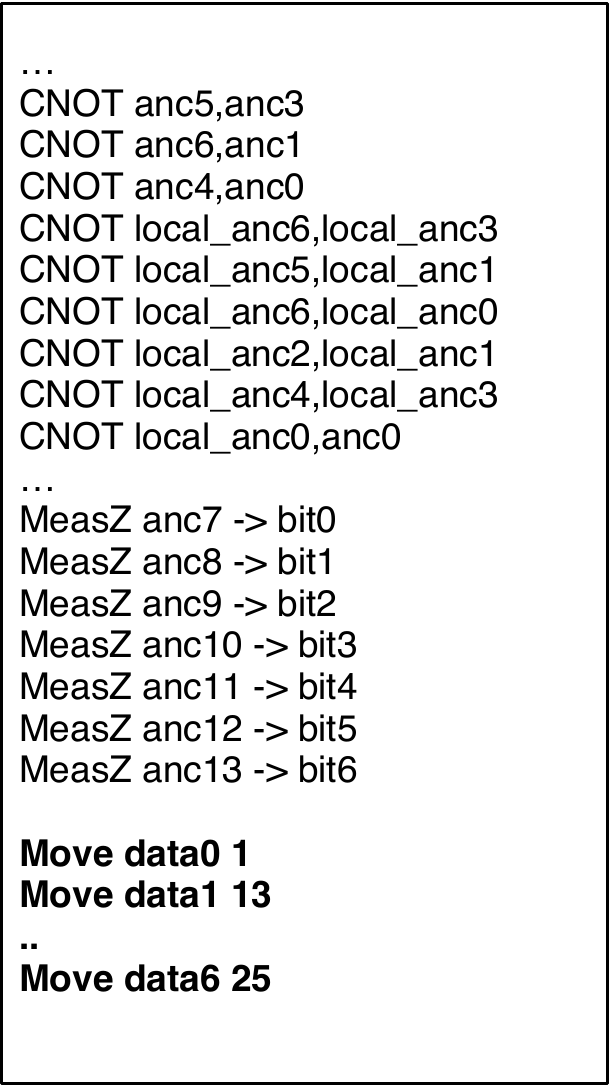}	
}
\caption
{
By adding the \emph{Move} instruction, conceptually the QASM for the protocol is transformed from (a) $\rightarrow$ (b) $\rightarrow$ (c) during the circuit synthesis.
(a) An example of a normal QASM, (b) The QASM with \emph{Move} instruction with the symbolic representation of the initial position of the data qubits and (c) The QASM with a specific value of the initial position of the data qubits.
As mentioned above, at the beginning of each SABRE iteration, the initial qubit mapping table is randomly generated.
The specific initial position of the data qubits, e.g. as shown in the figure \{$\cdots$, $data0^{init}$: 1, $data1^{init}$: 13, .., $data6^{init}$: 25, $\cdots$\}, is determined from the qubit mapping table.
}
\label{fig:qasm_moveback}
\end{figure*}

In the present work, we assume that after the syndrome measurement a logical gate acts on a logical qubit without additional operation.
For that, the arrangement of all the component qubits of a logical qubit after the syndrome measurement should be the same as the initial one of a logical qubit.
Therefore, after the main body of the syndrome measurement, a data qubit $data[i]$ of a logical qubit should be moved to its original location where it occupies in the logical qubit.
Since, when the circuit synthesis is conducted for the syndrome measurement, the specific location of $data[i]$ is not determined yet, it is described symbolically as $data[i]^{init}$ (see Figure~\ref{fig:DAG_move}).
In each SABRE iteration, an initial qubit mapping is randomly picked (see Figure~\ref{fig:sabre_architecture}), and therefore the initial position of $data[i]$ is determined at that time.
Therefore, after picking a random initial qubit mapping, the instruction ``\emph{Move} $data[i]$ $data[i]^{init}$" is translated to the instruction including a specific qubit index, for example ``\emph{Move} $data[i]$ $3$".
The circuit mapping algorithm then tries to move $data[i]$ from a current location to the physical qubit of index 3.
Figure~\ref{fig:qasm_moveback} shows the conceptual transformation of the protocol description during the circuit synthesis.

Sometimes, it may happen that the qubits already arrived at the designated positions may be moved to other places during treating the move of other qubits.
When such a situation happens, we insert the \emph{Move} about the qubit into FL forcibly and treat it again.
Since the \emph{Move-Back} is usually conducted at the last in the circuit, such forcible insertion into FL does not corrupt the logic of the circuit.
If all the data qubits arrive at the target positions, the \emph{Move-Back} for all the data qubits is completed.

We are now able to make a self-contained fault-tolerant quantum circuit for a single fault-tolerant quantum protocol.
However, it is still not enough for universal fault-tolerant quantum computing yet.
In the following subsection, we will see that which should be considered more for universal fault-tolerant quantum computing.

\subsection{Circuits for universal fault-tolerant quantum computing}\label{subsec:condition_3}

The third consideration for universal fault-tolerant quantum computing is that we need multiple fault-tolerant quantum computing protocols (and therefore circuits), for example, \emph{H}, \emph{T}, \emph{CNOT}, \emph{Encoder}, and Measurement in the \emph{Z} basis (\emph{MeasZ}).
All the protocols are differently designed for their own purpose but should act on the input having physically the same qubit arrangement, a logical qubit.
For that, the circuit synthesis of each protocol has to share the position of data qubits of a logical qubit.
Otherwise, if we focus on optimizing the circuit of each protocol separately, the initial qubit mapping for the circuit may be different from others, and therefore we need to perform the additional qubit movement conditioned on the situation (the last circuit and the next circuit).

To make all the circuits share the positions of the data qubits, we first need to make a reference initial qubit mapping for the protocols.
For that, we first divide all the protocols into two categories, \emph{pivot} protocols and \emph{non-pivot} protocols, and perform the circuit mapping on the pivot protocols first.
We then, in the circuit mapping on the non-pivot protocols, anchor the positions of the data qubits by following the results of the circuit mapping on pivot protocols when choosing a random initial mapping.

Then, with which criterion, the protocols can be categorized?
As mentioned above, the circuit of a pivot protocol is generated from the scratch without any limitation on the qubit allocation and therefore it can be highly optimized by the circuit mapping method itself.
But, the circuit of a non-pivot protocol is limited by the result of the pivot protocol.
In this circumstance, for better performance, it is reasonable to choose the most frequently executed protocols as the pivot, and such protocol is, needless to say, the syndrome measurement.
We therefore configure the logical qubit, the arrangement of physical qubits, from the initial qubit mapping of the circuit mapping on the syndrome measurement.

We first discuss 1-qubit logical operations: \emph{Encoder}, \emph{MeasZ}, and \emph{H} gate.
For the operations that can be implemented as transversal, the circuit mapping is very trivial.
Therefore, here we discuss how to implement the fault-tolerant circuit about \emph{Encoder}.
Encoder is the operation that makes the quantum state of a logical qubit, a code block, to a logical zero state $|0\rangle_L$ or a logical plus state $|+\rangle_L$ where $|+\rangle_L = (|0\rangle_L + |1\rangle_L)/\sqrt{2}$.
Before Encoder, all the component qubits of a logical qubit (or a code block) are not distinguished and have garbage quantum states.
However, after \emph{Encoder}, the data qubits of a logical qubit should be placed as determined in the circuit synthesis of the syndrome measurement .
See Figure~\ref{fig:steane_initial_mapping} as an example.
Therefore, after the main body of the \emph{Encoder}, the \emph{Move-Back} operation should be performed and the destination for each data qubit is, as mentioned above, specified from the configuration of a logical qubit not its initial position before $Encoder$.

\begin{figure*}[t]
\centering
\subfigure[]{
	\includegraphics[scale=0.39]{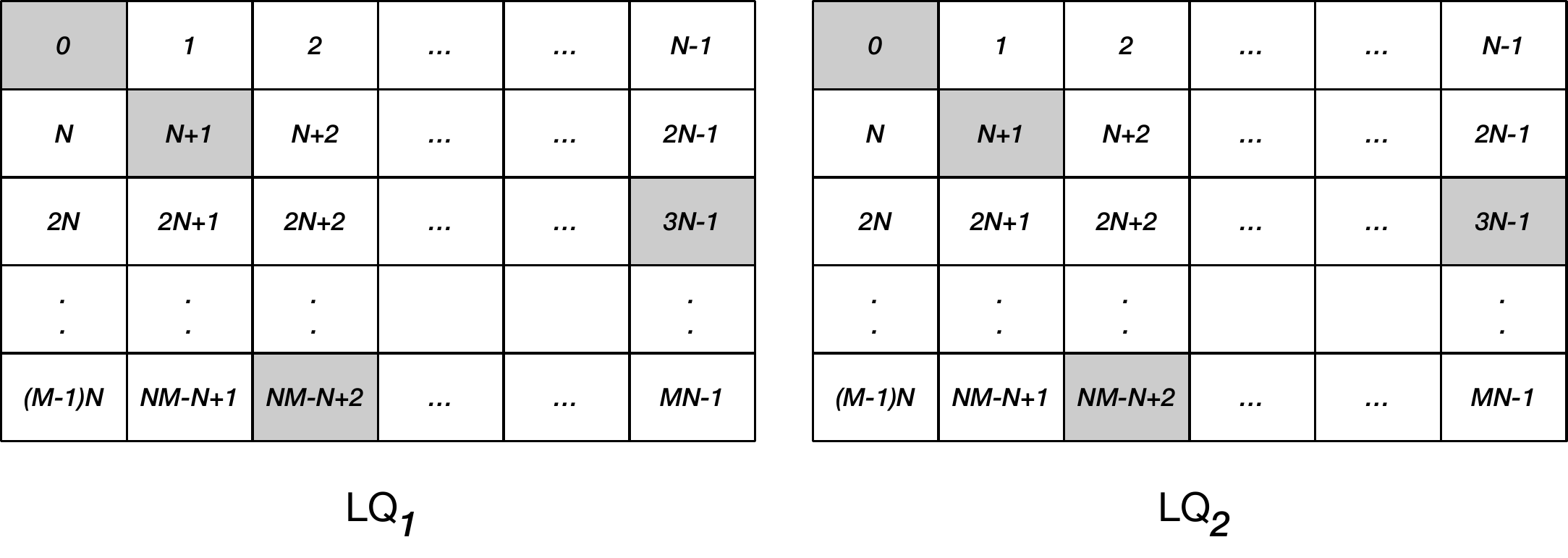}	
}
\subfigure[]{
	\includegraphics[scale=0.39]{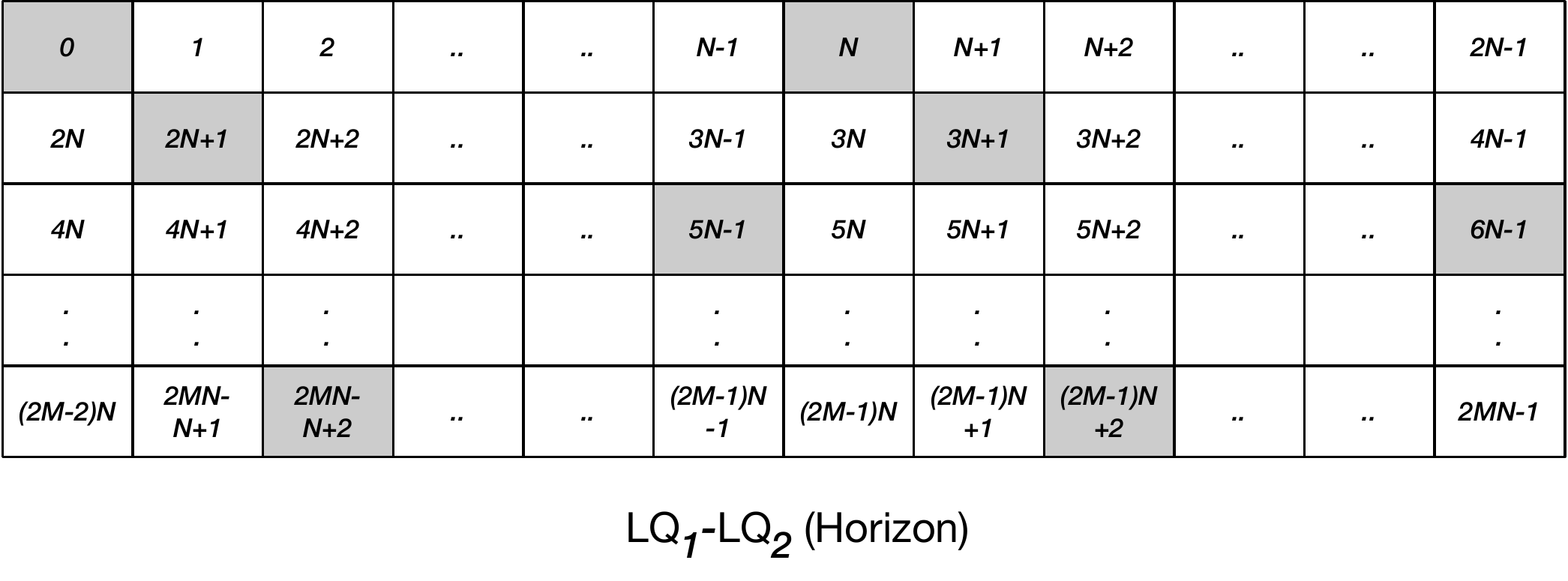}	
}
\caption
{
Example of the qubit layout extension.
(a) Two copies of a single logical data block of the size $m\times n$, and (b) The horizontally extended qubit layout, of the size $m\times 2n$.
The qubit layout can be extended in the vertical direction either.
Gray cells indicate the data qubits in the logical qubit block. 
The physical index of each data qubit should be relabelled in the extended qubit layout. }
\label{fig:extended_qubit_layout}
\end{figure*}

For 2-qubit logical operations such as \emph{CNOT} gate, the present work treats it as a protocol that acts on a single extended qubit layout.
For example, suppose that a logical qubit is defined on the 2-dimensional rectangular qubit layout of size $m \times n$.
We then extend the layout as $2m \times n$ for a vertical extension or $m \times 2n$ for a horizontal extension, and then allocate two logical qubits on it properly.
Please see Figure~\ref{fig:extended_qubit_layout}.
In doing so, all the physical operations for the 2-qubit logical operation act within two logical qubits.
On the other hand, for reference, Ref.~\cite{Svore.2007} uses the qubits outside the two logical qubits concerned as a communication channel.

\begin{figure}[ht]
\centering
\subfigure[]{
\mbox{ 
\Qcircuit @C=0.5em @R=1em {
\lstick{|\psi_L\rangle}	& \ctrl{1} 	&\gate{S_L} &\qw &\rstick{T_L|\psi\rangle_L}\\
\lstick{|A_L\rangle}	& \targ 	&\measureD{Z}  \cwx\\
}}} \vfill
\subfigure[]{
\mbox{ 
\Qcircuit @C=0.5em @R=1 em {
\lstick{|A_L\rangle}	& \ctrl{1} 	&\gate{X_LS_L} &\qw &\rstick{T_L|\psi\rangle_L}\\
\lstick{|\psi_L\rangle}	& \targ 	&\measureD{Z}  \cwx\\
}}}
\caption
{
Logical \emph{T} gate protocol that measures the magic state $|A\rangle_L$.
\emph{S} gate correction is required conditioned on the measurement outcome.
(a) Protocol that measures the magic state, and (b) Protocol that measures the logical data qubit~\cite{Zhou.2000}.
}
\label{fig:T_gate}
\end{figure}
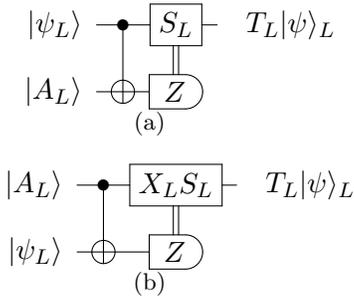

A certain error-correcting code implements a logical \emph{T} (and \emph{S}) gate as a 2-qubit gate protocol acting on a logical data qubit and a magic state (see Figure~\ref{fig:T_gate}).
In that case, the magic state preparation protocol (see Figure 13 of Ref.~\cite{Aliferis.2005}) should be synthesized first and the fault-tolerant circuit of the \emph{T} gate then is synthesized on the extended qubit layout.
The layout extension is based on the qubit mappings of both the logical qubit and the magic state.

The logical \emph{T} gate circuit can be synthesized in terms of two viewpoints, a combination of multiple basic quantum gates, and a single complex gate.
The first is, as shown in Figure~\ref{fig:T_gate}, we assemble the fault-tolerant circuits of logical \emph{CNOT}, \emph{MeasZ}, and \emph{S} gates.
For this, the magic (data) qubits of a magic state should be placed at the positions of the corresponding data qubits of a logical qubit to perform a logical \emph{CNOT} gate.
That is, the fault-tolerant circuit of a magic state preparation should include the \emph{Move-Back} operation for the magic qubits with reference to the logical qubit.
For that, the following instruction is required,
\emph{Move} $magic[i]$ $destination[i]$ where $destination[i]$ corresponds to the index of the $data[i]$ of the logical qubit in the extended layout.

On the other hand, the circuit can be synthesized as a single complex gate either.
In this case, unlike the first case, we don't need to consider the \emph{Move-Back} in the magic state preparation because it is possible to make the circuit based on the mapping of magic qubits $magic[i]$ just after the main body of the preparation protocol.
Besides, unlike the normal \emph{CNOT} gate including the \emph{Move-Back} of both the logical qubits, the magic qubits do not need to be back to the original position.
Therefore, this synthesis requires fewer qubit movements than the first synthesis approach.
To make a compact circuit, it is better to take the second approach.
In this regard, we prefer the protocol in (a) of Figure~\ref{fig:T_gate} to (b) of the same figure.
The \emph{Move-Back} operation of the latter requires more \emph{SWAP} gates than the former.
Figure~\ref{fig:T_circuit_compare} shows the comparison between both approaches.
Note that we applied this approach to the circuit synthesis for the magic state preparation.

\begin{figure}[t]
\centering
\includegraphics[scale=0.5]{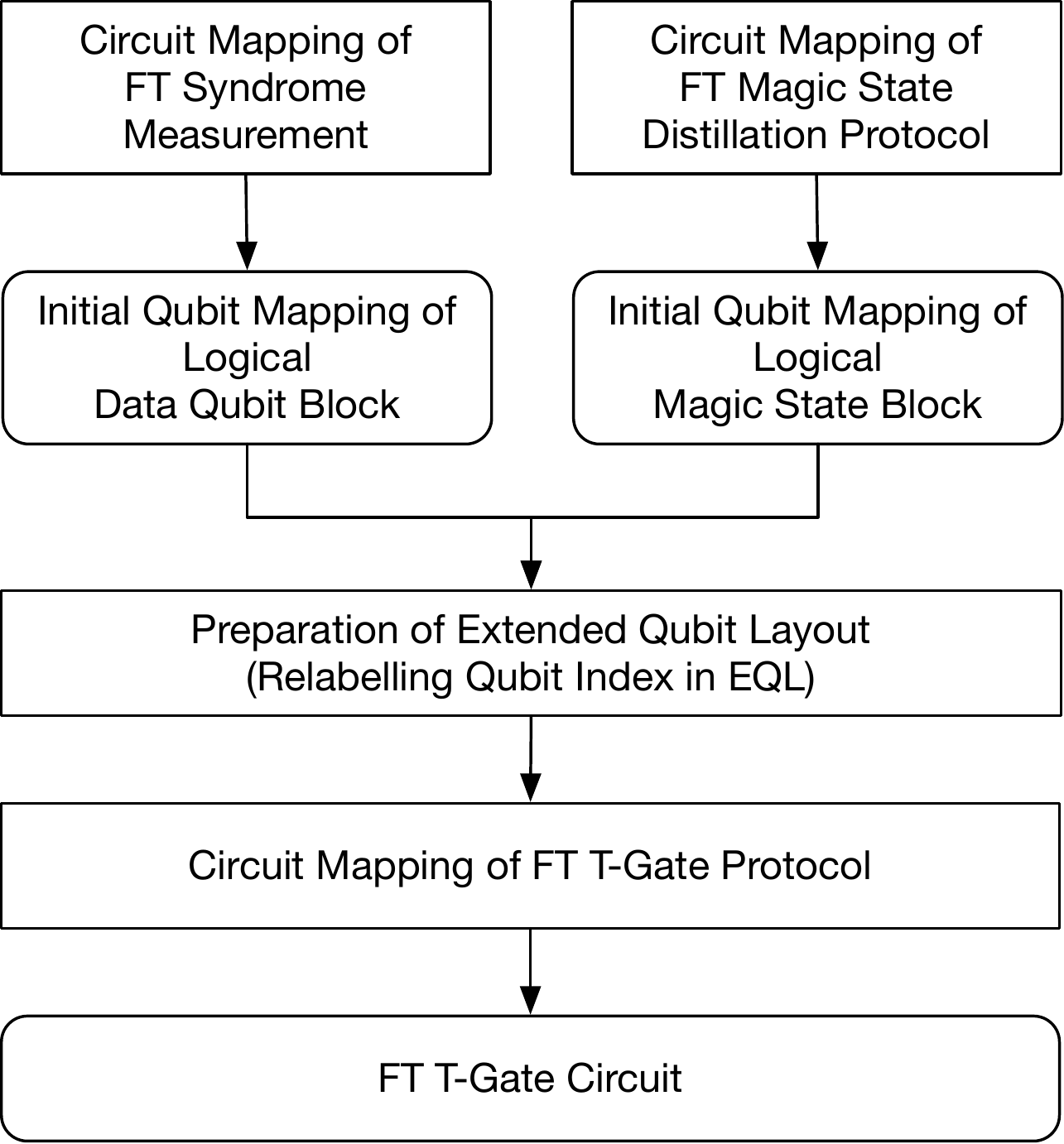}	
\caption{
Flow of the circuit synthesis the logical \emph{T} gate protocol.
After mapping the pivot protocols (syndrome measurement and state distillation) respectively, it is possible to determine the extended qubit layout for the logical \emph{T} gate protocol.
}
\label{fig:T_circuit_synthesis}
\end{figure}

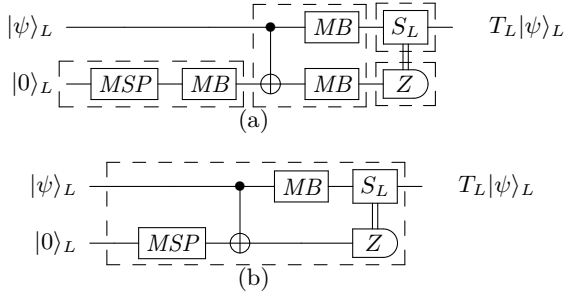
\begin{figure}[t]
\centering
\subfigure[]{
\mbox{ 
\footnotesize
\Qcircuit @C=1em @R=1 em {
\lstick{|\psi\rangle_L} & \qw  &\qw		& \ctrl{1} 	& \gate{\emph{MB}} & \gate{S_L} &\qw & \rstick{T_L |\psi\rangle_L}\\
\lstick{|0\rangle_L} & \gate{\emph{MSP}} 	& \gate{\emph{MB}} 	&  \targ & \gate{\emph{MB}} & \measureD{Z} \cwx
\gategroup{2}{1}{2}{3}{0.6em}{--}
\gategroup{1}{4}{2}{5}{0.6em}{--}
\gategroup{1}{6}{1}{6}{0.6em}{--}
\gategroup{2}{6}{2}{6}{0.6em}{--}\\
}
}}
\subfigure[]{
\mbox{ 
\footnotesize
\Qcircuit @C=1em @R=1 em {
\lstick{|\psi\rangle_L} 	& \qw & \qw  		& \ctrl{1} 	& \gate{\emph{MB}} & \gate{S_L} &\qw & \rstick{T_L |\psi\rangle_L}\\
\lstick{|0\rangle_L} & \qw & \gate{\emph{MSP}} &  \targ & \qw & \measureD{Z} \cwx
\gategroup{1}{2}{2}{6}{0.6em}{--}\\
}
}}
\caption
{
Comparison of the \emph{T} circuits according to the synthesis approach. (a) the combination of basic gates. (b) single complex gate.
Note that \emph{MSP} and \emph{MB} are respectively the abbreviations of the protocol of the magic state preparation and the Move-Back operation.
}
\label{fig:T_circuit_compare}
\end{figure}

\subsection{Circuit Partitioning}\label{subsec:circuit_partition}

In general, in the circuit synthesis of an ordinary quantum algorithm, all the operations are placed as forward as possible to minimize the length of the circuit.
However, to make an effective quantum circuit of a fault-tolerant protocol, we have to take its structure into account rather than unconditionally following the above-mentioned scheme.

A generic fault-tolerant quantum protocol begins by preparing ancilla qubits as a certain state and ends with the measurement of the ancilla qubits and the conditional processing based on the measurement outcomes.
In some cases, such a classical control operation is conducted once at the end of the protocol, but in other cases, performed multiple times within a single protocol.
In the latter case, if the quantum circuit after the circuit synthesis has a properly partitioned form on the basis of the turn of classical processing, a classical controller can execute the protocol easily.
Otherwise, the classical controller continuously needs to check the current progress of the circuit to find the turn for the classical control operations.
In doing so, it may make a decision for delaying some quantum operations for synchronization on the fly.
In what follows, we discuss how to make the quantum circuit of fault-tolerant quantum protocols partitioned without losing its logic.

For the circuit partitioning, we propose two techniques in this work.
First, one introduces \emph{barrier} statement~\cite{Cross.2017} to the protocol and the circuit synthesis treats it.
Second, one partitions the protocol beforehand, and performs the circuit synthesis individually but with sharing data common to all the tasks.
By the way, since the second method is very trivial, we skip the detailed explanation here but will be demonstrated as an example in Section~\ref{sec:golay}.

Suppose that the fault-tolerant protocol described in QASM includes the barrier statements at proper positions. 
Here, the proper position for the barrier statement is, on considering the structure of fault-tolerant quantum protocols, usually when just after all the ancilla qubits that were prepared within the protocol are measured.
Please see Figures~\ref{fig:steane_syndrome_measure_full} and~\ref{fig:qasm_steane_sm}.
Note that for the second method above we split the entire QASM up on the basis of the positions instead of inserting the \emph{barrier} statement.

Before going further, we need to say the following.
Please remind that for finding a fault-tolerant qubit move path we trace the qubit usage status throughout a protocol.
In this regard, the time when the usage status of all the ancilla qubits in the same category becomes \emph{inactivated} is the above-mentioned proper position.
Then, even without introducing the barrier statement, it is possible for the circuit synthesis algorithm to partition the circuit based on the usage status tracking.
But, the present work does not keep this in mind because it may happen that the architect of a fault-tolerant quantum protocol may not want to split the circuit up.
The circuit should work as the architect expects.

\begin{figure}[t]
\centering
\subfigure[]{
	\includegraphics[scale=0.5]{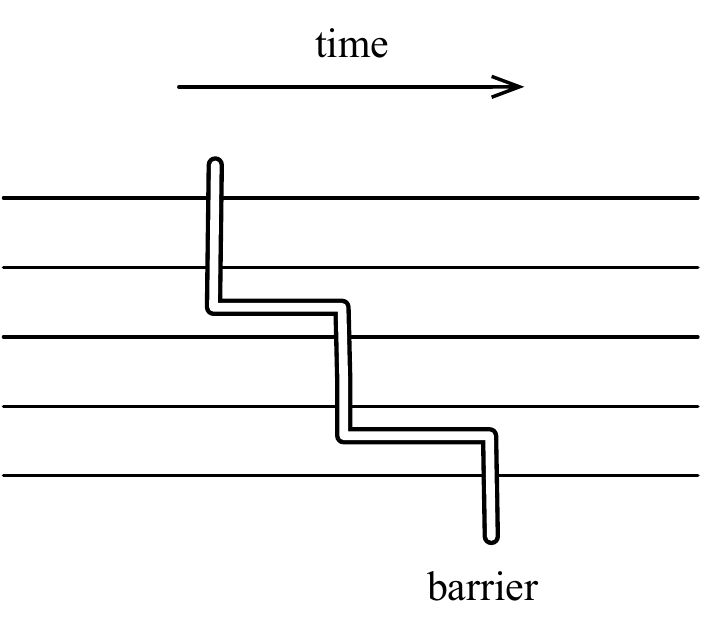}	
}
\subfigure[]{
	\includegraphics[scale=0.5]{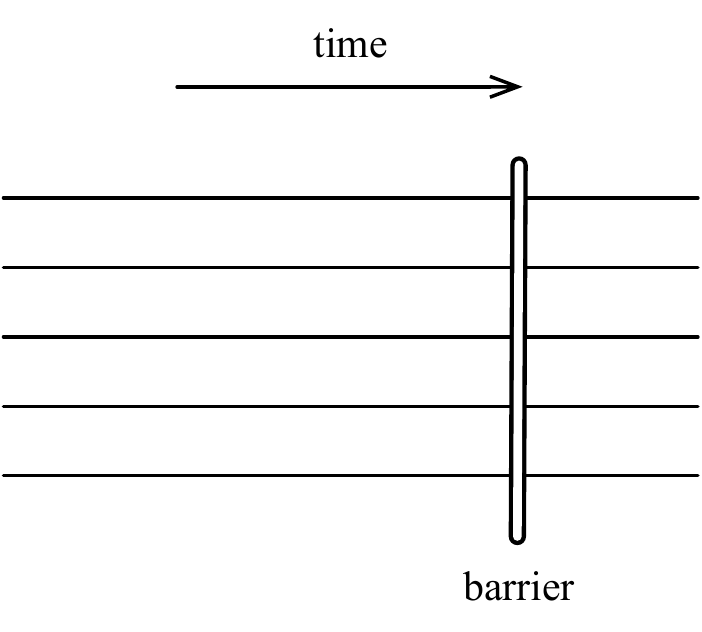}	
}
\caption
{
The effect of the circuit partitioning in the circuit synthesis for fault-tolerant quantum protocols.
(a) Before the circuit partitioning, two sub-circuits that need to be separated in time are in mixed in time flow.
(b) Partitioned circuit.
}
\label{fig:barrier_representation_effect}
\end{figure}

Back to the proposed method, let us say that the node including the barrier statement comes in FL after all of its preceding nodes are treated and exported to the quantum circuit.
Please note that each node in DAG includes a quantum instruction (see Figure~\ref{fig:DAG_move}).
If the barrier node is currently included in FL, we do not append the succeeding nodes (logically independent with the remaining nodes in FL) of the node just exported to a quantum circuit to FL, but keep them in a temporary list. 
If all the nodes except the barrier node are treated, that is, there remains only the barrier node in FL, all the nodes stored in the temporary list are moved to FL.
After that, the barrier node is removed from FL.
By doing so, the circuit can be logically partitioned.
The circuit synthesis basically cannot do anything for the nodes coming after the barrier node before processing the barrier.
Please see Figure~\ref{fig:barrier_representation_effect} for the effect of the circuit partitioning.	
In (a) of the figure, both the quantum instructions belonging to before and after the barrier are mixed up at one time, but in (b) they are separated.

\section{Example : Syndrome Measurement of $[[7,1,3]]$ Steane Code}\label{sec:steane_syndrome}

\begin{figure}[t]
\centering
\includegraphics[scale=0.5]{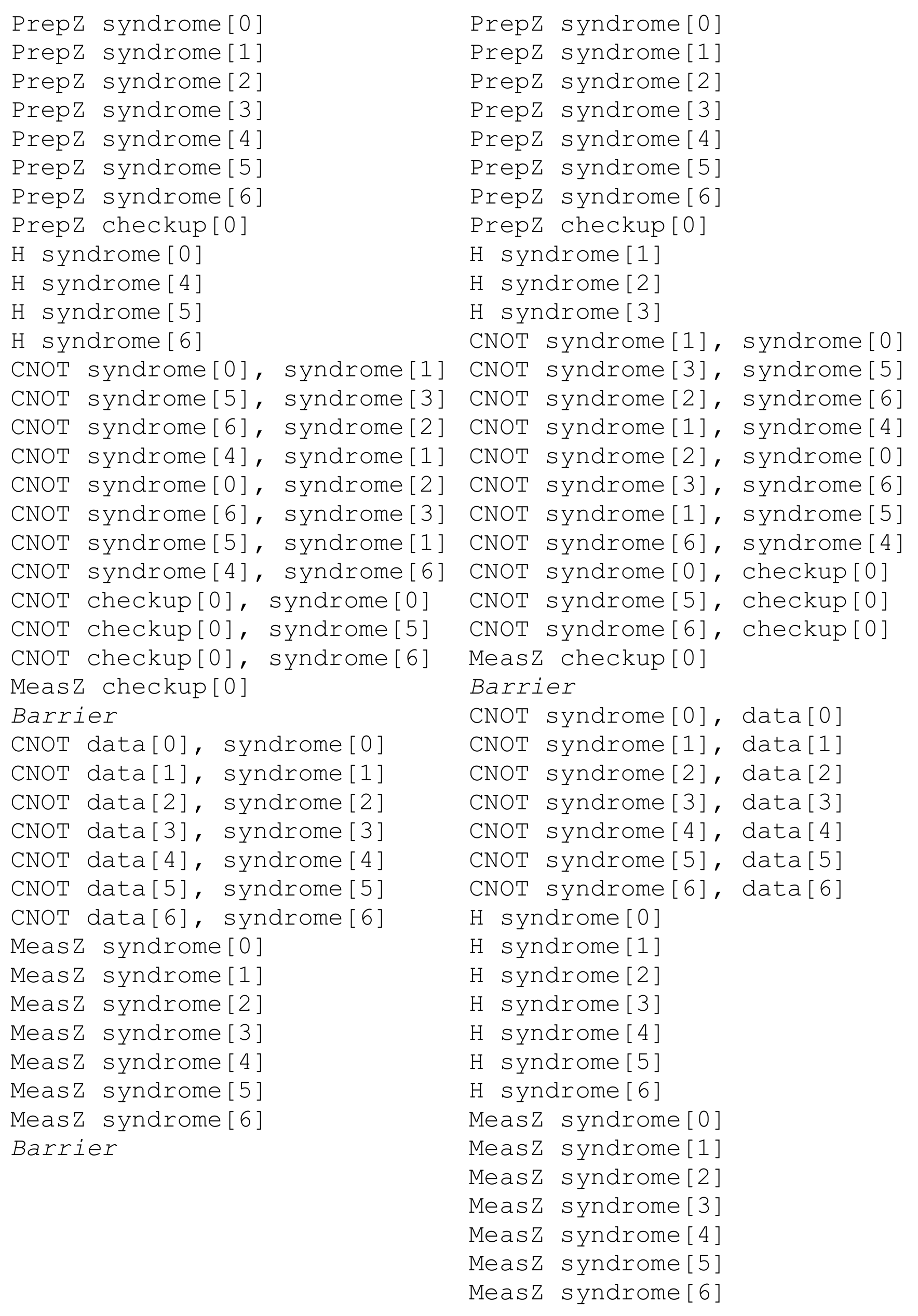}	
\caption{
QASM of the stabilizer measurement of $[[7,1,3]]$ Steane code~\cite{Steane.1997,Aliferis.2005,Weinstein.2015}.
Note that the declaration of qubits and classical bits are omitted.
The \emph{Z} (\emph{X})-type stabilizer measure is shown in the left (right) column.
From the first line to the line just before the \emph{barrier} statement the procedure of the logical ancilla state preparation (of Ref.~\cite{Goto.2016}) is described, and the main procedure of the syndrome measure is shown thereafter.
}
\label{fig:qasm_steane_sm}
\end{figure}

We show the fault-tolerant quantum circuit for the stabilizer measurement of $[[7,1,3]]$ Steane code.
We apply the Steane-EC method that consumes logical states ($|0\rangle_L$, $|+\rangle_L$) of the code~\cite{Steane.1997,Aliferis.2005,Weinstein.2015} (see Figure~\ref{fig:steane_syndrome_measure_full}).
The protocol is composed of the \emph{Z}-stabilizer measure and the \emph{X}-stabilizer measure in serial, and the \emph{Z} (\emph{X})-stabilizer measure is composed of the preparation of the logical plus (zero) state, and a sequence of \emph{CNOT} gates between data qubits and the syndrome qubits.
For the fault-tolerant preparation of a logical zero (plus) state, we apply Goto's single ancilla qubit verification method~\cite{Goto.2016} rather than the method comparing two copies of logical states~\cite{Aliferis.2005}.
In the method, a non-FT logical ancilla state is examined using a single checkup qubit.
If the verification succeeds, the main body of the syndrome measurement begins.
Otherwise, the logical ancilla state should be prepared again.
But, we do not take the repetition caused by the non-deterministic feature into account throughout the present work because the purpose of the present work is to generate a fault-tolerant circuit.

\begin{figure}[t]
\centering
\includegraphics[scale=0.6]{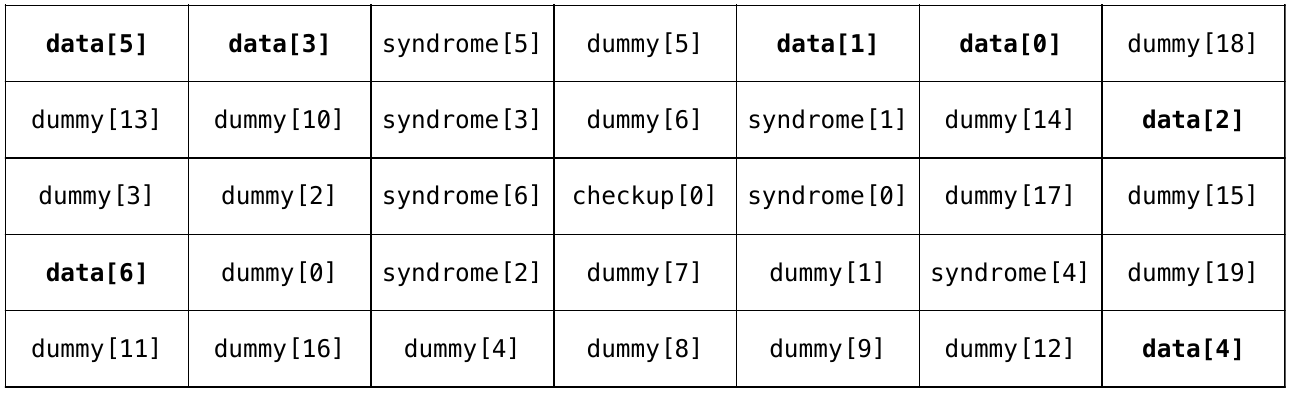}
\caption{
The initial qubit mapping for the fault-tolerant circuit of the syndrome measurement which is shown continuously over Figure~\ref{fig:steane_syndrome_measure_1}~$\sim$ \ref{fig:steane_syndrome_measure_7}.
}
\label{fig:steane_initial_mapping}
\end{figure}

Figure~\ref{fig:qasm_steane_sm} shows the QASM of the protocol which includes the barrier statements.
It is translated to DAG, and then the circuit synthesis is undergone.
For the 2-dimensional qubit layout of size $5\times 7$, we have obtained a fault-tolerant quantum circuit of depth 35.
Please see Figures~\ref{fig:steane_syndrome_measure_1}$\sim$~\ref{fig:steane_syndrome_measure_7}.
Each figure shows each part of the circuit: Preparation of $|+\rangle_L$, \emph{CNOT} ($data$[i], $syndrome$[i]), Preparation of $|0\rangle_L$ and \emph{CNOT} ($syndrome$[i], $data$[i]).
The initial qubit mapping for the circuit is shown in Figure~\ref{fig:steane_initial_mapping}.
Figure~\ref{fig:circuit_json_format} shows the text representation of the circuit in JSON format, which is generated from our implementation of the circuit synthesis algorithm.

By processing the \emph{barrier} statement, the entire circuit is separated into 4 sub-circuits in time flow even though it is not explicitly displayed there.
Besides, by the \emph{Move-Back} operation, it becomes that all the data qubits are placed at their initial positions after all the quantum operations.
Please compare Figure~\ref{fig:steane_initial_mapping} with Figure~\ref{fig:steane_syndrome_measure_7} (f).

\section{Fault-Tolerant Circuits of $[[7, 1, 3]]$ Steane code FTQC}\label{sec:steane_code_case}

This section shows how to prepare a full set of fault-tolerant quantum circuits for $[[7,1,3]]$ Steane code-based fault-tolerant quantum computing.
For that, we first need to determine which protocols are required and classify them into two categories, the \emph{pivot} protocols and the \emph{non-pivot} protocols.
We then perform the circuit synthesis for the pivot protocols and apply their results to the circuit synthesis of the non-pivot protocols.
Table~\ref{tab:ftqc_protocols} shows the protocols necessary for universal quantum computing divided into two categories.

\begin{table}[t]
\footnotesize
\caption{Protocols for Steane code based universal FTQC}
\centering
\begin{tabular}{c|c} \hline
Type & Protocols \\ \hline
\multirow{2}{*}{Pivot} & Syndrome Measurement, \\ 
&  Magic State Preparation \\ \hline
\multirow{2}{*}{Non-Pivot} & \emph{PrepZ} (or \emph{Encoder}), \emph{MeasZ}, \\ 
 &  \emph{H}, \emph{T}, \emph{S}, \emph{CNOT} \\ \hline
\end{tabular}
\label{tab:ftqc_protocols}
\end{table}

Before proceeding to the circuit synthesis for the pivot protocols, let us discuss the qubit layout first.
A qubit layout affects the cost and the performance of logical qubits, logical gates, and therefore, the entire quantum computing.
Therefore we need to pay special attention to it.
There exists an optimal qubit layout in terms of the circuit size defined as $\#qubit \times circuit\_depth$, as the size of the layout increases the number of qubits is increasing but the circuit depth may be decreased until the lower bound.
Therefore we need to compare the circuit sizes over various qubit layouts and select the best one.
We then apply the chosen qubit layout to generate the set of full fault-tolerant quantum circuits in the following subsection.

\subsection{Optimal Qubit Layout}\label{sec:layout_effect}

Over 2-dimensional qubit layouts of size $5\times 6 \sim 7\times 8$, we quantify the size of the fault-tolerant quantum circuit for the syndrome measurement based on two error correction methods, Shor EC and Steane EC.
For Steane EC~\cite{Steane.1997,Aliferis.2005,Weinstein.2015}, we exploit the fault-tolerant preparation of logical ancilla states $|0\rangle_L$ (and similarly $|+\rangle_L$) of Ref.~\cite{Goto.2016} (see Figure 1 therein).
For Shor EC~\cite{Shor.1996,Aliferis.2005,Weinstein.2015}, we apply the fault-tolerant preparation of the length-4 cat state described in Ref.~\cite{Aliferis.2005} (see Figure 6 therein).
To select the best qubit layout, we evaluate the circuits in terms of the circuit size, \emph{KQ}~\cite{Steane.2003}, where \emph{circuit~size} is defined as $circuit~depth$ $\times$ $\#qubits$.
We assume that in this work one ancilla state is generated in sequence, not multiple states in parallel.

\begin{table*}[htp]
\caption{The size of the fault-tolerant circuits about the syndrome measurement over various 2-dimensional qubit layouts. 
\emph{ideal} indicates the static analysis about the the protocol itself. 
The number in parentheses indicates the number of qubits of the corresponding qubit layout.
Note that the protocols based on Steane-EC and Shor-EC are composed of 15 and 12 qubits.
}
\footnotesize
\centering
\begin{tabular}{c||c||c|c|c|c|c|c|c|c|c} \hline \hline
\multirow{2}{*}{Method} & \multirow{2}{*}{item} &~\multirow{2}{*}{\emph{ideal}}~&~$5\times 6$~&~$5\times 7$~&~$6\times 6$~&~$5\times 8$~&~$6\times 7$~&~$6\times 8$~&~$7\times 7$~&~$7\times 8$ \\ 
 &			&  	& (30) & (35) & (36) & (40) & (42) & (48) & (49) & (56) \\ \hline \hline
\multirow{3}{*}{Steane-EC}&$depth$  		& 18 & 43 & 35 & 41 & 40 & 41 & 41 & 40 & 38 \\ \cline{2-11}
&$\# gates$ 	& 83 & 156 & 168 & 162 & 158 & 164 & 165 & 175 & 155 \\ \cline{2-11}
&$KQ$ 	    	& 270 & 1,290  & 1,225  & 1,476 & 1,600 & 1,722 & 1,968 & 1,960 & 2,128 \\ \hline \hline
\multirow{3}{*}{Shor-EC}&$depth$  	& 44 & 55 & 53 & 54 & 55 & 53 & 55 & 52 & 54 \\ \cline{2-11}
&$\# gates$ 					& 144 & 206 & 204 & 199 & 224 & 212 & 212 & 224 & 216 \\ \cline{2-11}
&$KQ$ 	    					& 528 & 1,650  & 1,855  & 1,944 & 2,200 & 2,226 & 2,640 & 2,548 & 3,025 \\ \hline \hline
\end{tabular}
\label{tab:layout_steane}
\end{table*}

Table~\ref{tab:layout_steane} shows the circuit sizes of fault-tolerant syndrome measurements over the qubit layouts by adopting \#\emph{gates} and \emph{circuit~depth} as the performance evaluation criteria.
It is well known that Shor-EC based protocol requires fewer qubits, but more quantum gates than Steane-EC based one~\cite{Aliferis.2005,Svore.2007}.
Therefore, for fault tolerance, the latter is preferred.
As shown in the Table, our circuit synthesis results also show the same tendency.
Based on the Table, we in the following section develop the circuits of fault-tolerant quantum protocols with the qubit layout of size $5\times7$.
For reference, in Ref.~\cite{Svore.2007} the qubit layout of size $6\times8$ is applied to optimize the threshold than to optimize the space-time resource in the local setting.

Before going further, we emphasize again that since our proposed algorithm is heuristic and non-deterministic, the results the table shows are not fixed for the layout.
They are just the most compact ones we have obtained so far, but there may exist more optimized circuits.

\subsection{Full Set of Fault-Tolerant Quantum Circuits}

We are going to perform the circuit mapping for the protocols listed in Table~\ref{tab:ftqc_protocols} in the following order.
The first is the syndrome measurement and the magic state preparation in the pivot protocols.
Since the fault-tolerant quantum circuit and the associated qubit mapping of the syndrome measurement are already described in Section~\ref{sec:steane_syndrome}, we discuss the magic state preparation only.

We take the protocol described in Ref.~\cite{Aliferis.2005} (see Fig. 13 therein).
It is composed of the repetition of the preparations of ancilla states (logical zero state and 7-qubit cat state), $\bar{T}\bar{X}\bar{T^{\dagger}}$ measurement and error detection.
Since the ancilla preparations and error detection include the selection of the next operation conditioned on the measurement outcomes, we apply the circuit partitioning we have discussed before.
For that, we insert the \emph{barrier} statements in the protocol, a total of 10 times.
The fault-tolerant preparations of a logical ancilla and a cat state respectively require a single checkup qubit.
Therefore, by resuing qubits, the protocol is composed of at least 15 qubits: 7 for data, 7 for ancilla, and 1 for the verification.
Finally, we emphasize that in the circuit synthesis of the magic state preparation (or distillation for other code), the final qubit mapping after the circuit execution is critical to the T gate circuit.

We go to the non-pivot protocols.
For the 1-qubit gate, only the circuit synthesis of the encoder is enough because the other gates can be implemented as the transversal gates, therefore there is nothing to do to realize a quantum circuit.
As mentioned several times, we take the protocol for that consuming a single checkup qubit protocol~\cite{Goto.2016}.
Like the magic state, in the circuit of the encoder, the final qubit mapping is so important.
It should be the configuration of the logical qubit.
In the initial mapping of the encoder, all the qubits are just dummy qubits that do not hold any meaningful quantum state.
The syndrome measurement and the magic state preparation include the encoder for the preparation of the logical ancilla qubits, but we don't apply the results there to the encoder and vice versa to make an optimized circuit for each respectively.

\begin{figure}[t]
\centering
\subfigure[]{
	\includegraphics[scale=0.5]{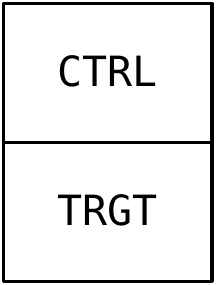}	
}
\subfigure[]{
	\includegraphics[scale=0.5]{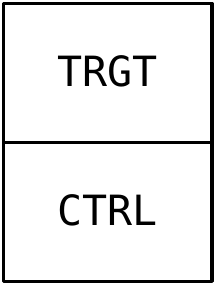}	
}
\subfigure[]{
	\includegraphics[scale=0.5]{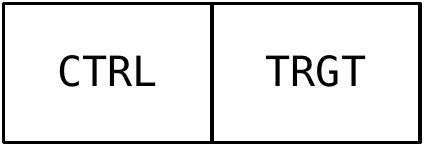}	
}
\subfigure[]{
	\includegraphics[scale=0.5]{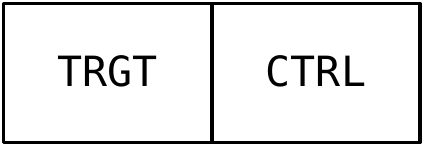}	
}
\caption
{
The relative arrangement of two logical qubits.
}
\label{fig:logical_qubit_layout}
\end{figure}

For the 2-qubit gates such as \emph{CNOT} and \emph{T}, we have to perform the circuit synthesis over all the cases of how two qubits are arranged on the logical qubit layout.
Please see Figure~\ref{fig:logical_qubit_layout}.
However, in the case of \emph{CNOT}, all the data qubits in both logical qubits have to move back to their initial positions after the operation.
Therefore, the circuits for the layouts (a) and (c) directly can be used for the layouts (b) and (d) by changing only the \emph{CNOT} directions.
On the other hand, for the \emph{T} gate, as mentioned before, the \emph{Move-Back} for the magic state is not required, and therefore the circuit for the layout (a) cannot be applied for the case (b).
Similarly for the layouts (c) and (d).
Therefore, we conduct the circuit synthesis of \emph{T} gate over all four cases, but only two for \emph{CNOT} gate.
Figures~\ref{fig:steane_cnot_vertical_1}~$\sim$~\ref{fig:steane_cnot_vertical_4} show all the snapshots of \emph{CNOT} gate for the qubits arranged vertically, ($ctrl_{n}$, $trgt_{s}$), and Figures~\ref{fig:steane_t_vertical_1}~$\sim$~\ref{fig:steane_t_vertical_7} show all the snapshots of \emph{T} gate for the qubits arranged vertically, ($data_{n}$, $magic_{s}$).

\begin{table*}[tp]
\scriptsize
\caption{
The static analysis of the fault-tolerant quantum circuits for universal fault-tolerant quantum computing.
The numbers of \emph{SWAP} gates and \emph{Barrier} indicate the quantities of those ones introduced in the circuit synthesis.
Note that the quantities of the other gates in the circuit are precisely the same as those included in the protocol.
Note that MB indicates the Move-Back operation.
}
\centering
\begin{tabular}{c|c|c|c|c|c|c|c|c} \hline \hline
\multirow{2}{*}{Operation} & \multicolumn{4}{c}{Protocol} \vline & \multicolumn{4}{c}{Circuit} \\  \cline{2-9}
 & Reference & \#Qubits & Depth & Gates & Depth & \#\emph{SWAP} & MB & \#Barrier \\ \hline
Syndrome &\multirow{2}{*}{\cite{Steane.1997,Aliferis.2005}} & \multirow{2}{*}{15} & \multirow{2}{*}{20} & \emph{CNOT}: 36, \emph{H}: 15,  & \multirow{2}{*}{35} & \multirow{2}{*}{80} & \multirow{2}{*}{O} & \multirow{2}{*}{3}\\ 
Measurement & &  &  & \emph{PrepZ}: 16, \emph{MeasZ}: 16 &  &  &  & \\ \hline 
\multirow{2}{*}{Encoder} & \multirow{2}{*}{\cite{Goto.2016}} & \multirow{2}{*}{8} & \multirow{2}{*}{8} & \emph{CNOT}: 11, \emph{H}: 3, & \multirow{2}{*}{18} & \multirow{2}{*}{31} & \multirow{2}{*}{O} & \multirow{2}{*}{-} \\ 
 &  &  &  & \emph{PrepZ}: 8, \emph{MeasZ}: 1 &  &  & &  \\ \hline 
 Magic State &\multirow{2}{*}{\cite{Aliferis.2005}} & \multirow{2}{*}{15} & \multirow{2}{*}{75} & \emph{CNOT}: 113, \emph{H}: 49, \emph{PepZ}: 49,  & \multirow{2}{*}{162} & \multirow{2}{*}{323} & \multirow{2}{*}{X} & \multirow{2}{*}{10}  \\
Preparation & &  &  & \emph{MeasZ}: 49, \emph{T}: 14, $T^{\dagger}$: 14 & & & &  \\ \hline 
h-\emph{CNOT} & Transversal & 14 & 1 & \emph{CNOT}: 7 & 21 & 136 & O & -  \\ \hline 
v-\emph{CNOT} & Transversal & 14 & 1 & \emph{CNOT}: 7 & 13 & 80 & O & - \\ \hline 
h-\emph{T} ($d_{e}$, $m_{w}$) & Figure~\ref{fig:T_circuit_compare} (b) & 14 & 3 & \emph{CNOT}: 7, \emph{MeasZ}: 7, \emph{S}: 7 & 22 & 96 & O & 1 \\ \hline 
h-\emph{T} ($d_{w}$, $m_{e}$) & Figure~\ref{fig:T_circuit_compare} (b) & 14 & 3 & \emph{CNOT}: 7, \emph{MeasZ}: 7, \emph{S}: 7 & 20 & 104 & O & 1 \\ \hline 
v-\emph{T} ($d_{n}$, $m_{s}$) & Figure~\ref{fig:T_circuit_compare} (b) & 14 & 3 & \emph{CNOT}: 7, \emph{MeasZ}: 7, \emph{S}: 7 & 17 & 71 & O & 1 \\ \hline 
v-\emph{T} ($d_{s}$, $m_{n}$) & Figure~\ref{fig:T_circuit_compare} (b) & 14 & 3 & \emph{CNOT}: 7, \emph{MeasZ}: 7, \emph{S}: 7 & 24 & 81 & O & 1 \\ \hline 
\end{tabular}
\label{tab:list_performance_steane_ftqc}
\end{table*}

To conclude this section, we list the setting and the result of the circuit synthesis over all the protocols shown in Table~\ref{tab:ftqc_protocols}.
Please see Table~\ref{tab:list_performance_steane_ftqc}.
Note that for the S gate correction in the T gate, we apply the transversal gate rather than the state injection method like the T gate protocol.
For the fault tolerance, the state injection method is used but for the purpose of this work, the transversal gate is enough.

\section{$[[23,1,7]]$ Golay Code}\label{sec:golay}

$[[23, 1, 7]]$ Golay code is well known as a high error correction threshold~\cite{Aliferis.2005,Cross.2009}, but it has been rarely studied how to realize universal fault-tolerant quantum computing with that code.
Due to its large code block (logical qubit), it is a challenging problem to find fault-tolerant quantum circuits~\cite{Paetzick.2012,Zheng.2018}.

It is also tricky and time-consuming to find the circuits with the proposed method.
Therefore, in this work, we apply the \emph{divide-and-conquer} approach to find the fault-tolerant quantum circuit of the syndrome measurement efficiently.
That is, we divide the entire protocol into several sub-protocols, obtain the fault-tolerant circuit for each sub-protocol by applying the proposed algorithm, and finally combine all the sub-circuits.
We hire the protocol of preparing a logical zero state of Ref.~\cite{Paetzick.2012}, and divide it into the non-FT preparation stage of $|0\rangle_L$ and the verification stage (see Figures 3 and 4 therein).
We then perform the circuit synthesis of the former and apply the result to the synthesis of the latter.
Please note that the fact that the former acts in a non-fault-tolerant manner does not indicate that its non-FT circuit is enough for that.
By taking the divide-and-conquer method, the entire circuit can be naturally partitioned, and classical control operations can be conducted between the sub-circuits.

\begin{figure}[t]
\centering
\includegraphics[scale=0.4]{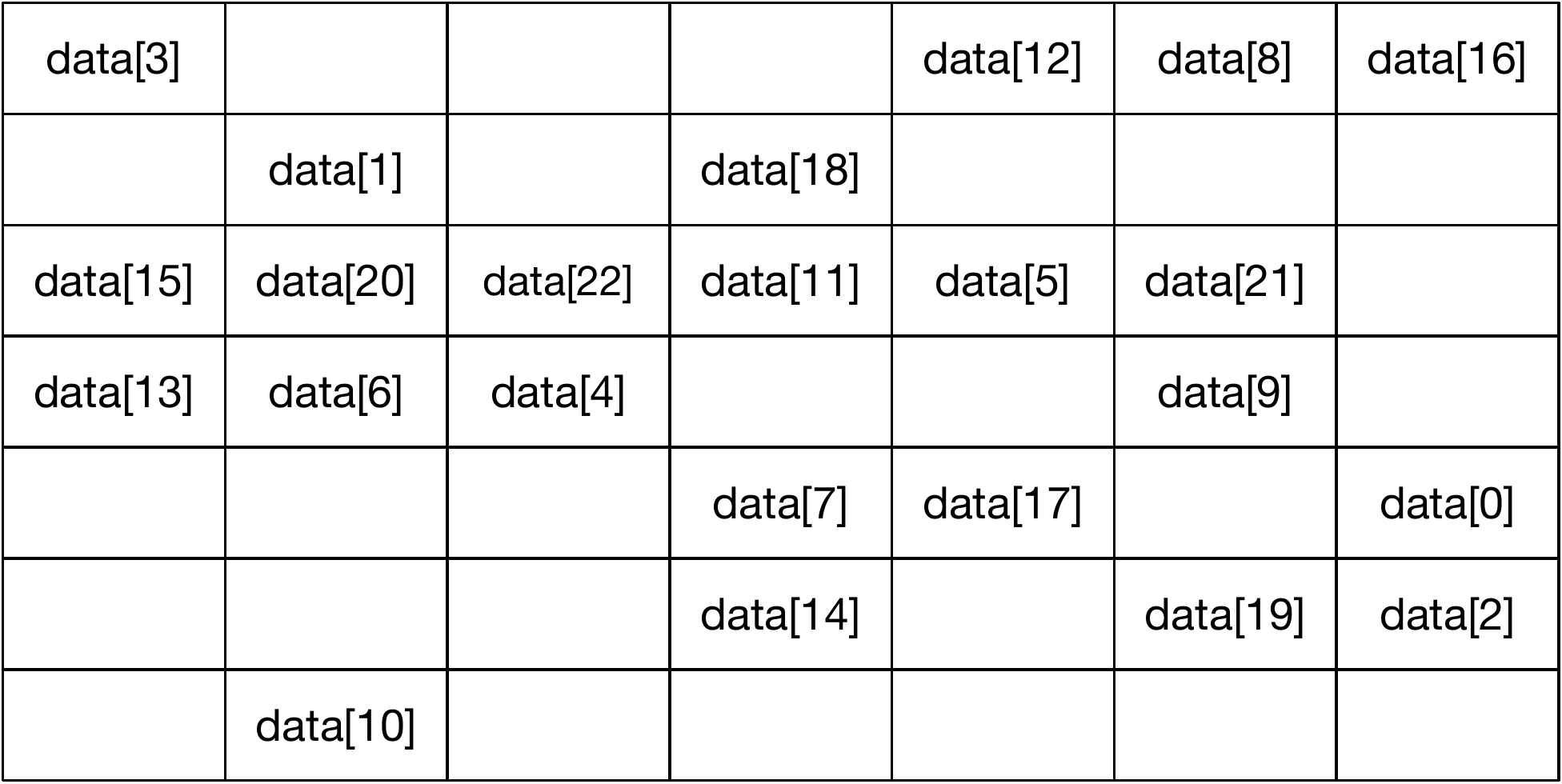}	
\caption{
Qubit layout of size $7 \times 7$ for a logical zero state $|0\rangle_L$ in $[[23, 1, 7]]$ Golay code of Ref.~\cite{Paetzick.2012}.
Based on this layout of 49 qubits, we obtain a fault-tolerant circuit of the preparation protocol (Figure 4 therein).
The circuit includes 269 \emph{SWAP} gates to make 57 \emph{CNOT} gates executable on the qubit layout without noisy \emph{SWAP} gates between data qubits.
Empty cells indicate dummy qubits working as the communication channels.
}
\label{fig:golay_qubit_layout}
\end{figure}

\begin{figure}[t]
\centering
\subfigure[]{
	\includegraphics[scale=0.4]{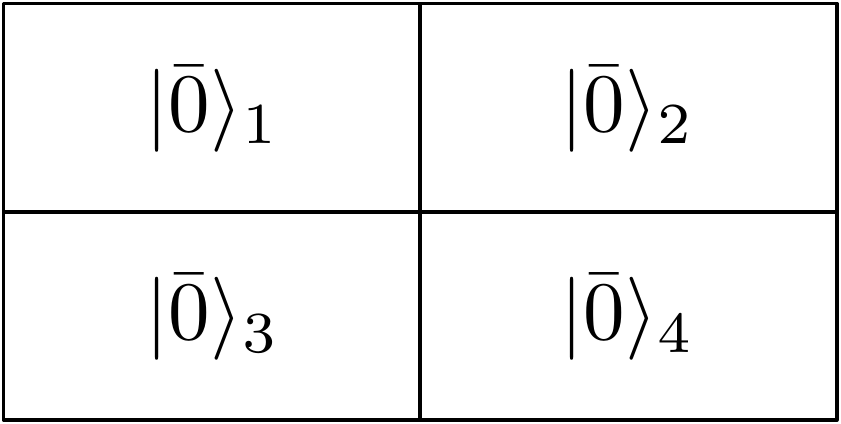}	
}
\subfigure[]{
	\includegraphics[scale=0.4]{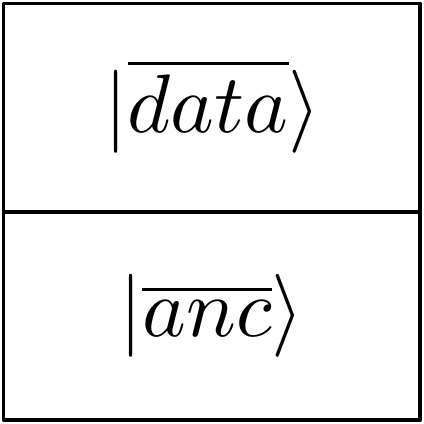}	
}
\caption{
Qubit Layouts for $[[23, 1, 7]]$ Golay code.
(a) Extended qubit layout for encoding $|0\rangle_L$ in the fault-tolerant manner~\cite{Paetzick.2012} (see Figure 3 therein).
Each cell, shown in Figure~\ref{fig:golay_qubit_layout}, indicates the code block for the corresponding logical zero state $|\bar{0}\rangle_i$ prepared in the non-fault-tolerant manner.
(b) Extended qubit layout for a logical qubit composed of the code blocks of data qubits $|\overline{data}\rangle$ and logical ancilla $|\overline{anc}\rangle$. Each cell is arranged as the code block shown in Figure~\ref{fig:golay_qubit_layout}, but the roles of the qubits in the block differ.
}
\label{fig:golay_extended_block}
\end{figure}

Let us assume that the size of the qubit layout for the above non-FT preparation circuit is $m \times n$.
Then for the fault-tolerant circuit for the verification part, we need an extended qubit layout of size $2m \times 2n$ (see Figure~\ref{fig:golay_extended_block}).
The initial qubit mapping is constituted of 4 copies of the final mapping of the non-FT preparation circuit (see Figure~\ref{fig:golay_qubit_layout}).
How to extend qubit layout was discussed in Section~\ref{subsec:condition_3}.
We then apply the circuit synthesis algorithm to find the fault-tolerant circuit functioning for the verification, and then combine the resulting circuit with four copies of the non-FT circuit for preparing $|0\rangle_L$.
We now have a fault-tolerant circuit for the preparation of a logical zero state.
Since the code is self-dual, the fault-tolerant circuit for the preparation of $|+\rangle_L$ is almost the same as the FT circuit for preparing $|0\rangle_L$.

For $[[7,1,3]]$ Steane code FTQC, we determined the configuration of a logical qubit from the circuit synthesis result of the syndrome measurement protocol.
We applied the Steane-EC method that requires logical states $|0\rangle_L$ and $|+\rangle_L$ as ancilla states.
For the fault-tolerant preparation of logical states, we refer to Ref.~\cite{Goto.2016} that is composed of the non-FT preparation of one logical state and the verification of it by use of 1 physical qubit.
For the Golay code, we take the same approach, but the difference is the complexity of the fault-tolerant preparation of logical states as mentioned above.
We can say that the non-FT preparation of logical states $|0\rangle_L$ and $|+\rangle_L$ is the dominant part of the syndrome measurement protocol in the Golay code QEC.
Therefore, in this work, we determine the configuration of a logical qubit from the circuit synthesis of the non-FT preparation protocol.

As mentioned several times, the fault-tolerant preparation of the logical states in the Golay code takes much time and space (qubits).
In this work, we assume logical ancilla states are supplied from an ancilla factory, not generated at site.
Therefore, with the 2-dimensional qubit layout of size 14 x 7, we define a logical qubit made of the 23 data qubits, the 23 ancilla qubits for a logical ancilla state, and some dummy qubits for the communication channel. 
We leave a detailed method about how to supply the logical ancilla states practically as future work.

The code distance of the Golay code is 7 and therefore it is possible to correct arbitrary 3-qubit errors.
Therefore, as mentioned above, it is possible to allow a one-time noisy \emph{SWAP} gate between data-type qubits during the syndrome measurement, $\lfloor (7-1)/4\rfloor$ = 1.
Please note that in the fault-tolerant preparation of logical ancilla states, we do not permit any noisy \emph{SWAP} gate between the data-type qubits.
Table~\ref{tab:procedure_circuit_synthesis_golay} describes the procedure and the static data of the circuit synthesis of the fault-tolerant preparation of the logical zero states and the syndrome measurement.

\begin{table*}[t]
\footnotesize
\caption{Process and static analysis for the circuit synthesis of $[[23, 1, 7]]$ Golay code. 
The entire circuit synthesis is undergone in sequence: 1) Non-FT preparation of $|0\rangle_L$ ($|+\rangle_L$), 2) Verification and 3) Syndrome Measurement.
Note that \emph{Move-Back*} is a partial move-back operation for the data qubits in $|\bar{0}\rangle_1$, not for all.
Note that the value in parentheses in Circuit Depth indicates the circuit depth at the protocol level without the locality effect of the qubit layout.
The circuit depth includes the physical preparation of $|0\rangle$.}
\centering
\begin{tabular}{c||c|c|c} \hline \hline
 & \multicolumn{3}{c}{Protocol} \\ \cline{2-4}
 & Non-FT preparation of & \multirow{2}{*}{Verification} & \multirow{2}{*}{Syndrome Measurement} \\ 
 & $|0\rangle_L$ ($|+\rangle_L$) & &   \\ \hline
\multirow{2}{*}{Qubit Mapping} & \multirow{2}{*}{$M_{init}\rightarrow \cdots \rightarrow M_{LQ}$} & $M_{LQ} \rightarrow \cdots \rightarrow M_{LQ}$ & $M_{LQ} \rightarrow \cdots \rightarrow M_{LQ}$ \\ 
 &  & (\emph{Move-Back*}) & (\emph{Move-Back}) \\ \hline
Size of Qubit Layout & $m \times n$ & $2m \times 2n$ & $2m \times n$ (or $m \times 2n$) \\ \hline
Allowable Noisy \emph{SWAP}& \multirow{2}{*}{0} & \multirow{2}{*}{0} & \multirow{2}{*}{1}\\ 
between Data Qubits &  &  & \\ \hline
Circuit Depth (Ideal Depth) & 51 (9) & 21 (4) & 11 (5) \\ \hline
$\#$ \emph{SWAP} & 269 & 296 & 50 \\ \hline \hline
\end{tabular}
\label{tab:procedure_circuit_synthesis_golay}
\end{table*}

\section{Discussion}\label{sec:discussion}

We leave some comments for the readers who are going to implement our algorithm or develop new ones.

The circuit synthesis for an ordinary quantum algorithm always succeeds in finding a quantum circuit if the size of a quantum chip (qubit layout) is not less than the number of qubits in the quantum algorithm.
According to a target case (quantum algorithm, qubit connectivity, and an initial qubit mapping), the quality of the resulting circuit or the efficiency of the synthesis procedure may be very poor, but nothing makes the circuit mapping impossible fundamentally.

On the other hand, for a fault-tolerant quantum protocol, the situation differs because in the circuit the qubits have to move in a fault-tolerant manner.
If the size of a target quantum chip equals to or is bigger modestly than the number of qubits in the fault-tolerant protocol, the circuit synthesis may fail to find a fault-tolerant circuit.
Particularly, the proposed algorithm tries to find a fault-tolerant quantum circuit based on a randomly picked initial qubit mapping, such a phenomenon that a circuit algorithm cannot generate a fault-tolerant circuit happens more frequently when the size of the quantum chip is not greater remarkably.
In our implementation, we apply the time limit for SABRE traversal, the amount of which is proportional to the size of the protocol.
If the SABRE iteration is not completed within the time limit, we forcibly turn off the current SABRE iteration and begin the next iteration.

In the present work, we have not discussed a lower bound on the size of a quantum chip for an $n$-qubit fault-tolerant quantum protocol.
Finding such a bound theoretically is combinatorially extremely challenging.
It depends on the initial qubit mapping, the property of a quantum protocol (highly parallel or serial), the shape of the qubit layout, and so on.
For the purpose of reference, we have observed that for the Steane-EC based syndrome measurement protocol the proposed algorithm succeeds in finding a fault-tolerant quantum circuit acting on the qubit layout of 4x5, but of longer depth 72.
Since the protocol itself is designed as using 15 qubits, the minimum requirement for the fault-tolerant communication may not be so large but at cost of the circuit depth.

The proposed algorithm continuously selects a locally optimal \emph{SWAP} gate.
In that case, even though a selected \emph{SWAP} gate was evaluated as optimal at a cost evaluation at the time, it may not play a significant role in the resulting circuit finally.
For example, it may happen that occasionally consecutive \emph{SWAP} gates for a pair of qubits are involved in the circuit.
In this work, even though we have not represented it explicitly, we deal with the problem in two stages.
First, during the circuit mapping, if a selected \emph{SWAP} gate at the current cost evaluation is the same as the previous one, then we reject the selection and take a random one.
Please note that here the selected \emph{SWAP} gate should work on a pair of a data-type qubit and a non-data-type qubit in general. 
Otherwise, we refuse to take it.
Second, after each SABRE iteration, as post-processing for the resulting circuit, we cancel out the quantum instructions repeated consecutively on the same qubits.

We now discuss future works caused by our report.
This works defines the qubit configuration of a logical qubit from the circuit synthesis of the syndrome measurement, and then obtain compact fault-tolerant quantum circuits based on that configuration.
For the 2-qubit gates, we have considered all the possible relative arrangements of logical qubits.
As shown in Table~\ref{tab:list_performance_steane_ftqc}, the sizes of the circuits are different according to the arrangement.
This is because both the layout for a logical qubit and the arrangement of physical qubits in the logical qubit are not symmetric.
Even though an individual quantum circuit is optimized in terms of the circuit depth, operating universal fault-tolerant quantum computing may be tricky because according to the qubit arrangement \emph{CNOT} (\emph{T}) gate works differently.
In this regard, we need to consider how to operate universal fault-tolerant quantum computing efficiently with the quantum circuits of non-uniform performance.

\section{Conclusion}\label{sec:conclusion}

We summarize the present work. 
To date, various concatenated quantum codes and their fault-tolerant protocols have been proposed theoretically, but their realization in the local setting has been rarely discussed.
Only for a few well-known codes, their fault-tolerant circuits obtained by hand works have been reported.
Those results or methodologies can not be directly applied to different qubit layouts or concatenated quantum codes.

In the present work, to automate the fault-tolerant quantum circuit synthesis, we raised four requirements, presented our approaches for them, and described how to implement them with the existing heuristic quantum circuit mapping algorithm.
As a result, it is now possible to obtain a set of fault-tolerant quantum circuits for an arbitrary concatenated quantum code by running the algorithm.

The proposed algorithm does not work deterministically, and therefore the presented static analysis data about the circuits are not fixed for the protocol and the qubit layout.
To get the most optimized circuits, we need to iterate the synthesis (SABRE) rounds as much as possible.

\begin{acknowledgements}
This work was partly supported by Institute for Information \& communications Technology Promotion (IITP) grant funded by the Korea government (MSIT) (No. 2019-0-00003,
Research and Development of Core technologies for Programming, Running, Implementing and Validating of Fault-Tolerant Quantum Computing System) and the National Research Foundation of Korea (NRF) grant funded by the Korea government (MSIT) (No. NRF-2019M3E4A1080146).
\end{acknowledgements}

\bibliographystyle{unsrtnat}
\bibliography{reference} 

\onecolumn
\newpage
\appendix

\begin{figure*}[t]
\centering
\subfigure[Step 1]{
	\includegraphics[scale=0.35]{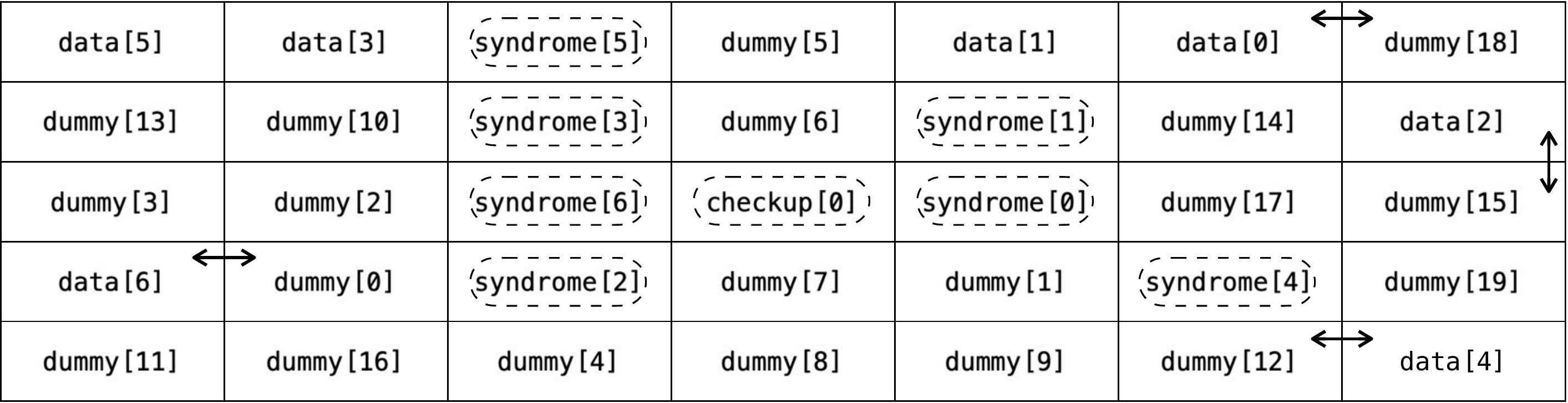}
}
\subfigure[Step 2]{
	\includegraphics[scale=0.35]{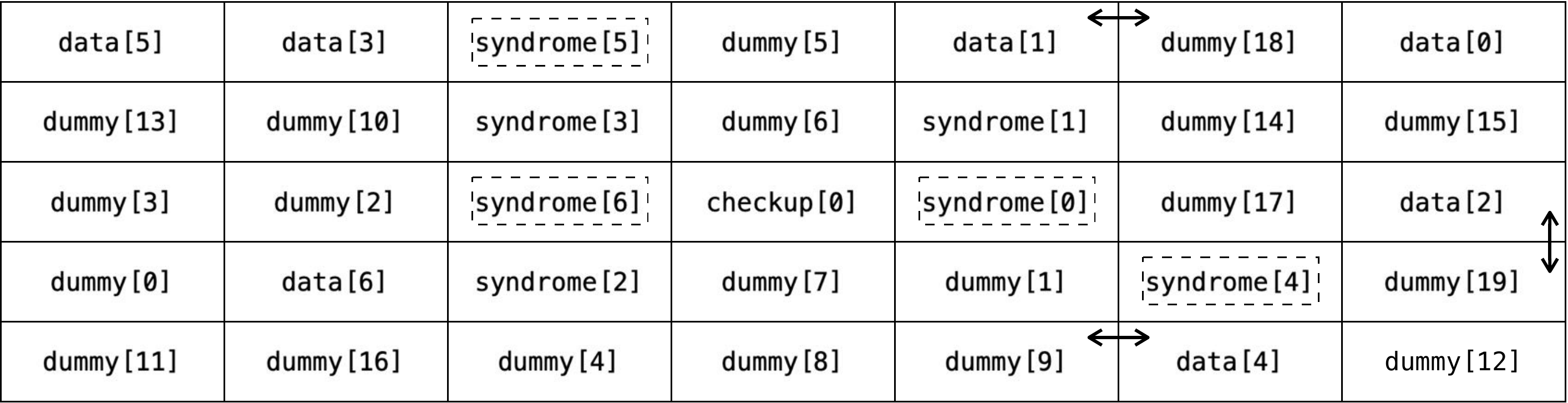}
}
\subfigure[Step 3]{
	\includegraphics[scale=0.35]{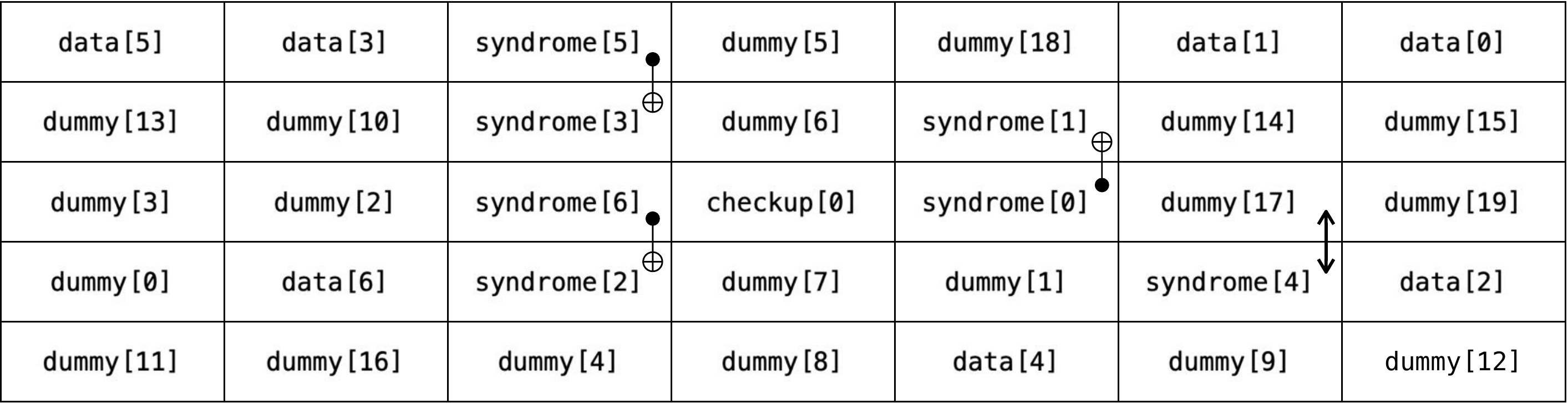}
}
\subfigure[Step 4]{
	\includegraphics[scale=0.35]{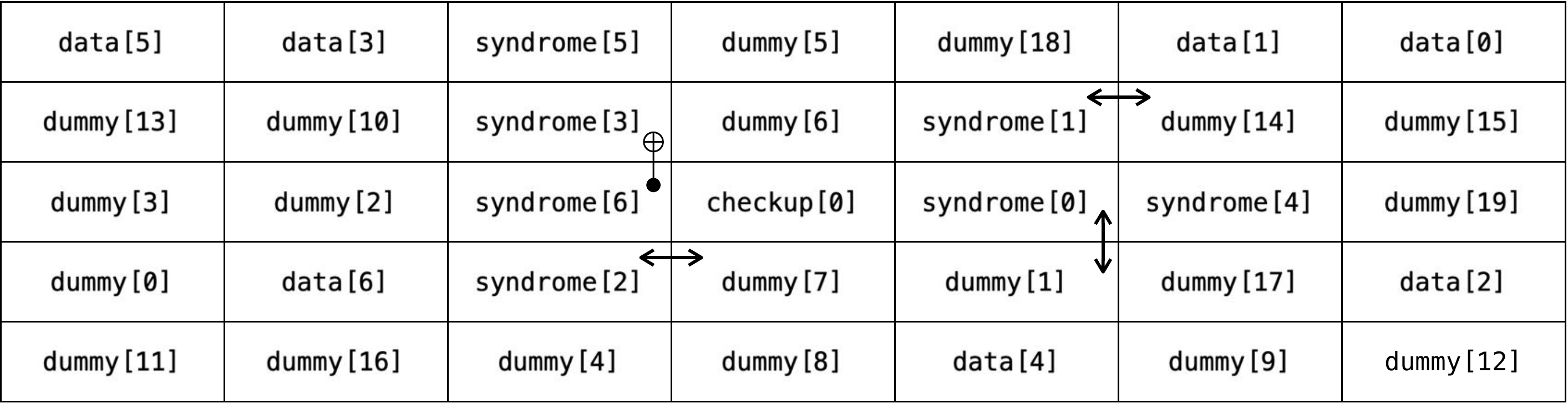}
}
\subfigure[Step 5]{
	\includegraphics[scale=0.35]{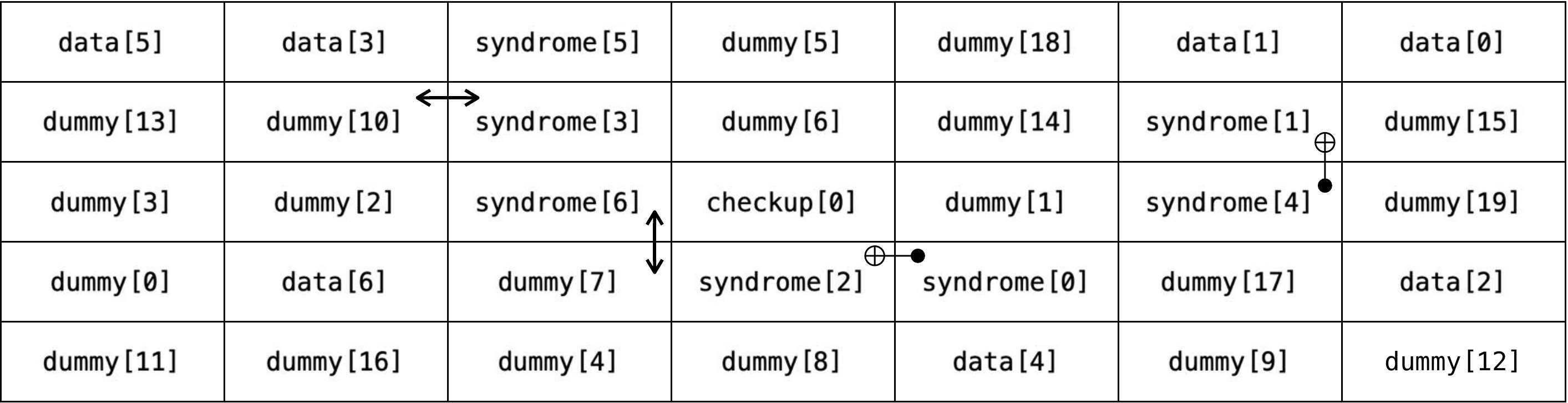}
}
\subfigure[Step 6]{
	\includegraphics[scale=0.35]{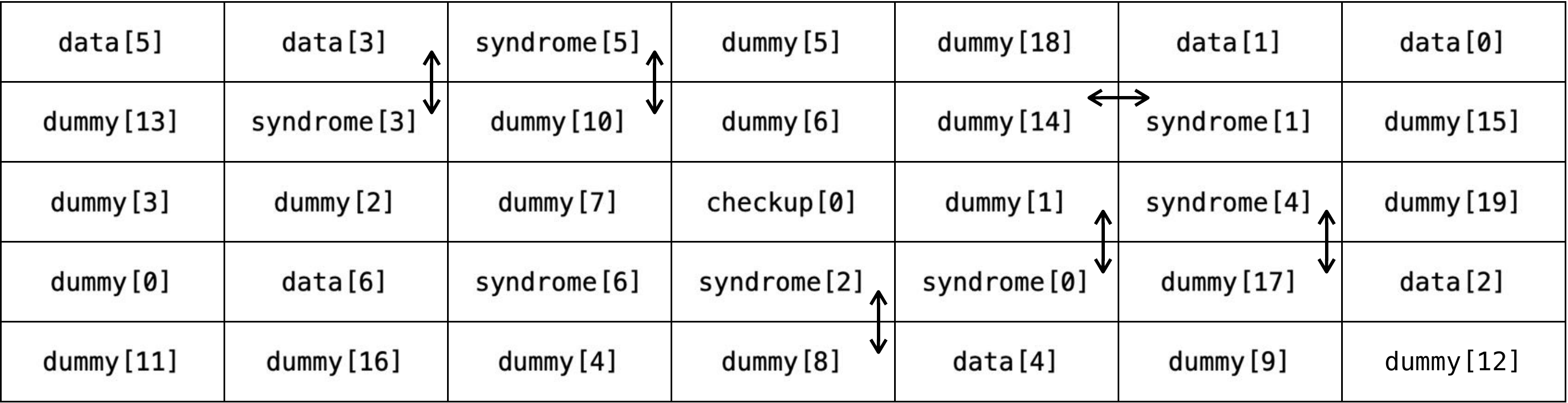}
}
\caption
{
The first part of the fault-tolerant quantum circuit of stabilizer measurement of $[[7, 1, 3]]$ Steane code: Fault-tolerant preparation of the logical state $|+\rangle_L$.
Note that the rectangles, rounded rectangles, hexagons and bi-directed arrow respectively indicate \emph{H}, \emph{PrepZ}, \emph{MeasZ} and \emph{SWAP} gates.
}
\label{fig:steane_syndrome_measure_1}
\end{figure*}

\begin{figure*}[t]
\centering
\subfigure[Step 7]{
	\includegraphics[scale=0.35]{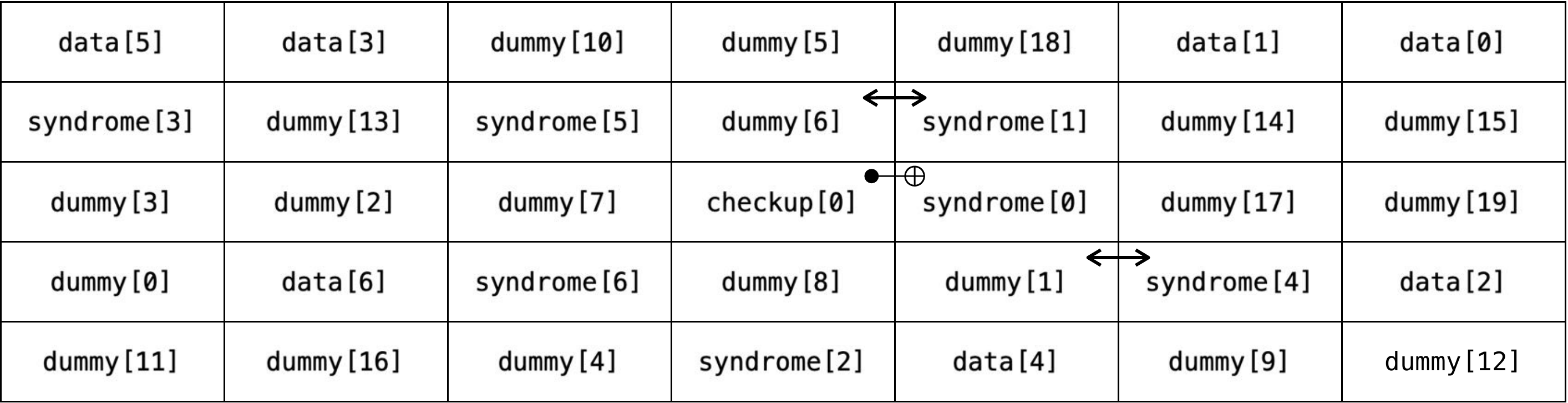}
}
\subfigure[Step 8]{
	\includegraphics[scale=0.35]{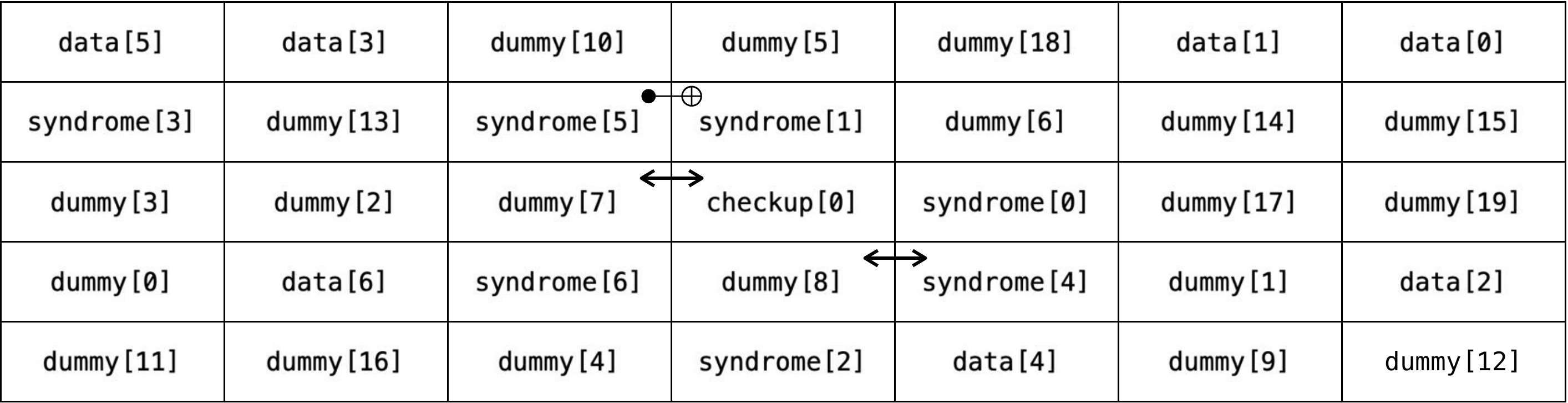}
}
\subfigure[Step 9]{
	\includegraphics[scale=0.35]{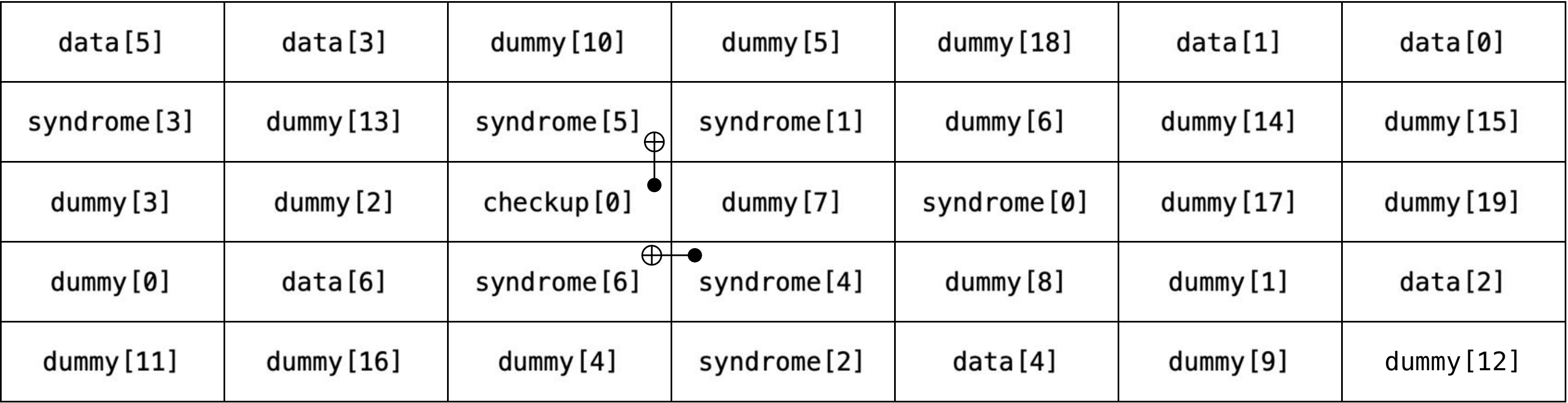}
}
\subfigure[Step 10]{
	\includegraphics[scale=0.35]{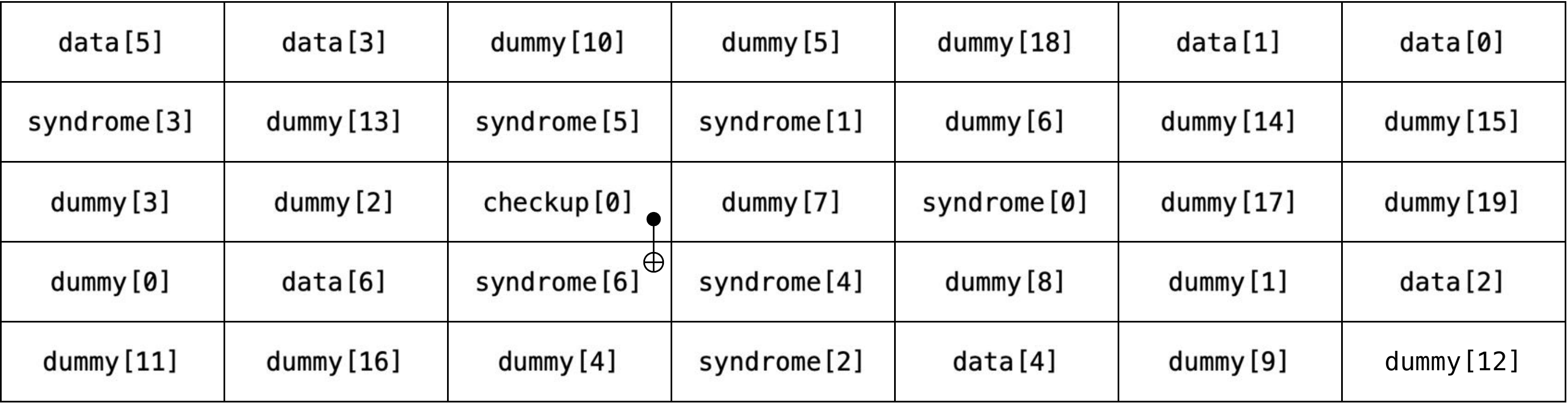}
}
\subfigure[Step 11]{
	\includegraphics[scale=0.35]{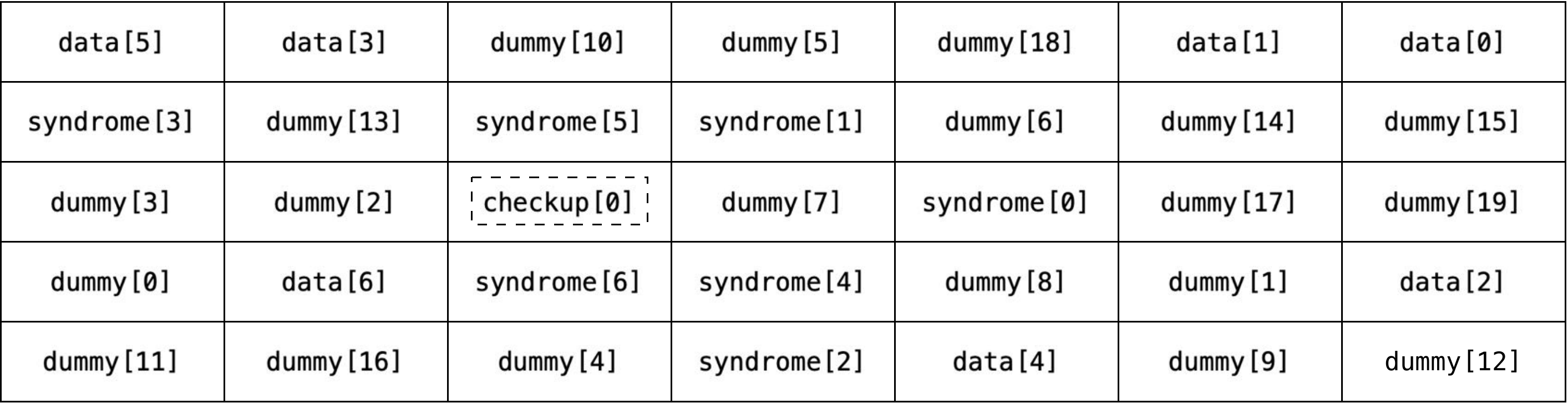}
}
\subfigure[Step 12 (Barrier)]{
	\includegraphics[scale=0.35]{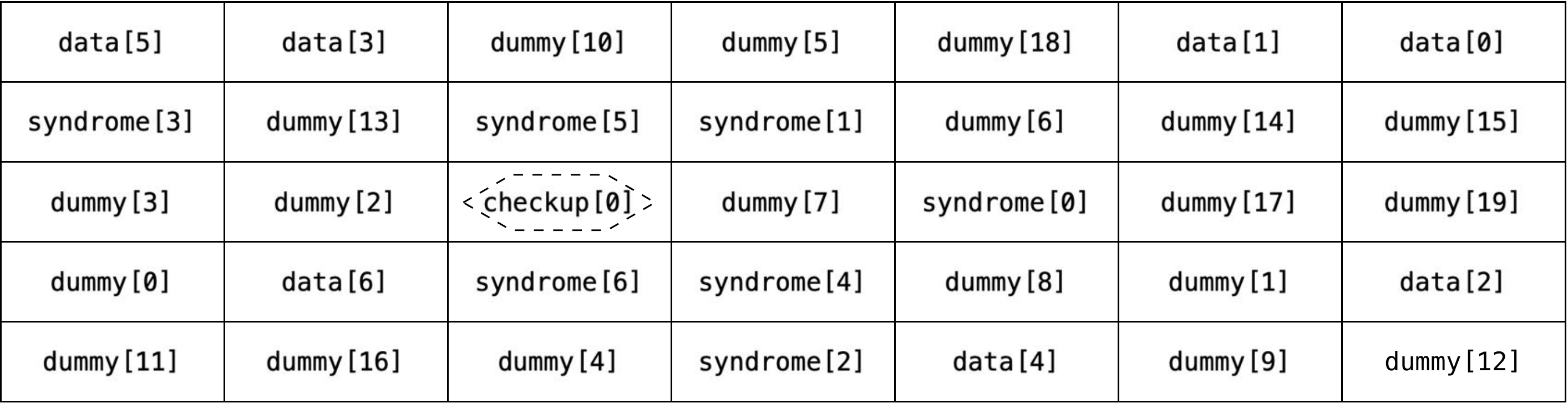}
}
\caption
{
(Continued from Figure~\ref{fig:steane_syndrome_measure_1}) The first part of the fault-tolerant quantum circuit of stabilizer measurement of $[[7, 1, 3]]$ Steane code: Fault-tolerant preparation of the logical state $|+\rangle_L$.
Note that the rectangles, rounded rectangles, hexagons and bi-directed arrow respectively indicate \emph{H}, \emph{PrepZ}, \emph{MeasZ} and \emph{SWAP} gates.
}
\label{fig:steane_syndrome_measure_2}
\end{figure*}

\begin{figure*}[t]
\centering
\subfigure[Step 13]{
	\includegraphics[scale=0.35]{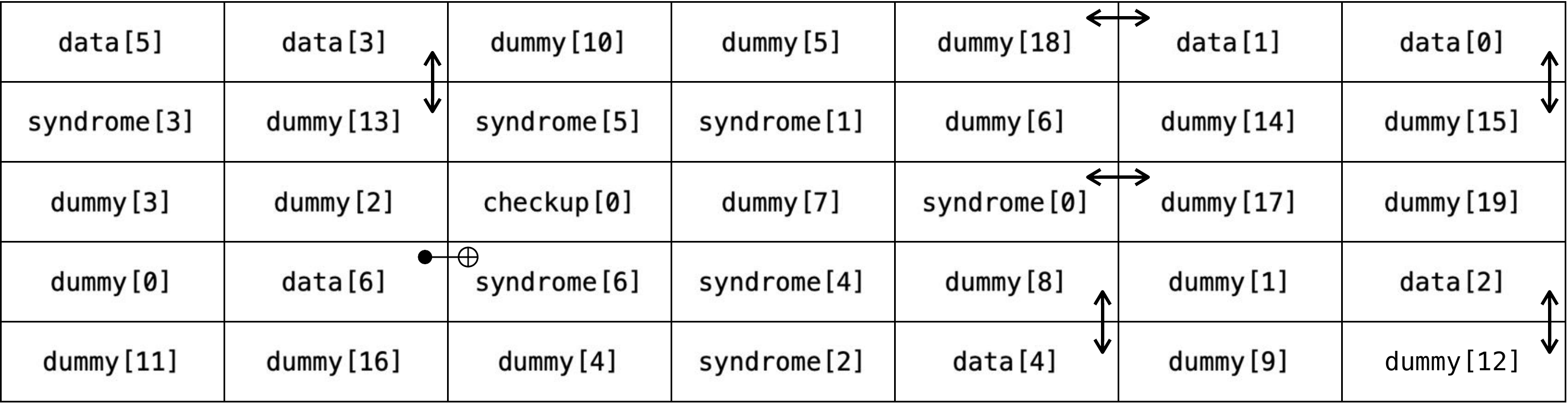}
}
\subfigure[Step 14]{
	\includegraphics[scale=0.35]{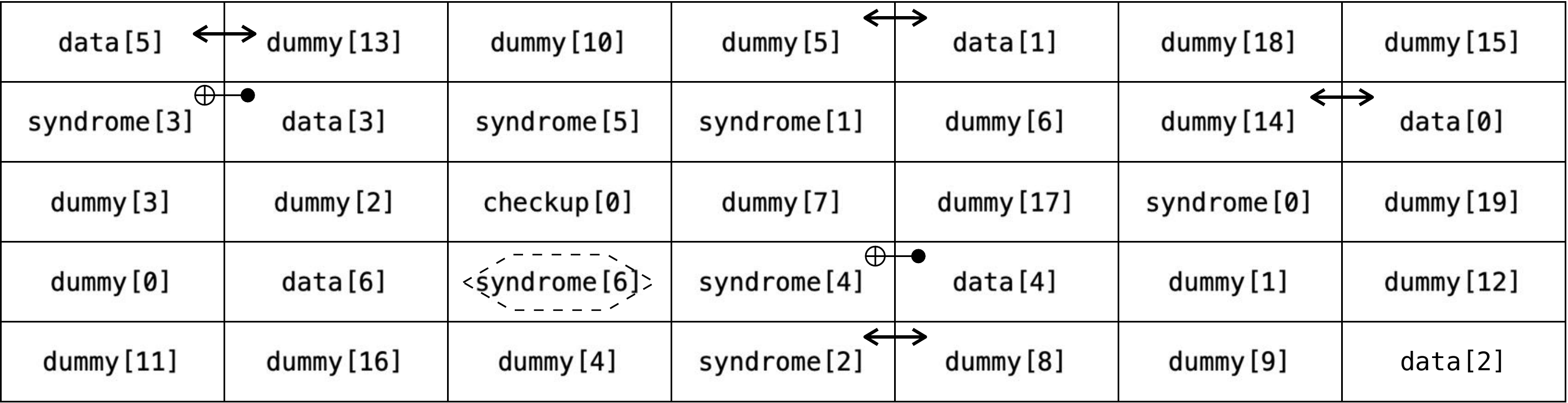}
}
\subfigure[Step 15]{
	\includegraphics[scale=0.35]{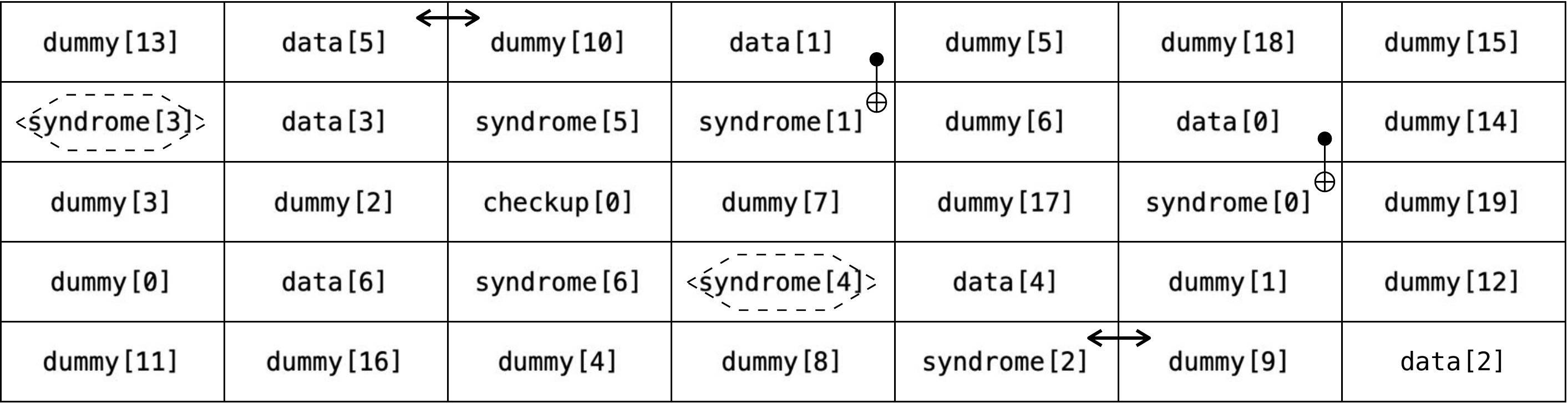}
}
\subfigure[Step 16]{
	\includegraphics[scale=0.35]{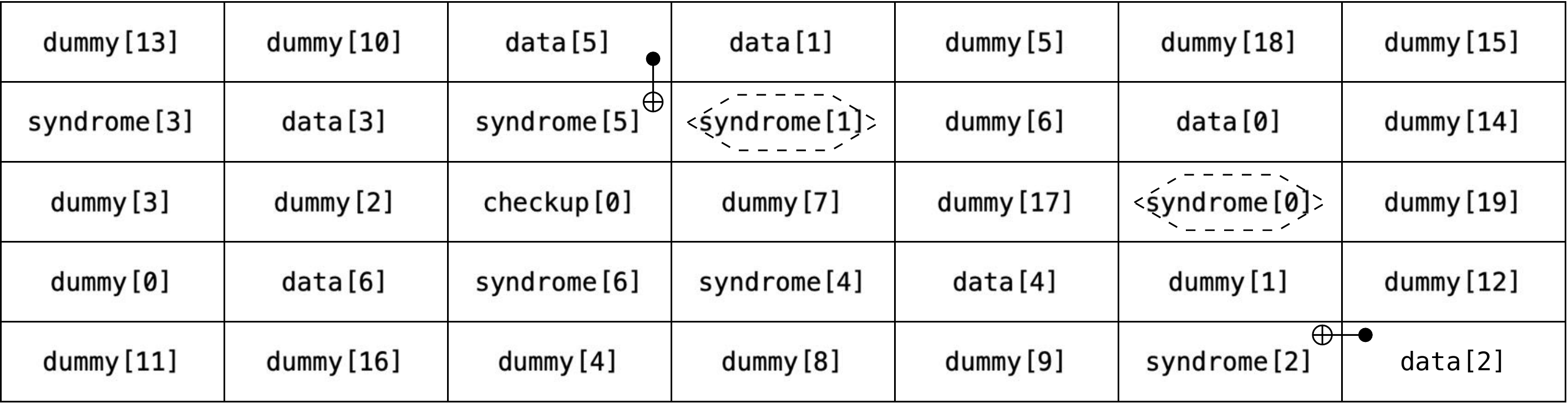}
}
\subfigure[Step 17 (Barrier)]{
	\includegraphics[scale=0.35]{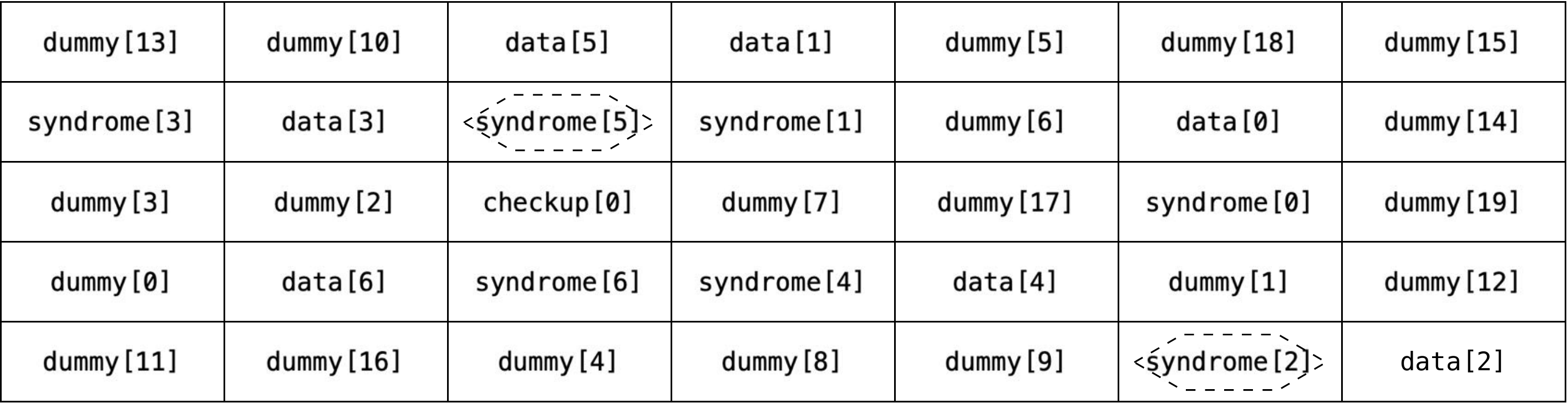}
}
\caption
{
The second part of the fault-tolerant quantum circuit of stabilizer measurement of $[[7, 1, 3]]$ Steane code: Transversal \emph{CNOT} between data qubits and syndrome qubits.
Note that the rectangles, rounded rectangles, hexagons and bi-directed arrow respectively indicate \emph{H}, \emph{PrepZ}, \emph{MeasZ} and \emph{SWAP} gates.
}
\label{fig:steane_syndrome_measure_3}
\end{figure*}

\begin{figure*}[t]
\centering
\subfigure[Step 18]{
	\includegraphics[scale=0.35]{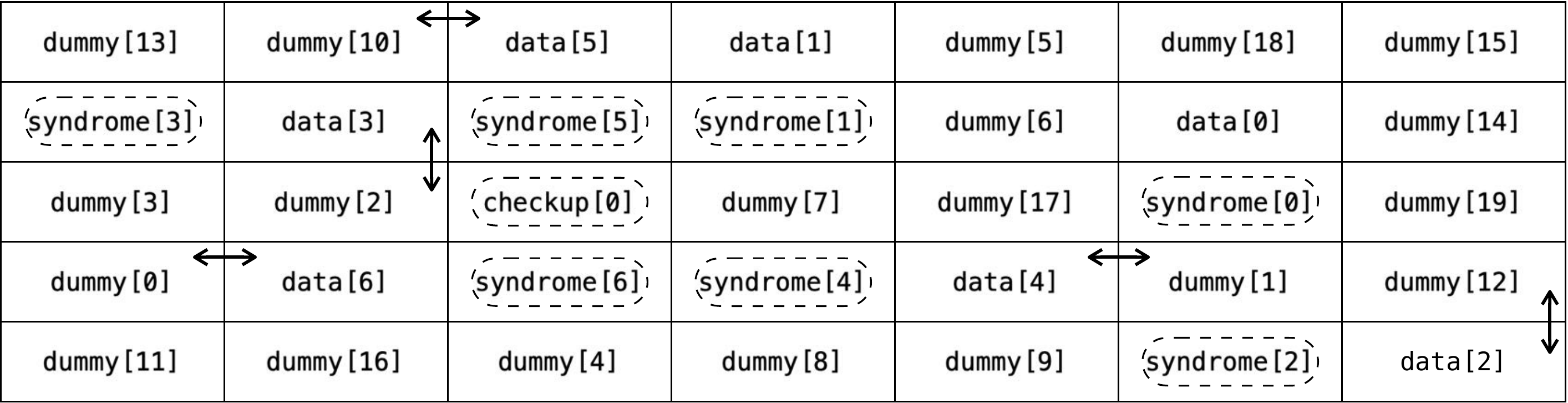}
}
\subfigure[Step 19]{
	\includegraphics[scale=0.35]{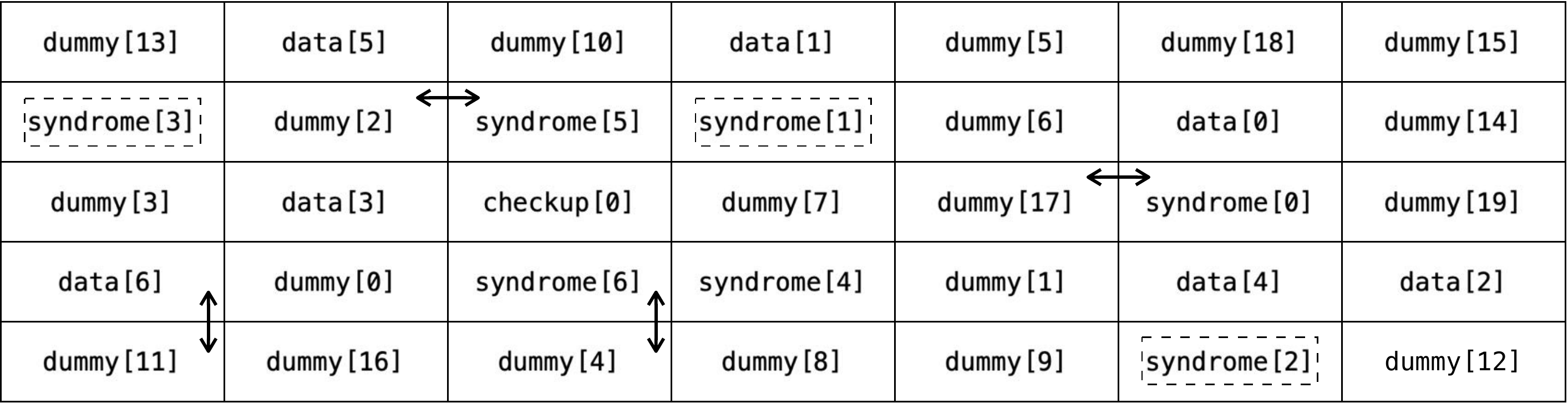}
}
\subfigure[Step 20]{
	\includegraphics[scale=0.35]{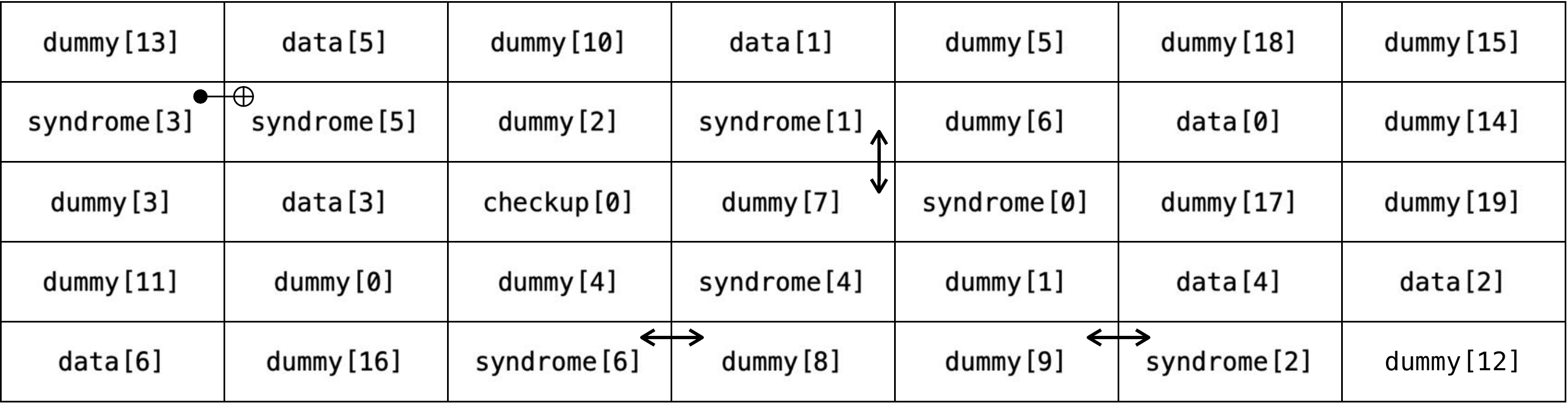}
}
\subfigure[Step 21]{
	\includegraphics[scale=0.35]{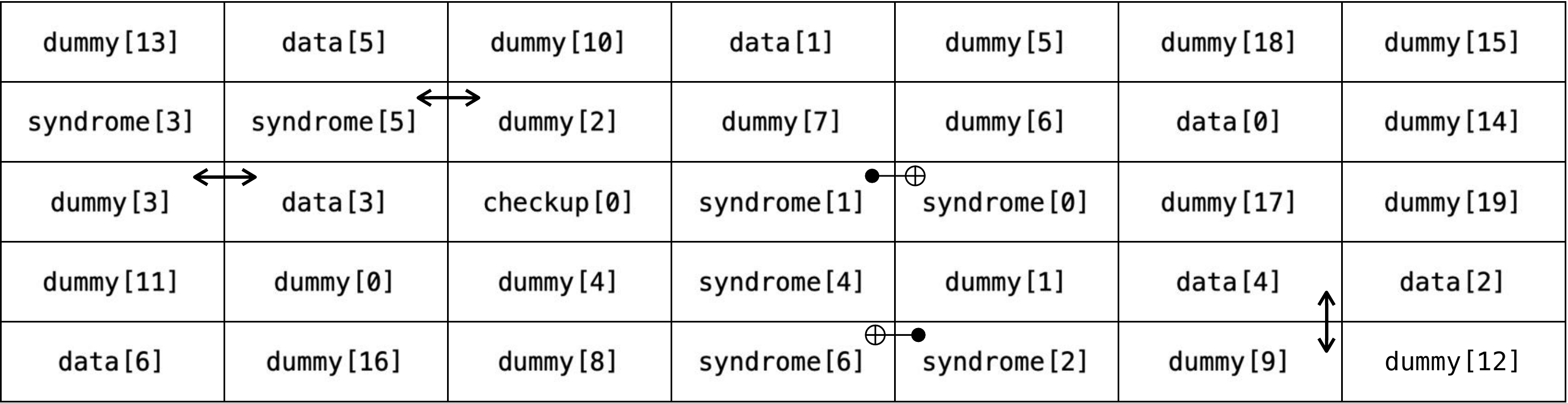}
}
\subfigure[Step 22]{
	\includegraphics[scale=0.35]{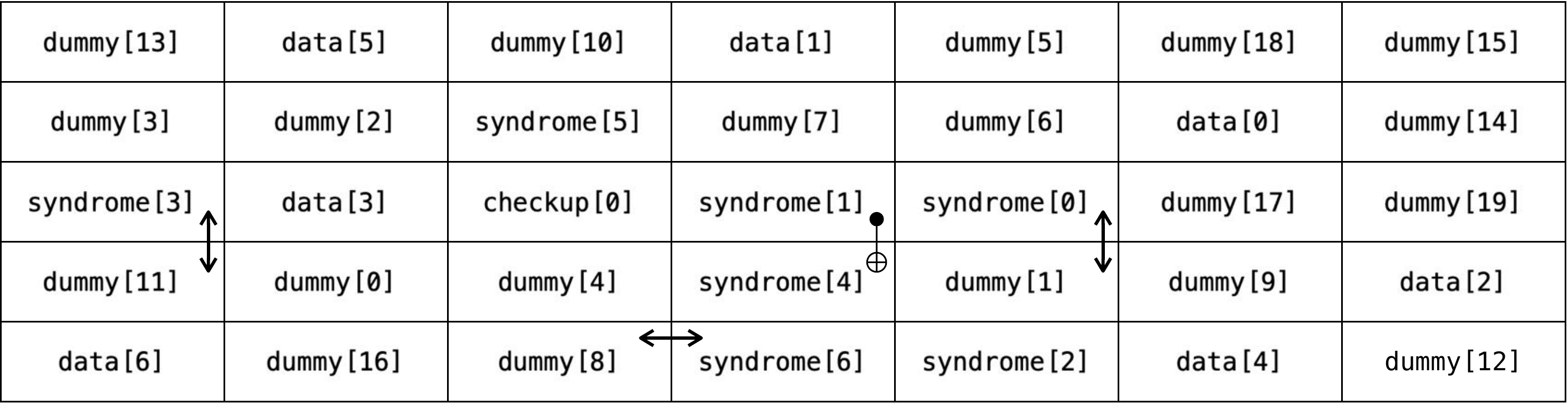}
}
\subfigure[Step 23]{
	\includegraphics[scale=0.35]{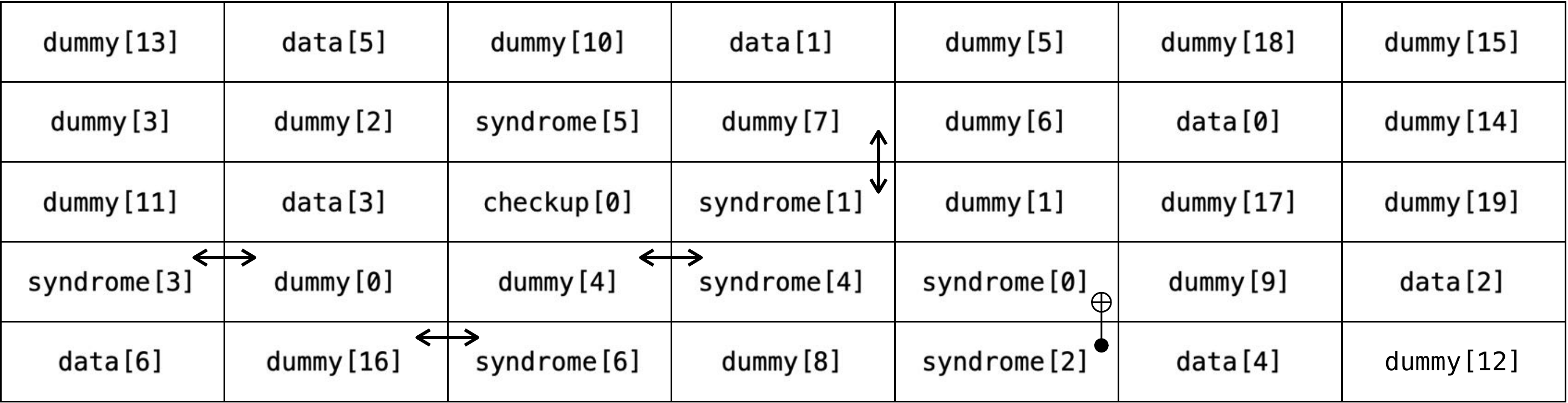}
}
\caption
{
The third part of the fault-tolerant quantum circuit of stabilizer measurement of $[[7, 1, 3]]$ Steane code: Fault-tolerant preparation of the logical state $|0\rangle_L$.
Note that the rectangles, rounded rectangles, hexagons and bi-directed arrow respectively indicate \emph{H}, \emph{PrepZ}, \emph{MeasZ} and \emph{SWAP} gates.
}
\label{fig:steane_syndrome_measure_4}
\end{figure*}

\begin{figure*}[t]
\centering
\subfigure[Step 24]{
	\includegraphics[scale=0.35]{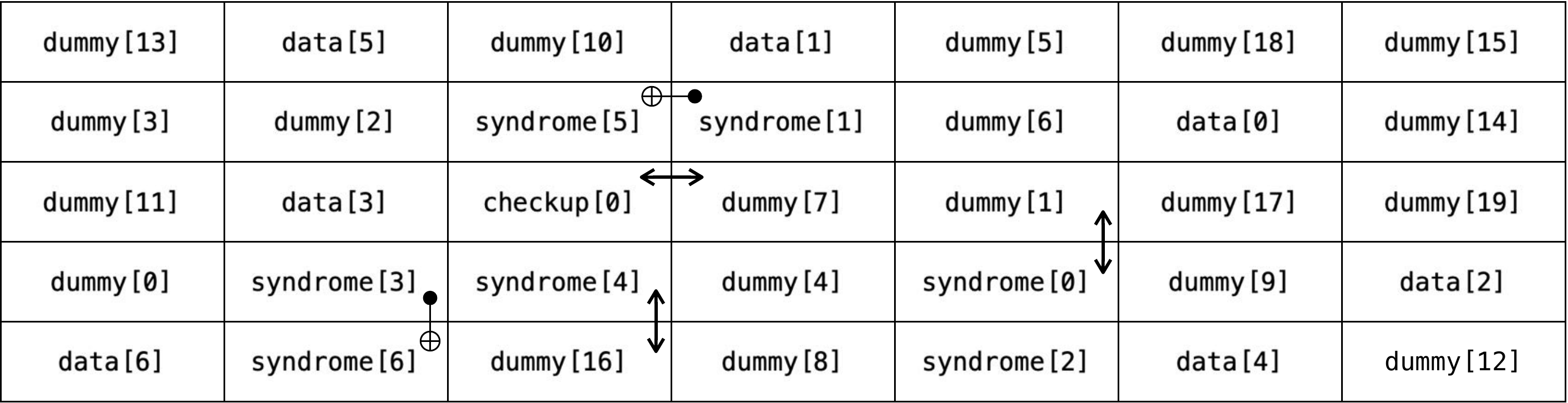}
}
\subfigure[Step 25]{
	\includegraphics[scale=0.35]{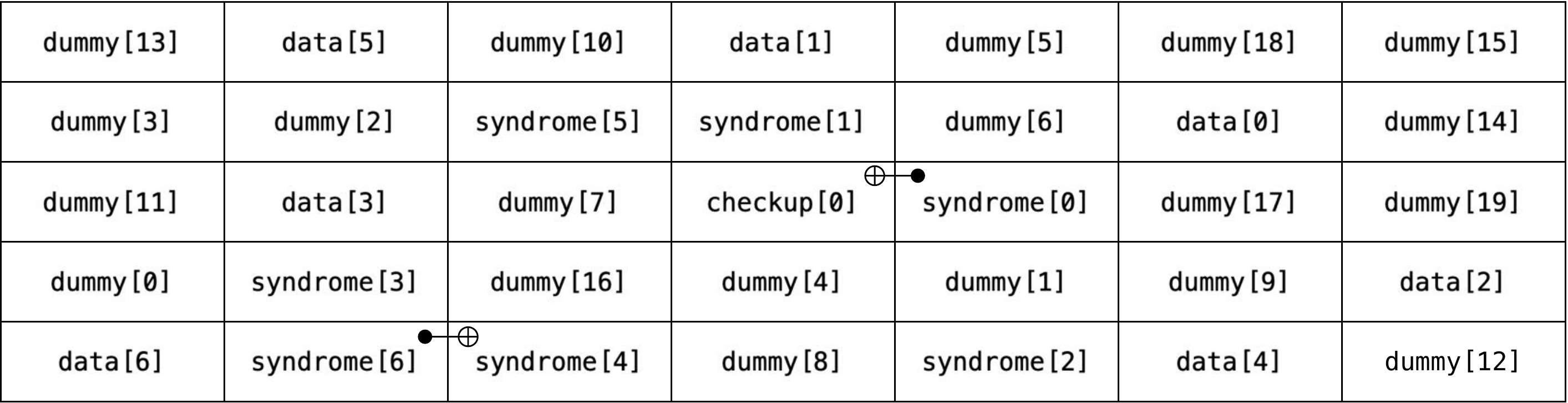}
}
\subfigure[Step 26]{
	\includegraphics[scale=0.35]{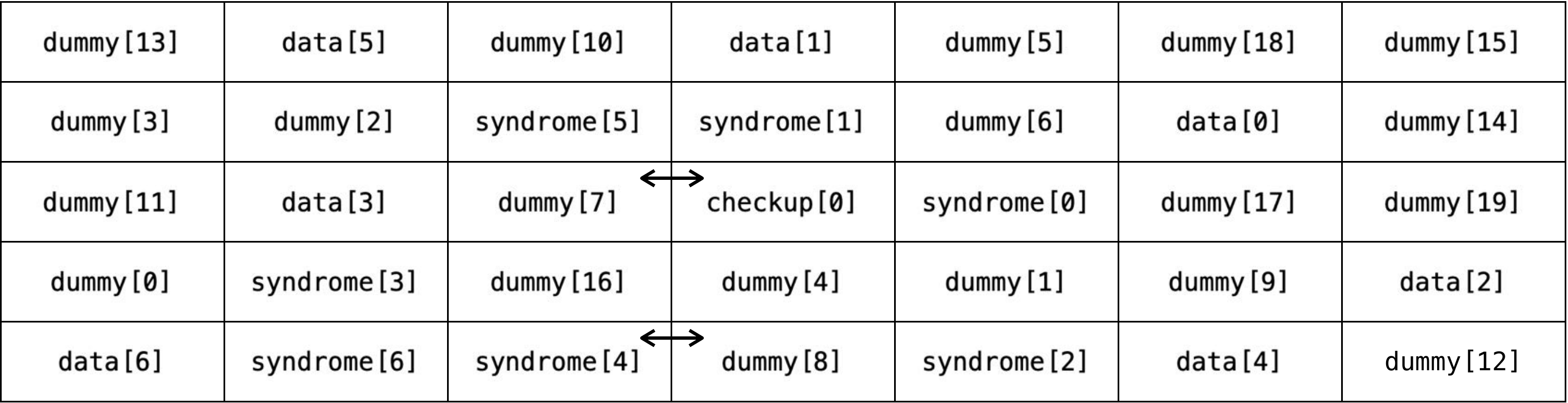}
}
\subfigure[Step 27]{
	\includegraphics[scale=0.35]{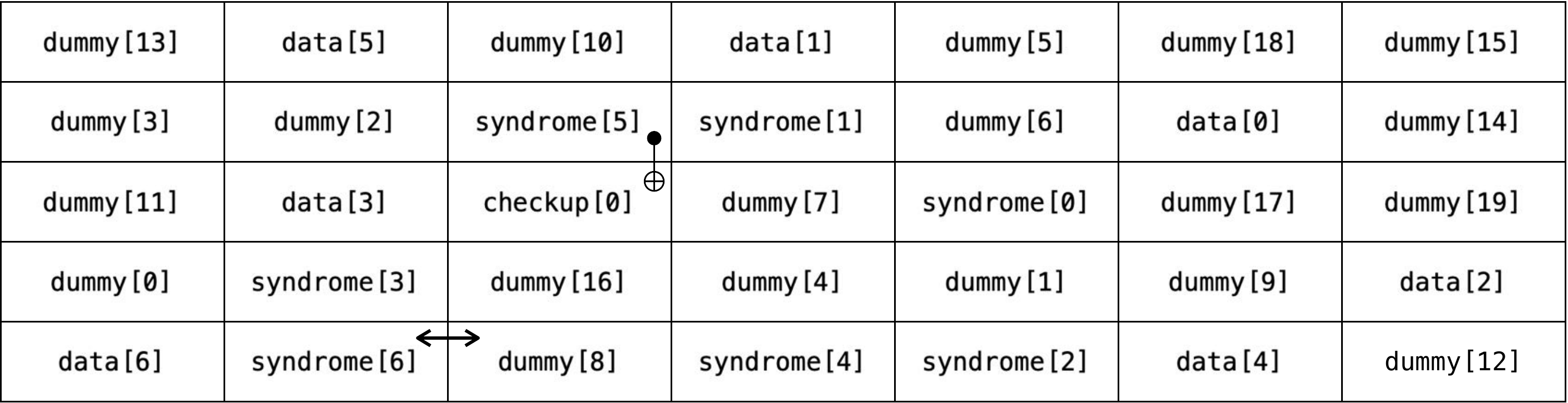}
}
\subfigure[Step 28]{
	\includegraphics[scale=0.35]{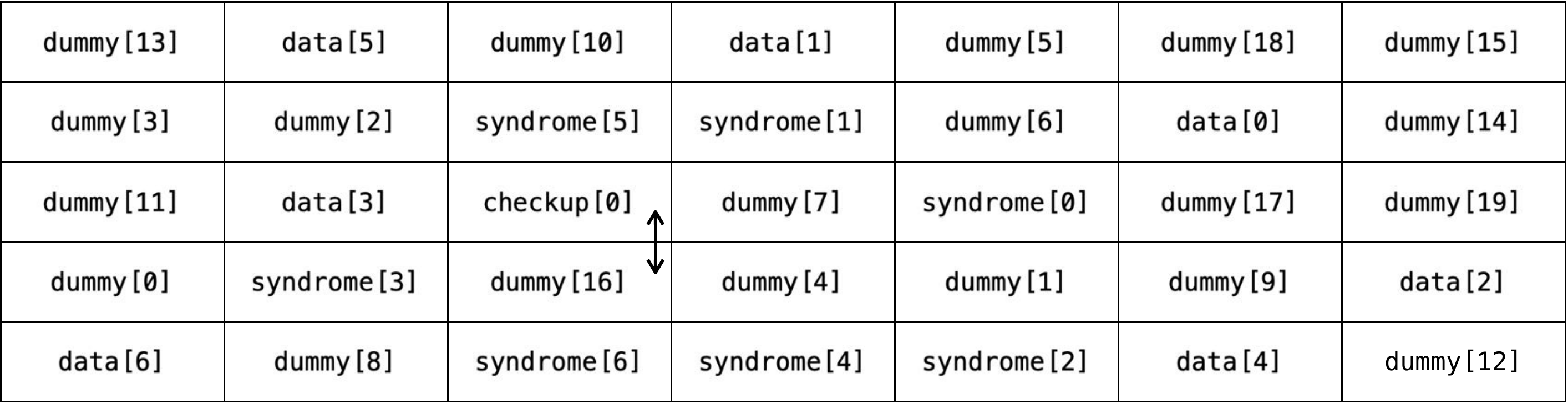}
}
\subfigure[Step 29]{
	\includegraphics[scale=0.35]{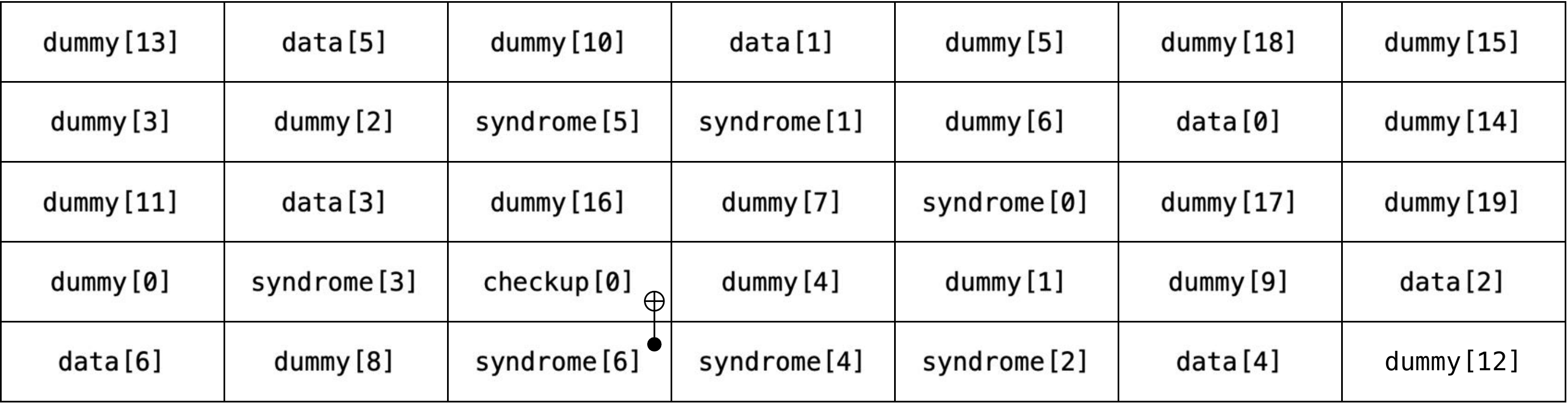}
}
\caption
{
(Continued from Figure~\ref{fig:steane_syndrome_measure_4}) The third part of the fault-tolerant quantum circuit of stabilizer measurement of $[[7, 1, 3]]$ Steane code: Fault-tolerant preparation of the logical state $|0\rangle_L$.
Note that the rectangles, rounded rectangles, hexagons and bi-directed arrow respectively indicate \emph{H}, \emph{PrepZ}, \emph{MeasZ} and \emph{SWAP} gates.
}
\label{fig:steane_syndrome_measure_5}
\end{figure*}

\begin{figure*}[t]
\centering
\subfigure[Step 30 (Barrier)]{
	\includegraphics[scale=0.35]{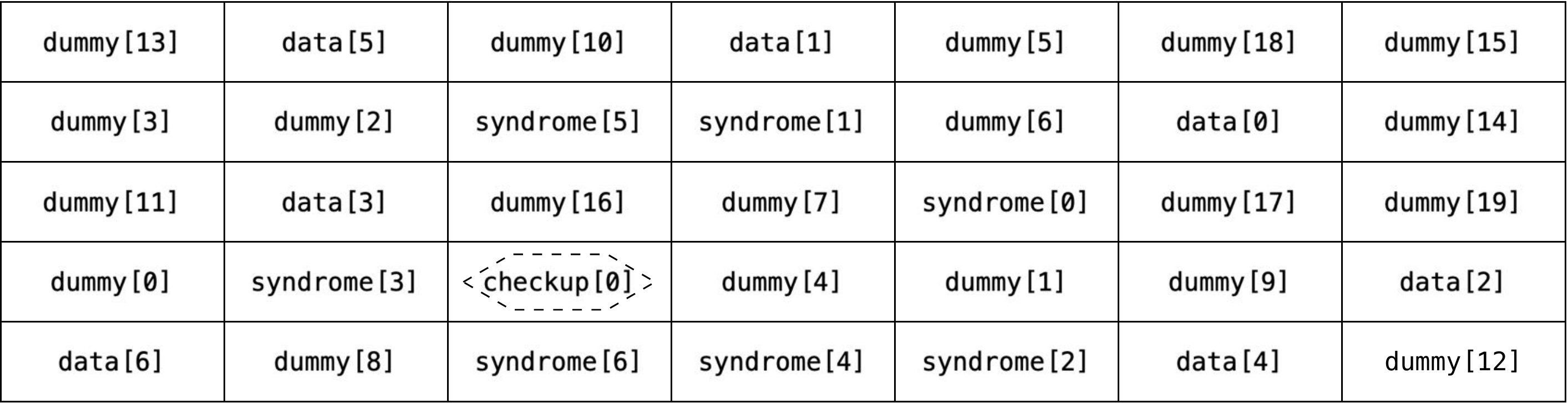}
}
\caption
{
(Continued from Figure~\ref{fig:steane_syndrome_measure_5}) The third part of the fault-tolerant quantum circuit of stabilizer measurement of $[[7, 1, 3]]$ Steane code: Fault-tolerant preparation of the logical state $|0\rangle_L$.
Note that the rectangles, rounded rectangles, hexagons and bi-directed arrow respectively indicate \emph{H}, \emph{PrepZ}, \emph{MeasZ} and \emph{SWAP} gates.
}
\label{fig:steane_syndrome_measure_6}
\end{figure*}

\begin{figure*}[t]
\centering
\subfigure[Step 31]{
	\includegraphics[scale=0.35]{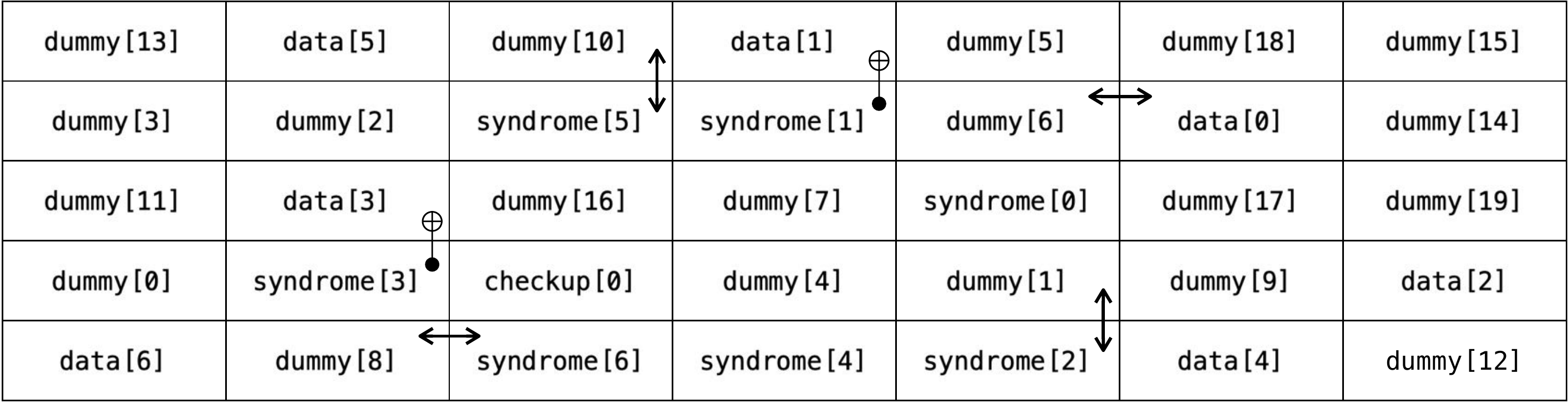}
}
\subfigure[Step 32]{
	\includegraphics[scale=0.35]{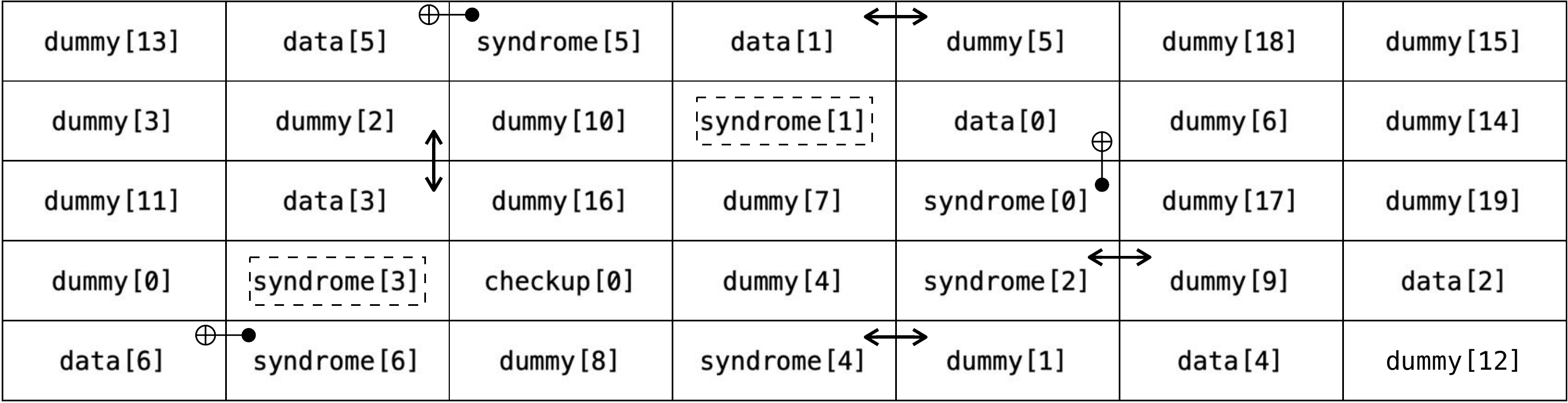}
}
\subfigure[Step 33]{
	\includegraphics[scale=0.35]{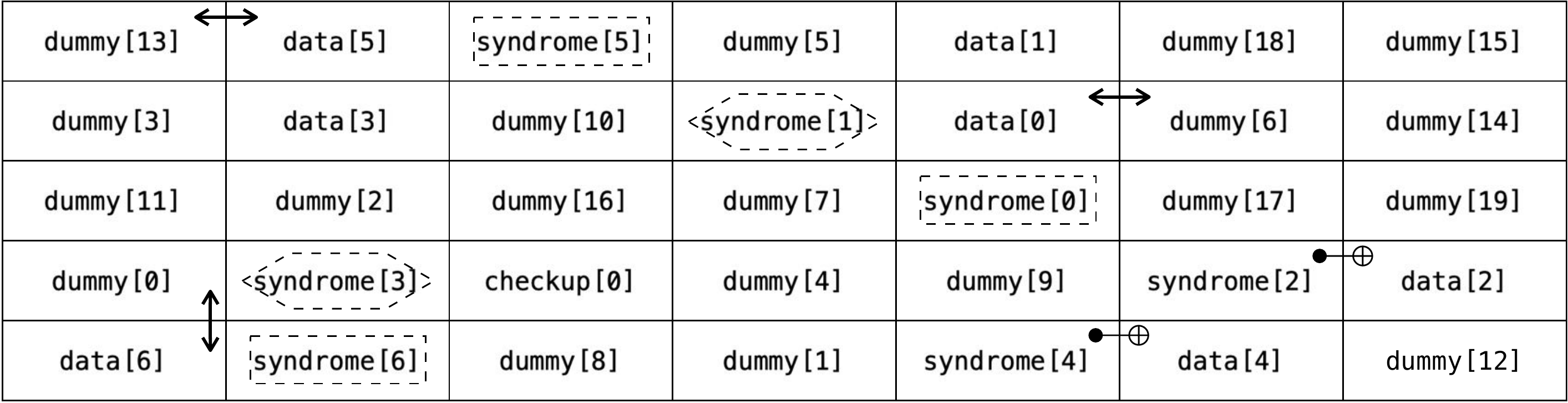}
}
\subfigure[Step 34]{
	\includegraphics[scale=0.35]{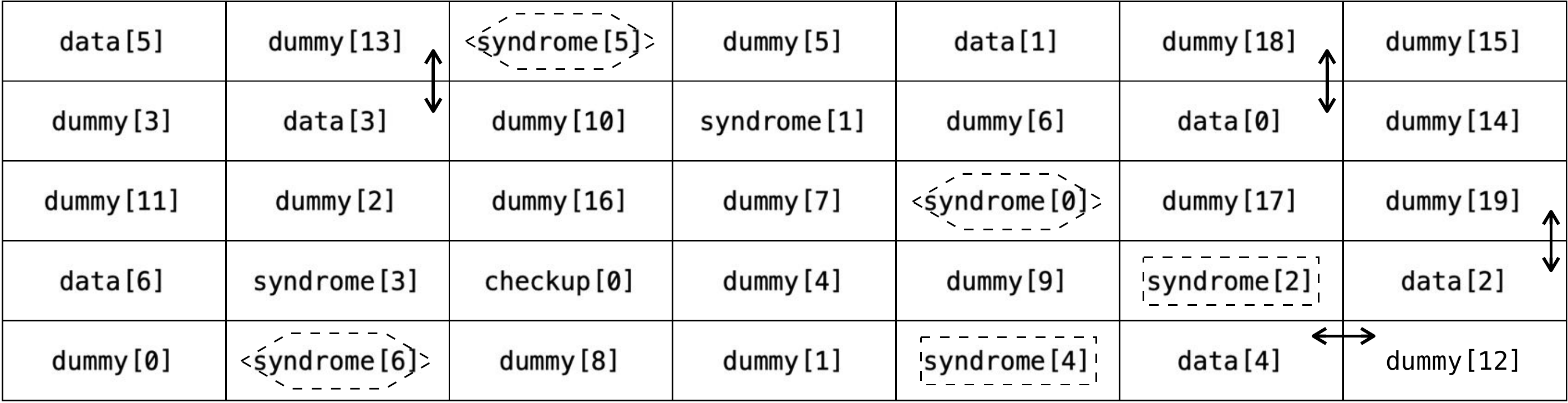}
}
\subfigure[Step 35]{
	\includegraphics[scale=0.35]{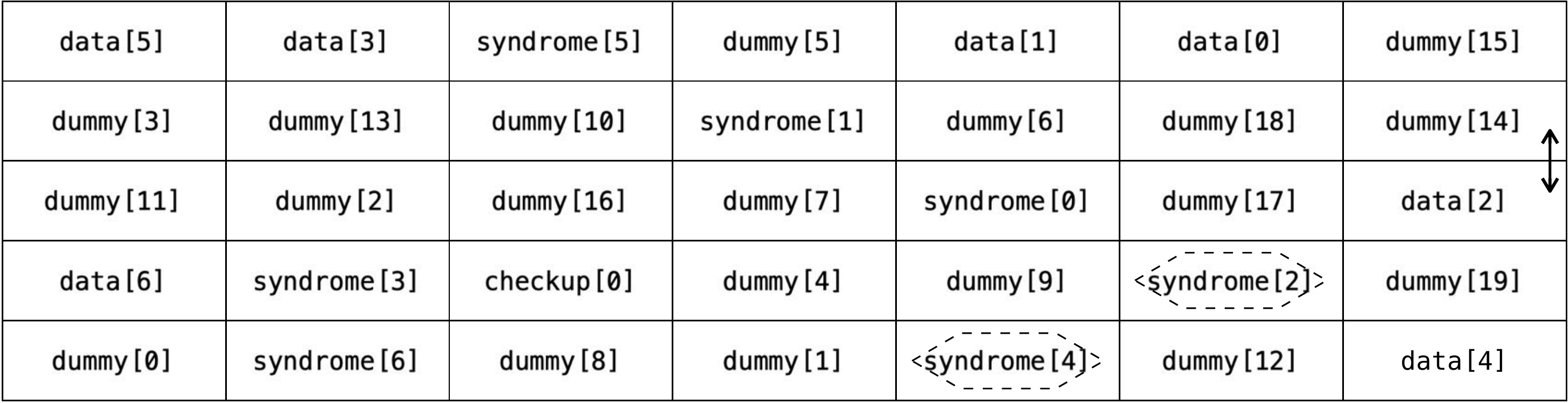}
}
\subfigure[Final Qubit Mapping]{
	\includegraphics[scale=0.35]{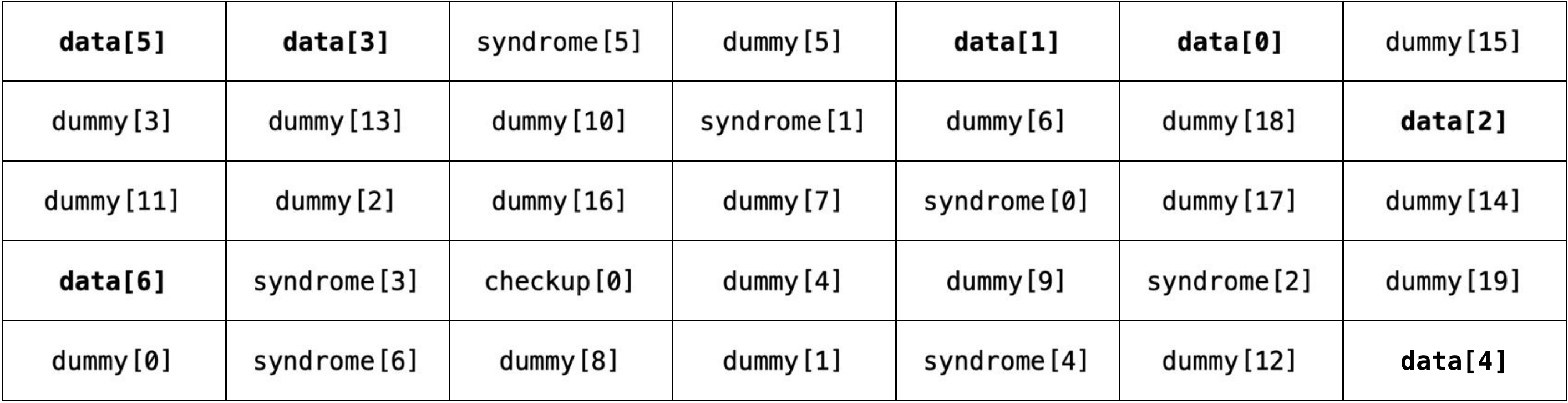}
}
\caption
{
The fourth part of the fault-tolerant quantum circuit of stabilizer measurement of $[[7, 1, 3]]$ Steane code: Transversal \emph{CNOT} between data qubits and syndrome qubits.
By the \emph{Move-Back} operations, the data qubits after all the quantum operations are placed in their initial positions (see Figure~\ref{fig:steane_initial_mapping}).
Note that the rectangles, rounded rectangles, hexagons and bi-directed arrow respectively indicate \emph{H}, \emph{PrepZ}, \emph{MeasZ} and \emph{SWAP} gates.
}
\label{fig:steane_syndrome_measure_7}
\end{figure*}

\begin{figure*}[t]
\centering
\includegraphics[scale=0.6]{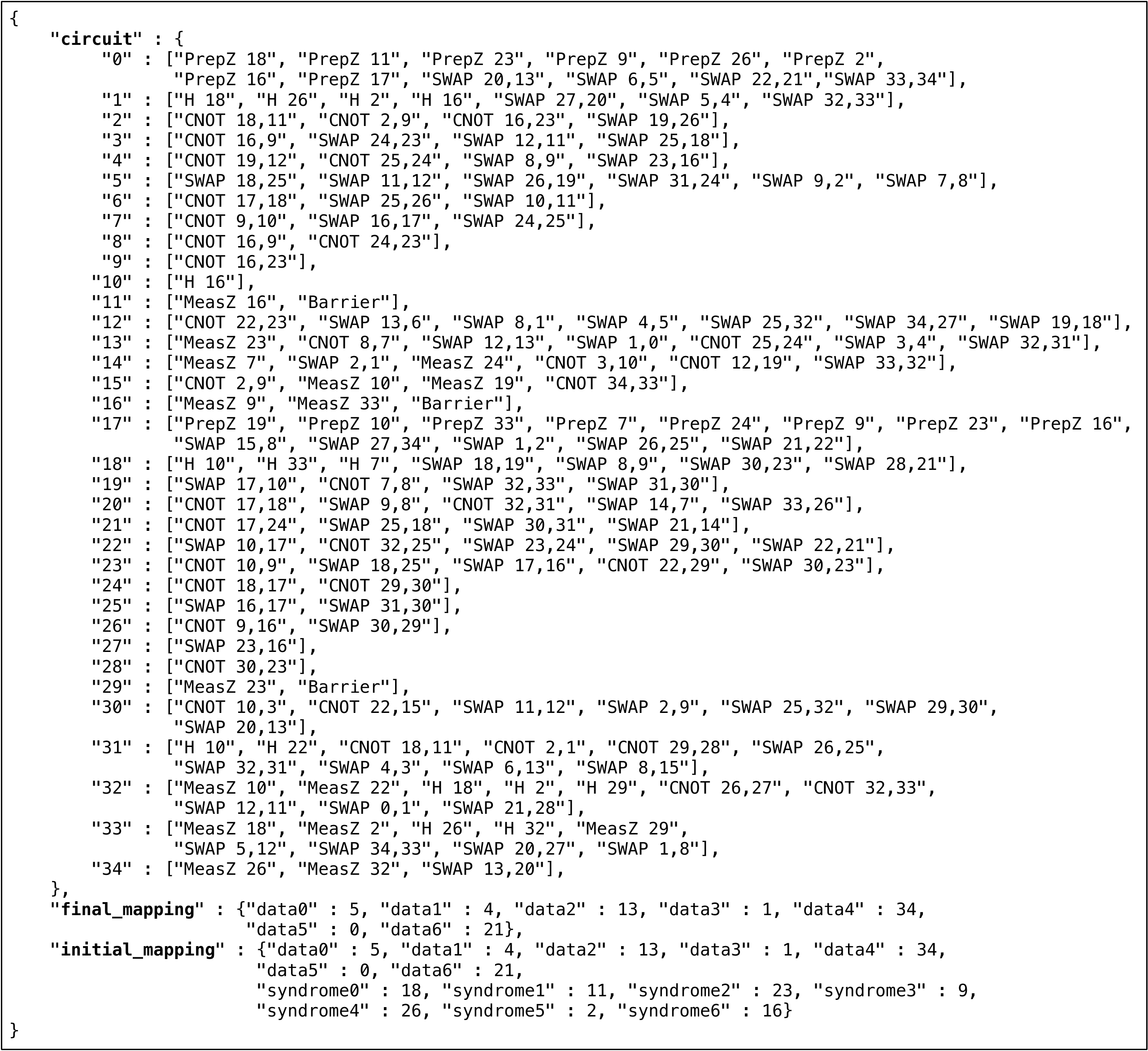}	
\caption{
The JSON format representation of the fault-tolerant quantum circuit shown in Figures~\ref{fig:steane_syndrome_measure_1}~$\sim$~\ref{fig:steane_syndrome_measure_7}.
}
\label{fig:circuit_json_format}
\end{figure*}

\begin{figure*}[t]
\centering
\subfigure[Initial Mapping based on Vertically Extended Layout]{
	\includegraphics[scale=0.35]{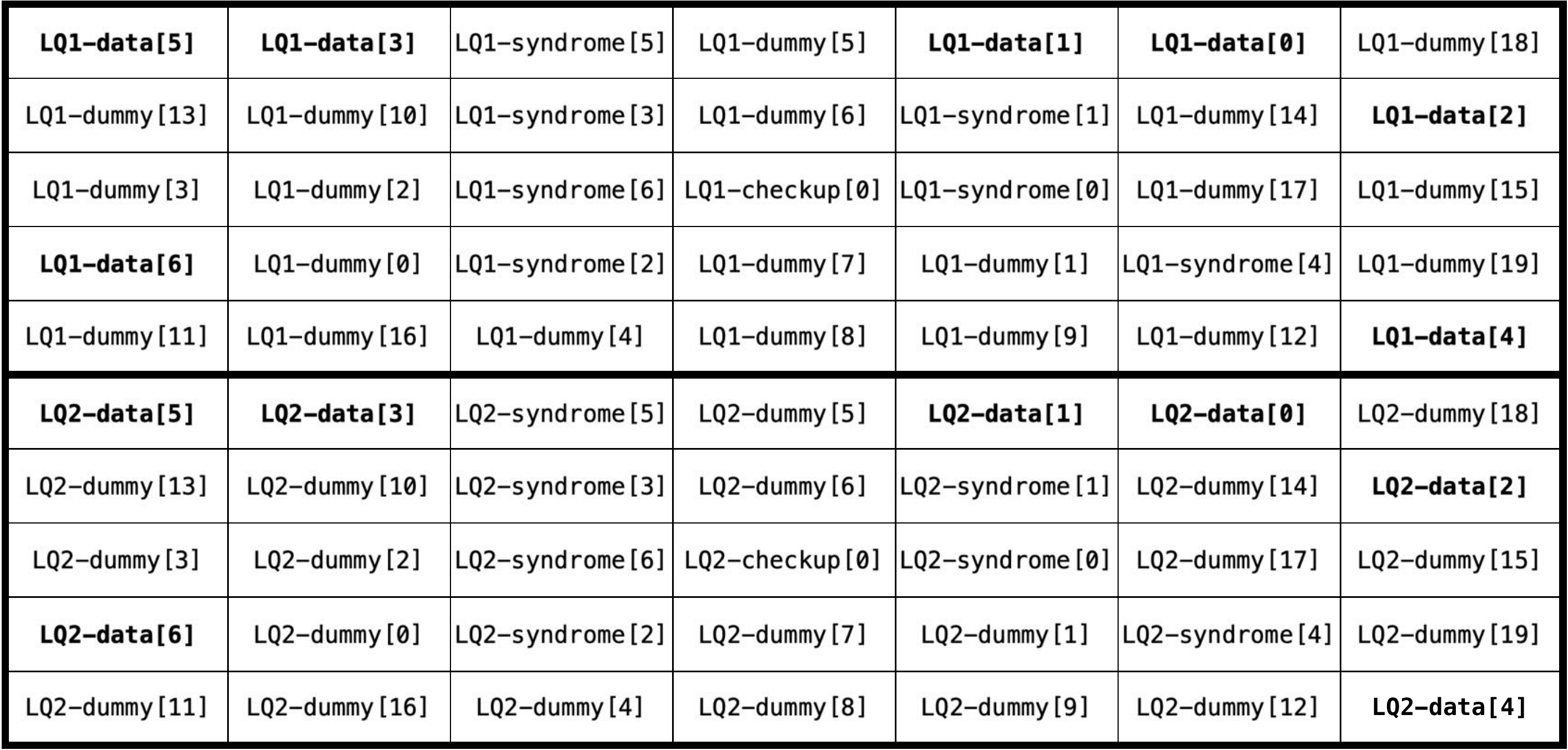}
}
\subfigure[Step 1]{
	\includegraphics[scale=0.35]{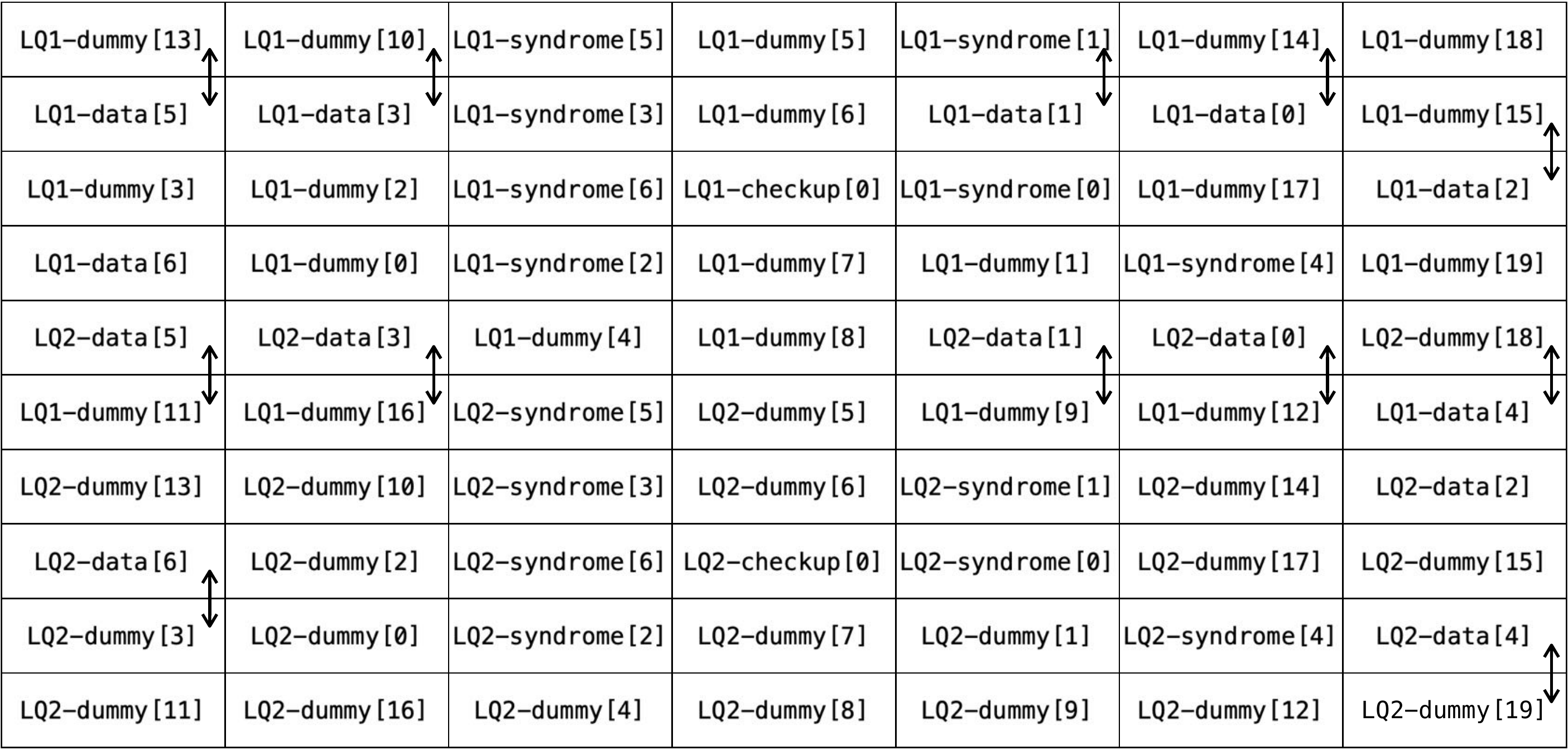}
}
\subfigure[Step 2]{
	\includegraphics[scale=0.35]{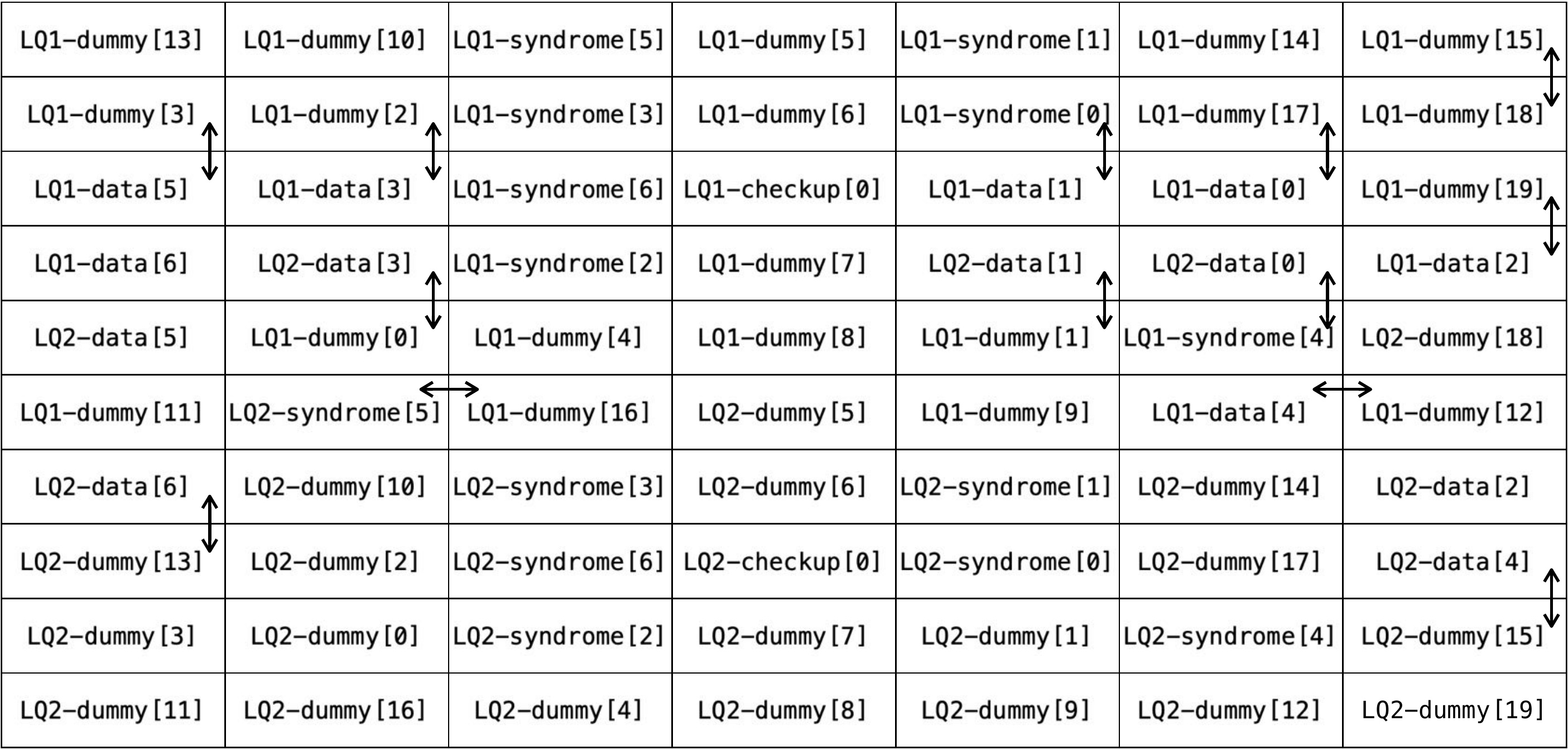}
}
\subfigure[Step 3]{
	\includegraphics[scale=0.35]{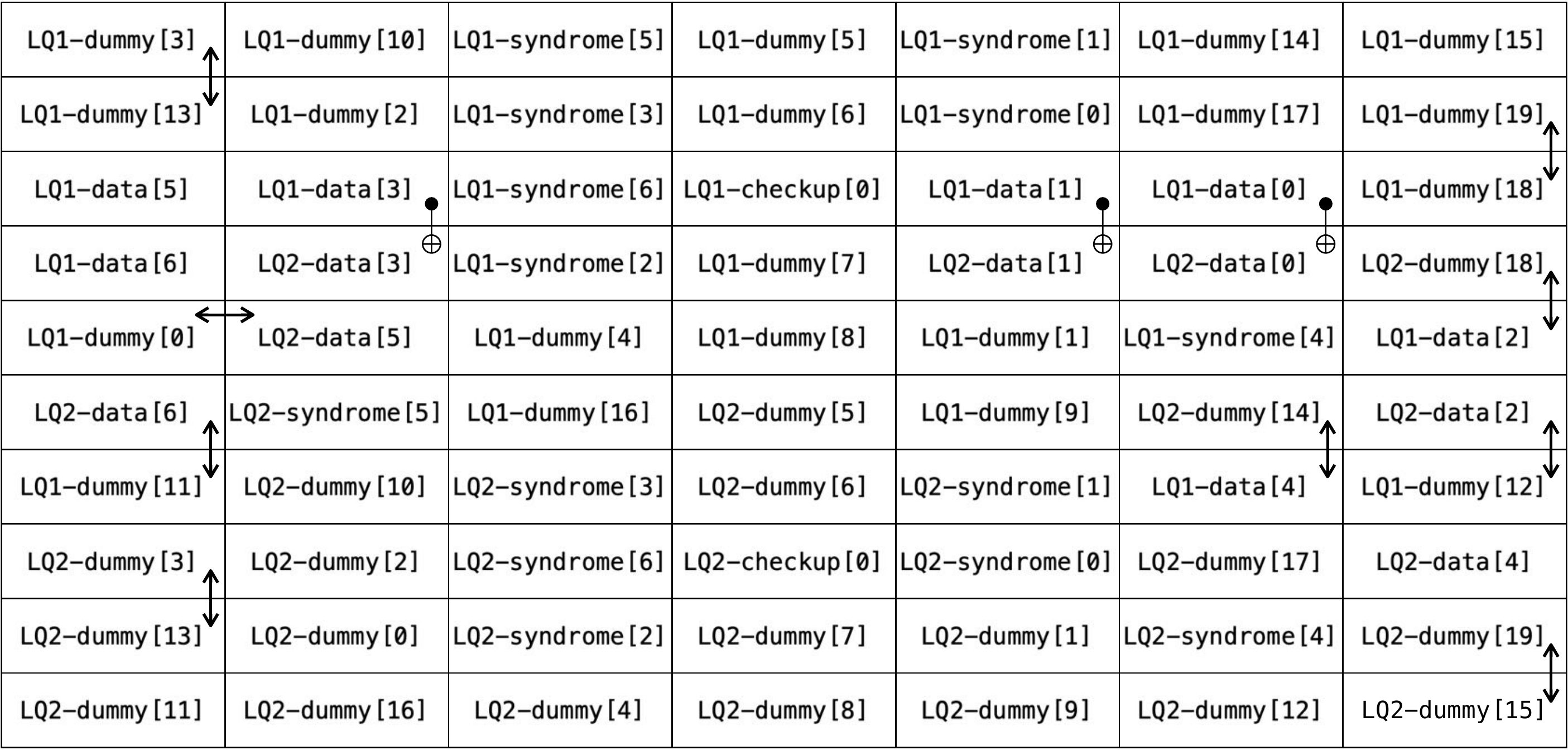}
}
\caption
{
Logical \emph{CNOT} gate operations for qubits arranged vertically.
(a) The vertically extended qubit layout of logical qubits, LQ1 and LQ2.
Note that the bi-directed arrow indicates \emph{SWAP} gate.
}
\label{fig:steane_cnot_vertical_1}
\end{figure*}

\begin{figure*}[t]
\centering
\subfigure[Step 4]{
	\includegraphics[scale=0.35]{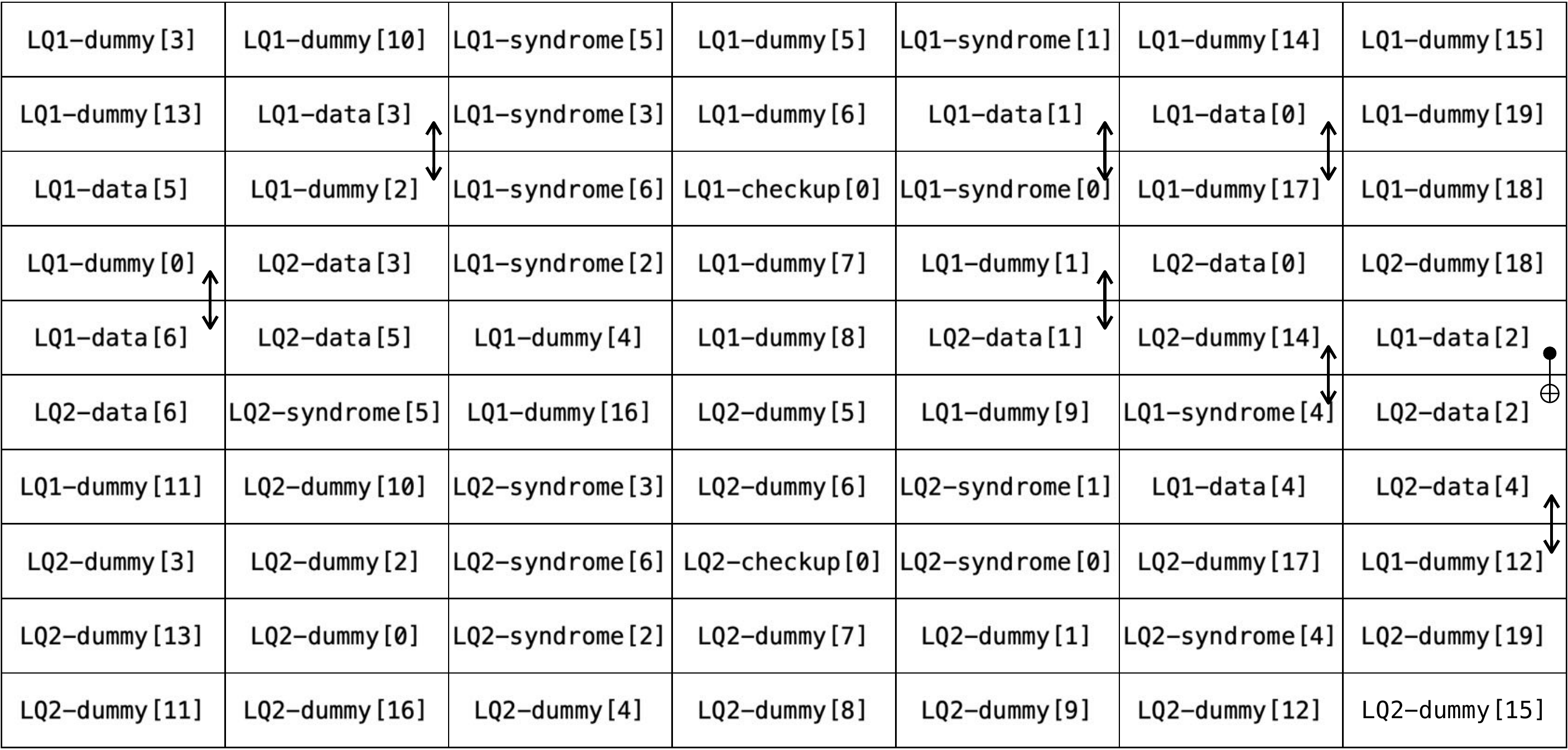}
}
\subfigure[Step 5]{
	\includegraphics[scale=0.35]{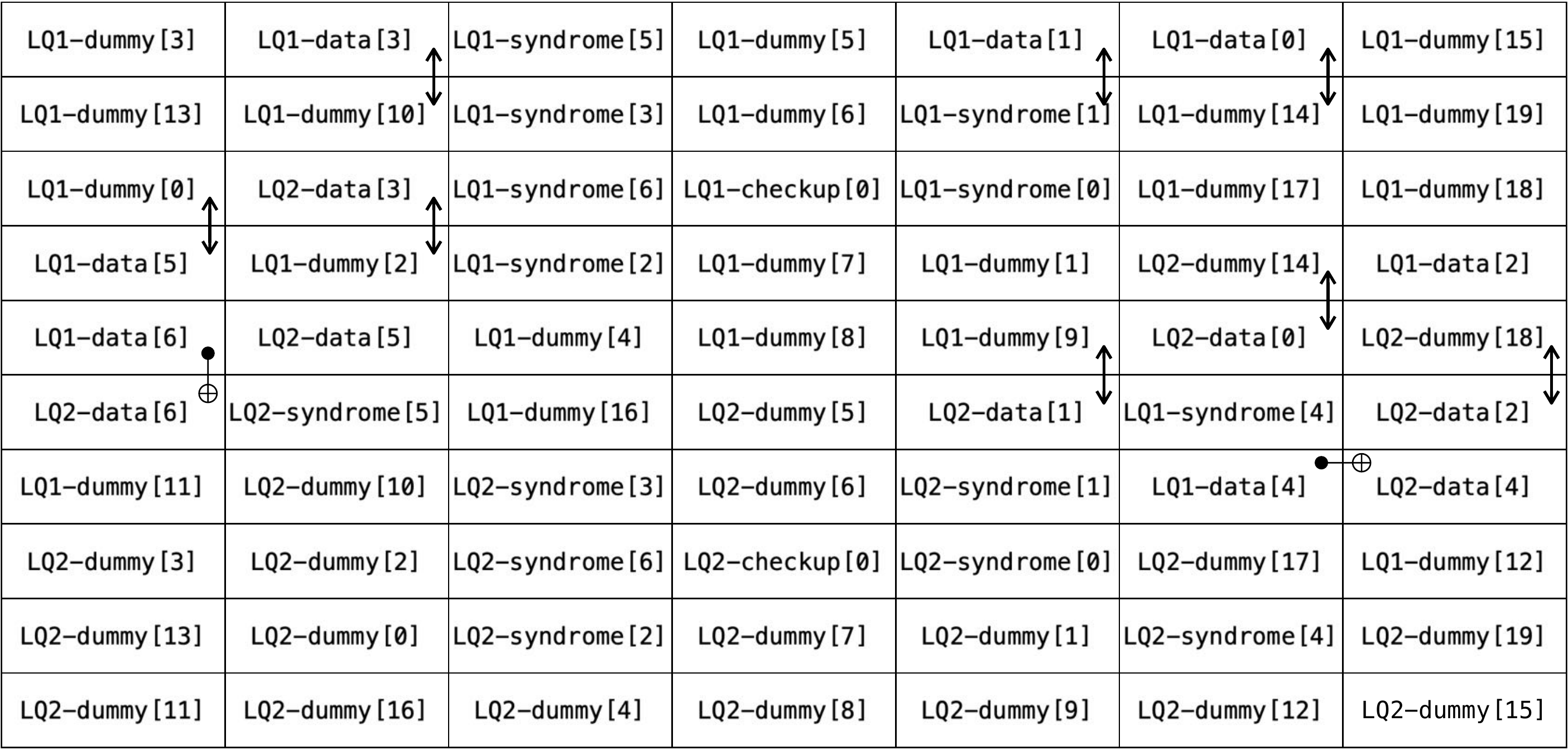}
}
\subfigure[Step 6]{
	\includegraphics[scale=0.35]{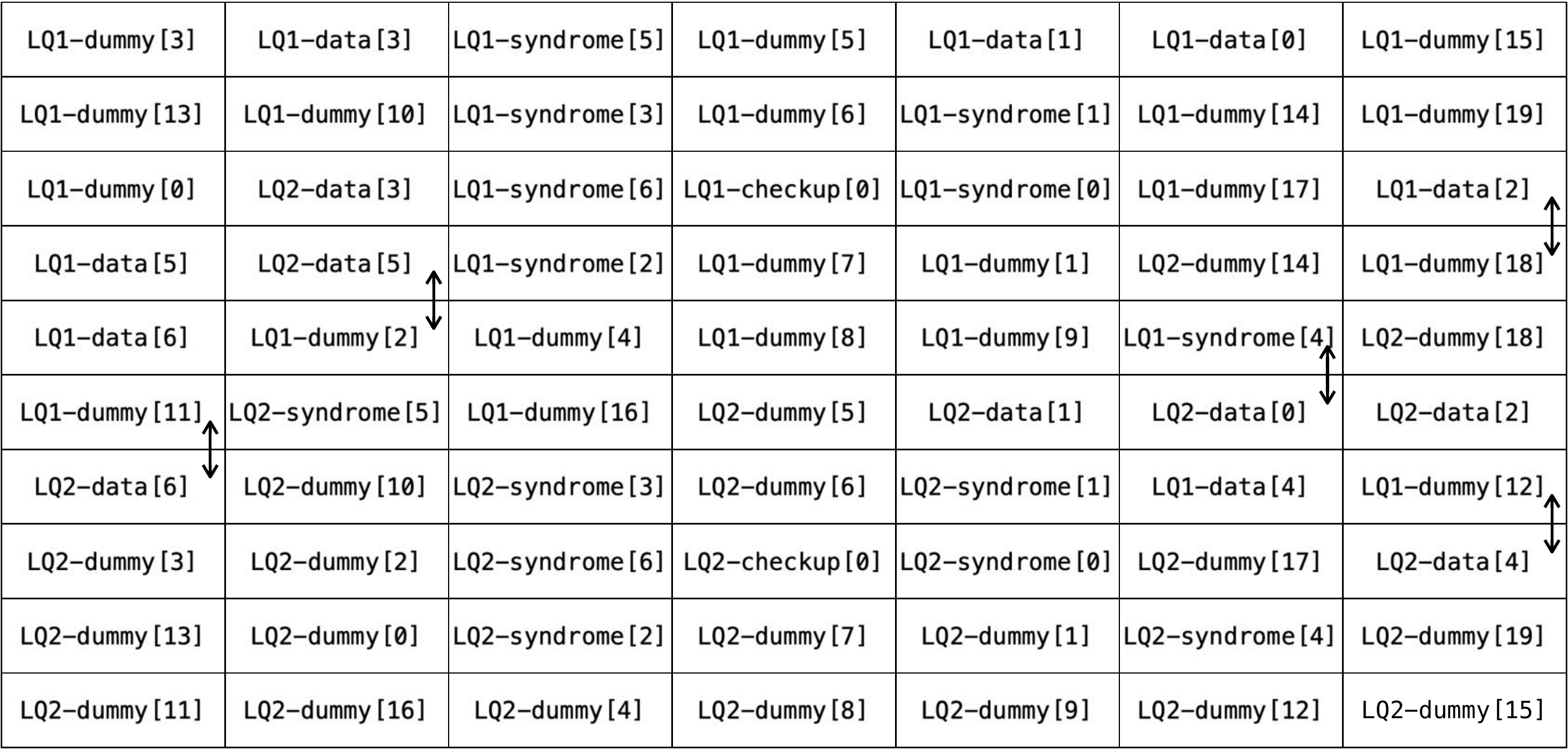}
}
\subfigure[Step 7]{
	\includegraphics[scale=0.35]{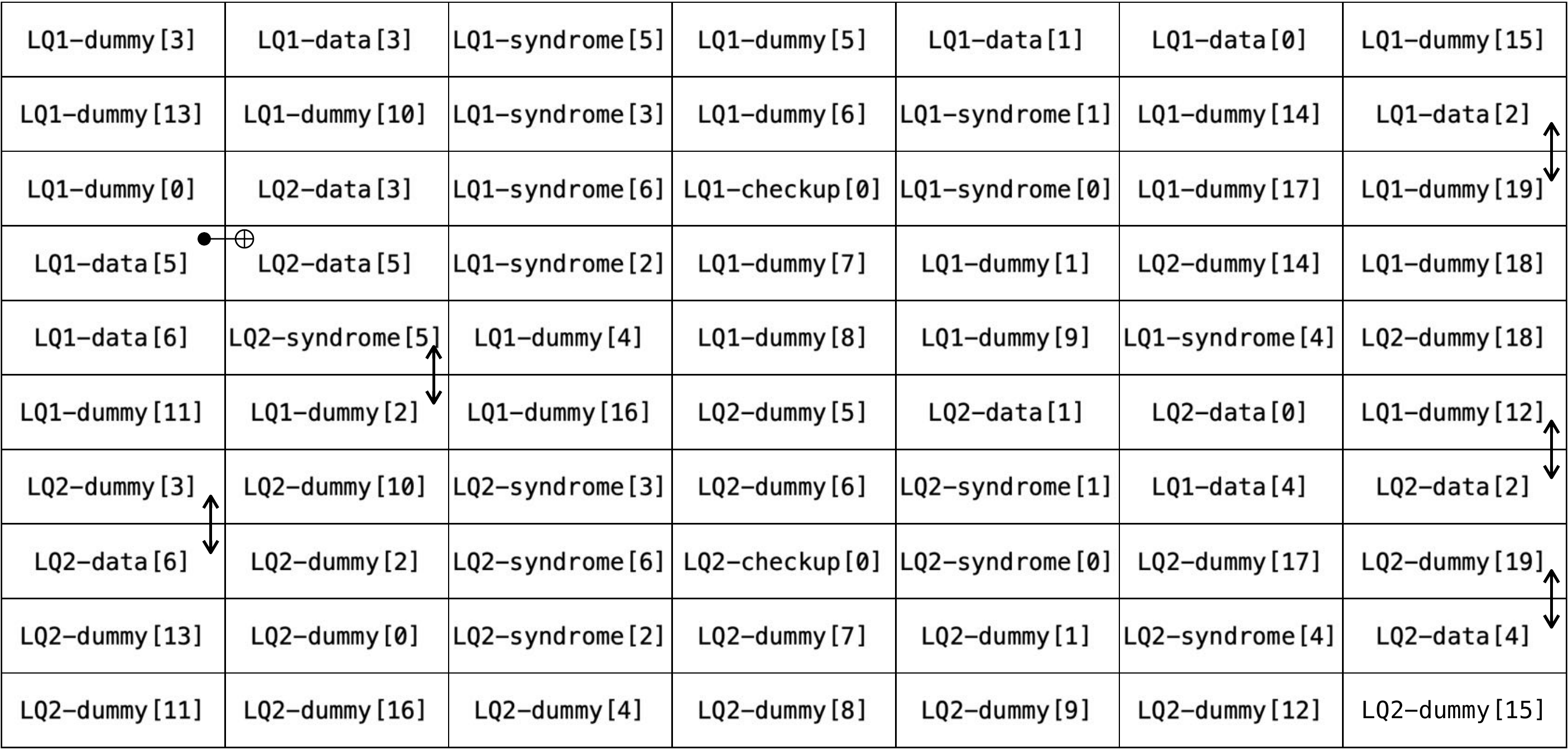}
}
\caption
{
(Continued from Figure~\ref{fig:steane_cnot_vertical_1}) Logical \emph{CNOT} gate operations for qubits arranged vertically.
Note that the bi-directed arrow indicates \emph{SWAP} gate.
}
\label{fig:steane_cnot_vertical_2}
\end{figure*}

\begin{figure*}[t]
\centering
\subfigure[Step 8]{
	\includegraphics[scale=0.35]{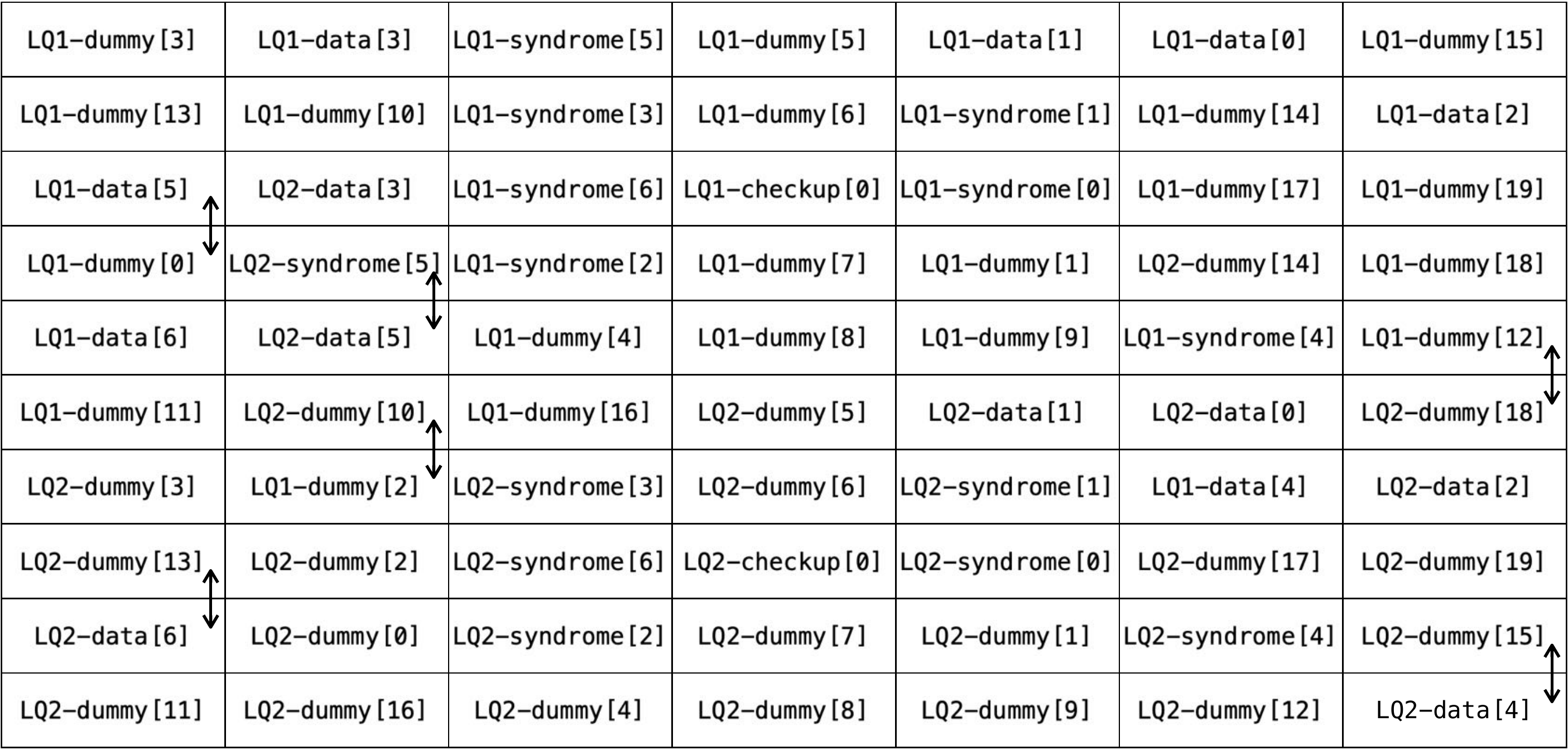}
}
\subfigure[Step 9]{
	\includegraphics[scale=0.35]{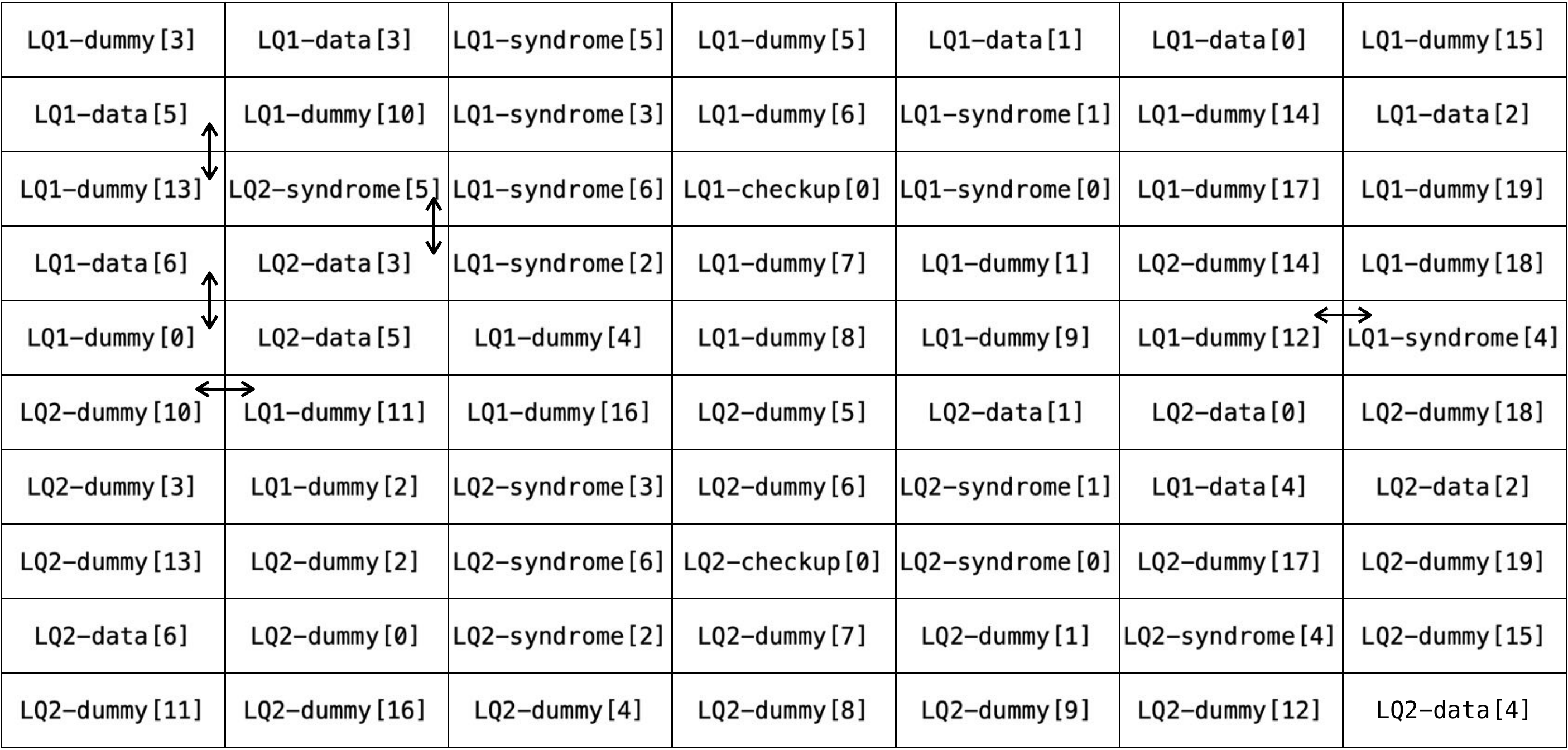}
}
\subfigure[Step 10]{
	\includegraphics[scale=0.35]{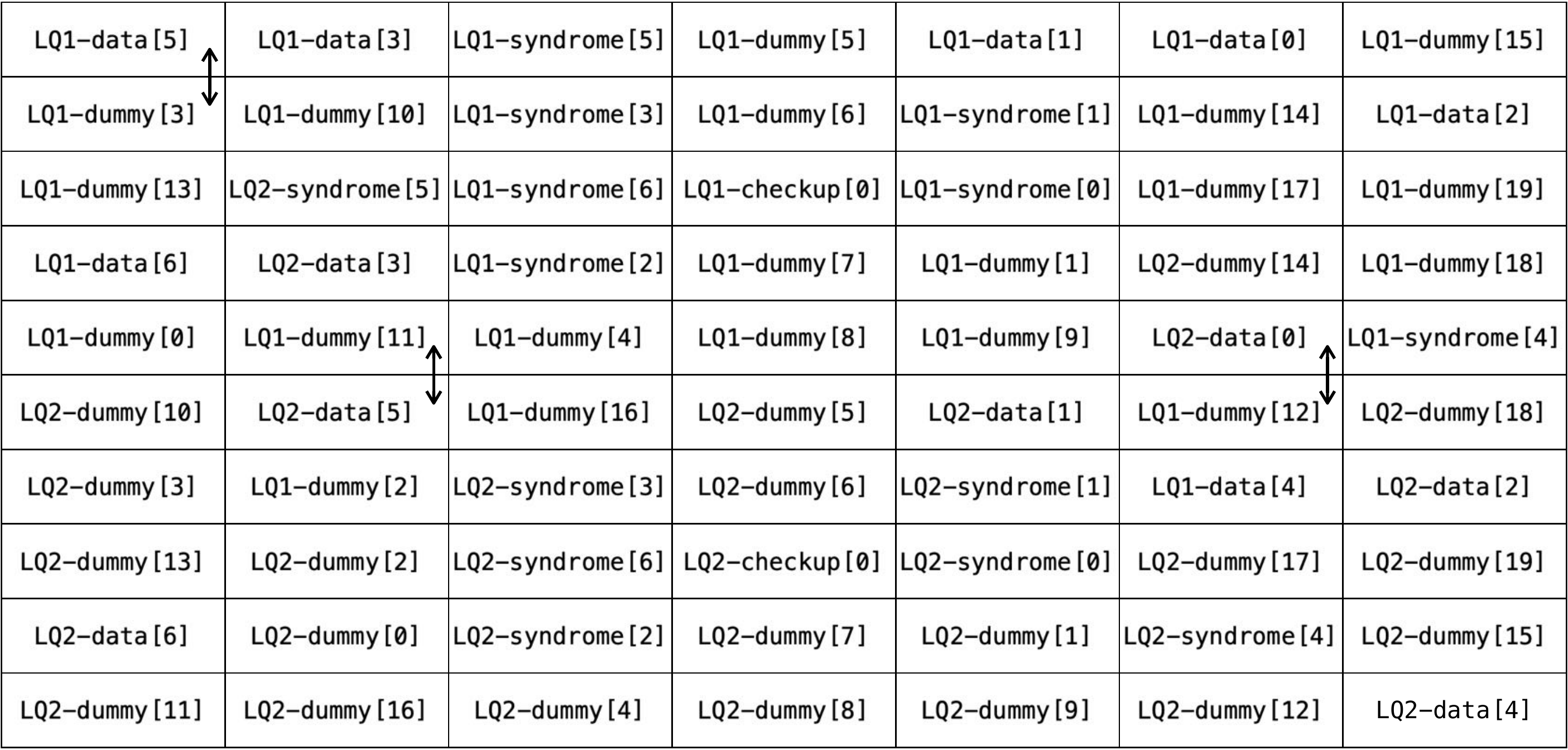}
}
\subfigure[Step 11]{
	\includegraphics[scale=0.35]{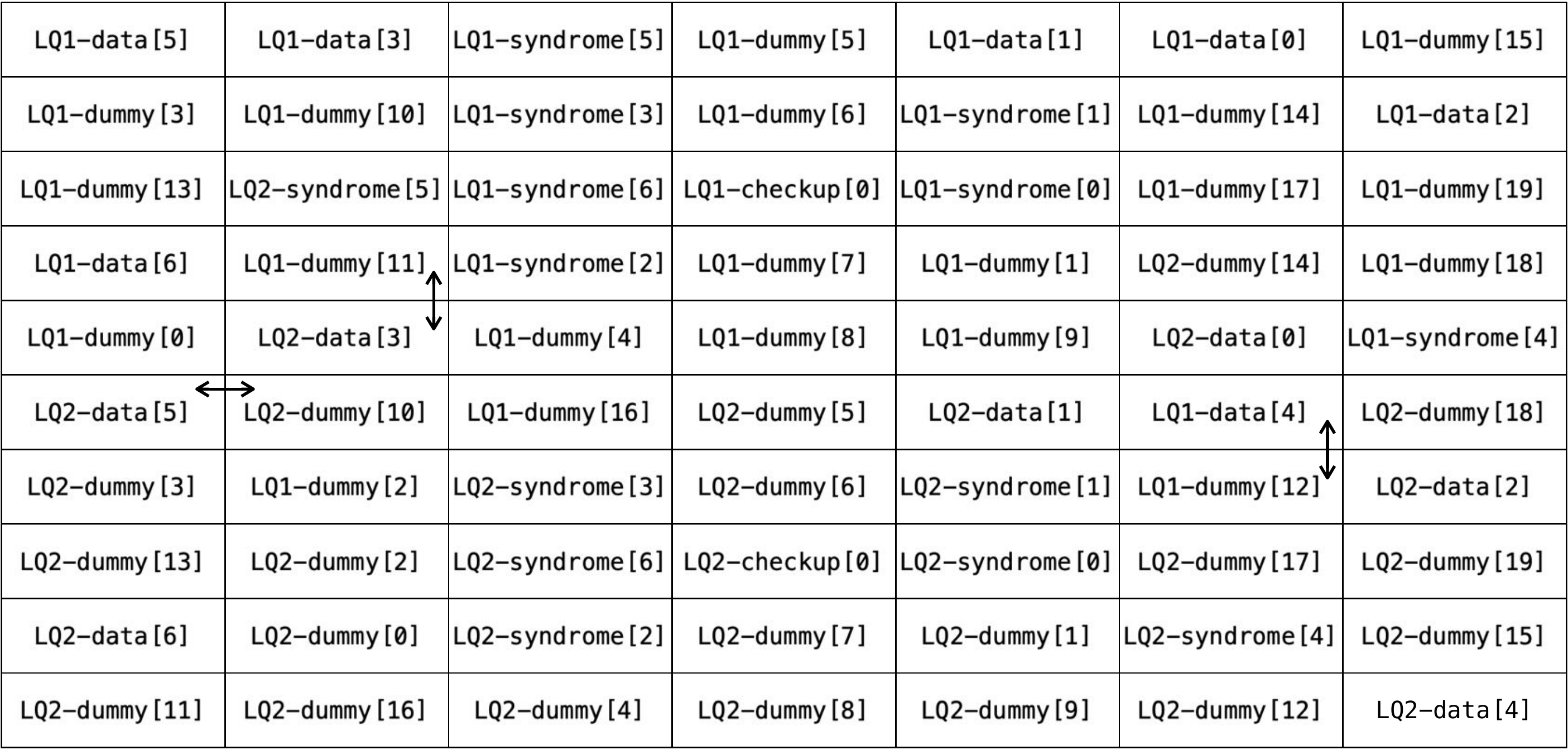}
}
\caption
{
(Continued from Figure~\ref{fig:steane_cnot_vertical_2}) Logical \emph{CNOT} gate operations for qubits arranged vertically.
Note that the bi-directed arrow indicates \emph{SWAP} gate.
}
\label{fig:steane_cnot_vertical_3}
\end{figure*}

\begin{figure*}[t]
\centering
\subfigure[Step 12]{
	\includegraphics[scale=0.35]{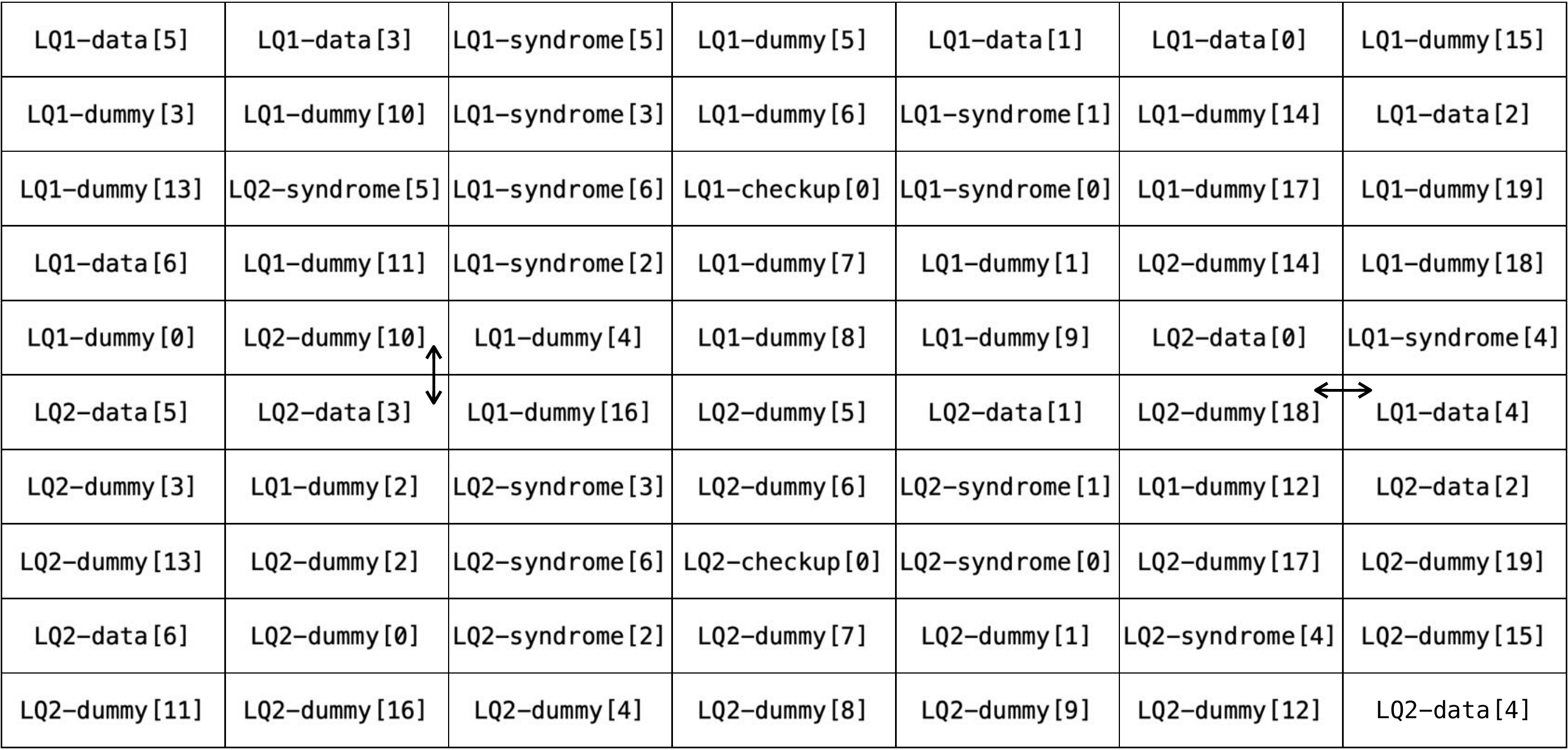}
}
\subfigure[Step 13 (Final Qubit Mapping)]{
	\includegraphics[scale=0.35]{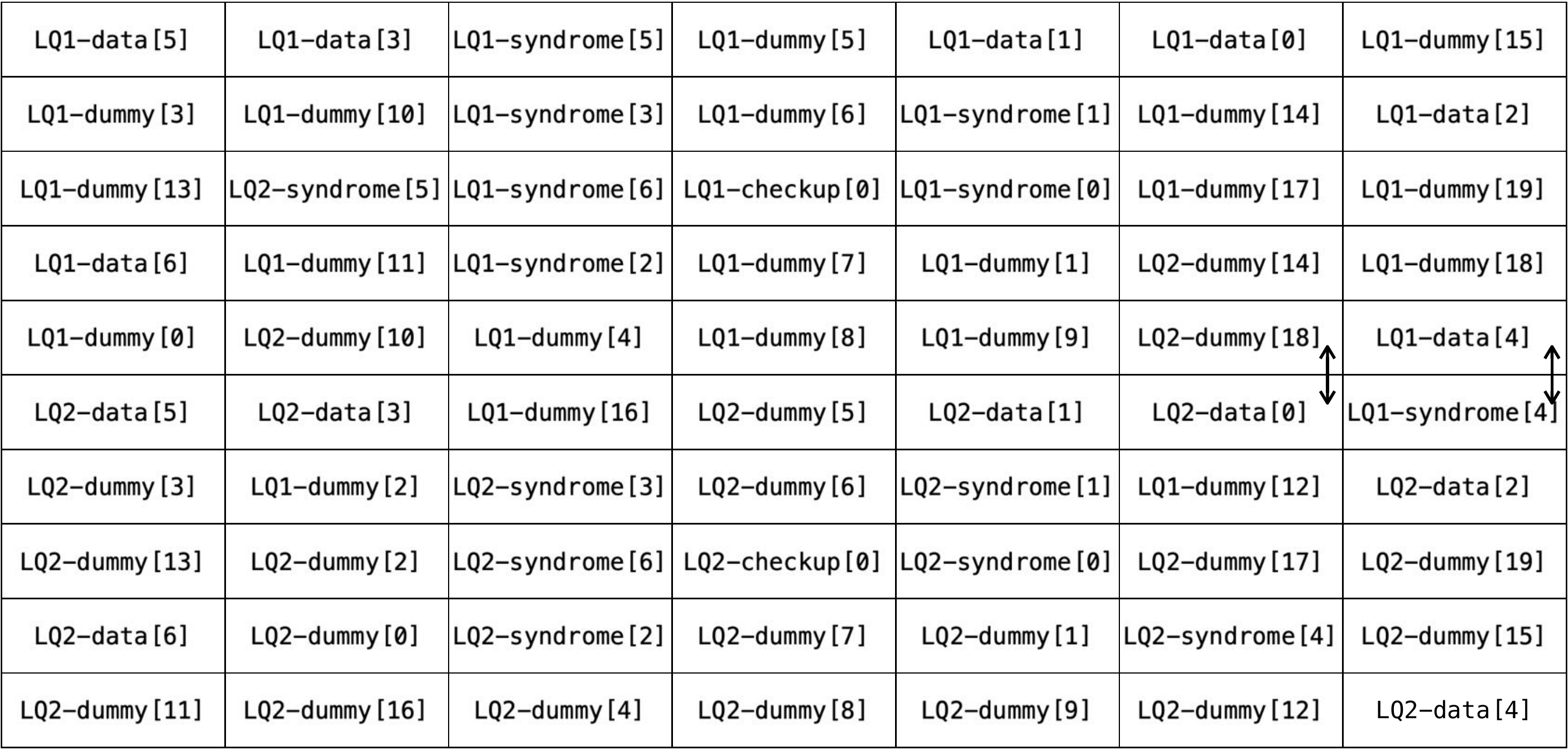}
}
\caption
{
(Continued from Figure~\ref{fig:steane_cnot_vertical_3}) Logical \emph{CNOT} gate operations for qubits arranged vertically.
By the \emph{Move-Back} operations, the data qubits after all the quantum operations are placed in their initial positions (compare (b) and Figure~\ref{fig:steane_cnot_vertical_1} (a)).
Note that the bi-directed arrow indicates \emph{SWAP} gate.
}
\label{fig:steane_cnot_vertical_4}
\end{figure*}

\begin{figure*}[t]
\centering
\subfigure[Initial Mapping based on Vertically Extended Layout, $(data_{n}, magic_{s})$]{
	\includegraphics[scale=0.35]{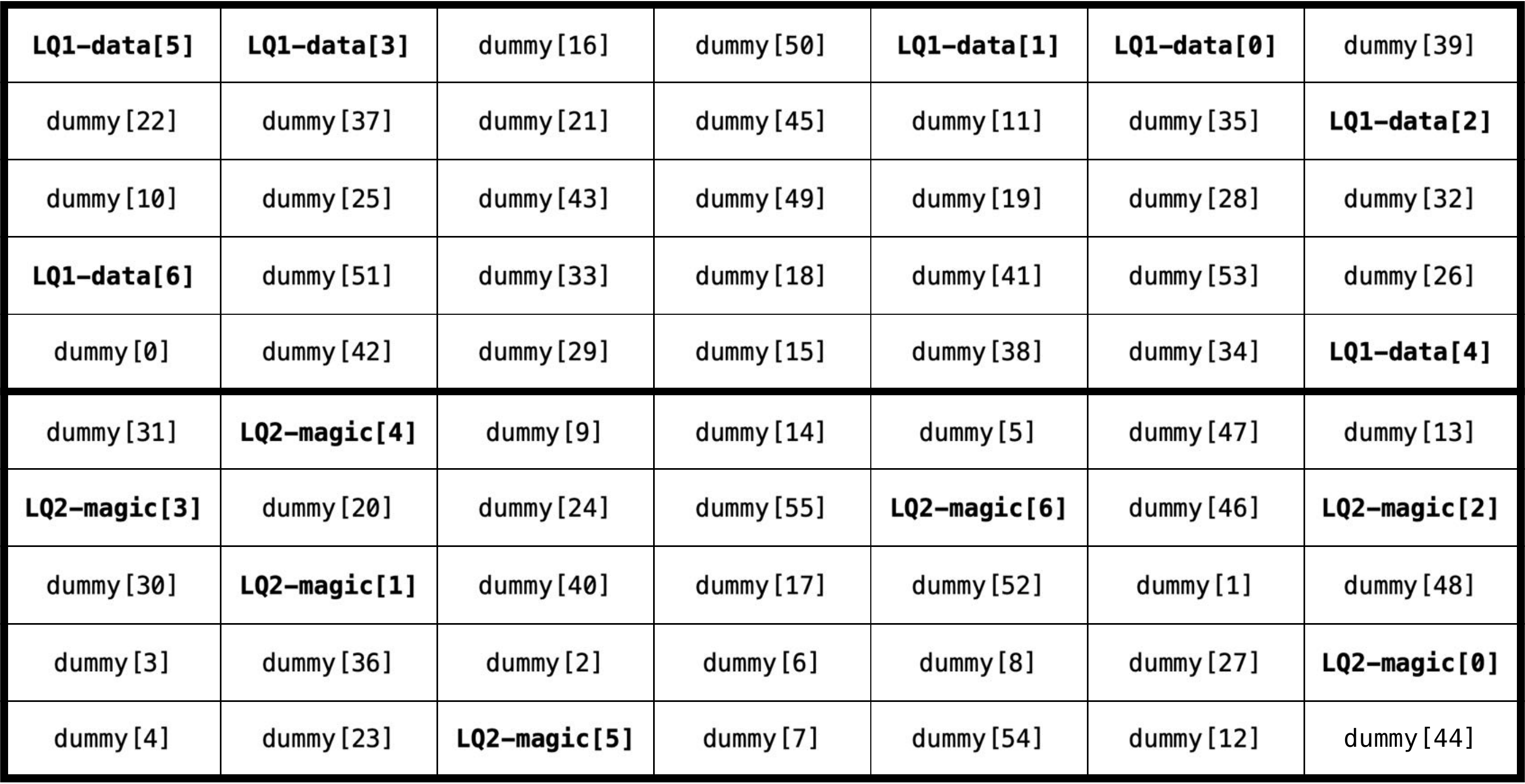}
}
\subfigure[Step 1]{
	\includegraphics[scale=0.35]{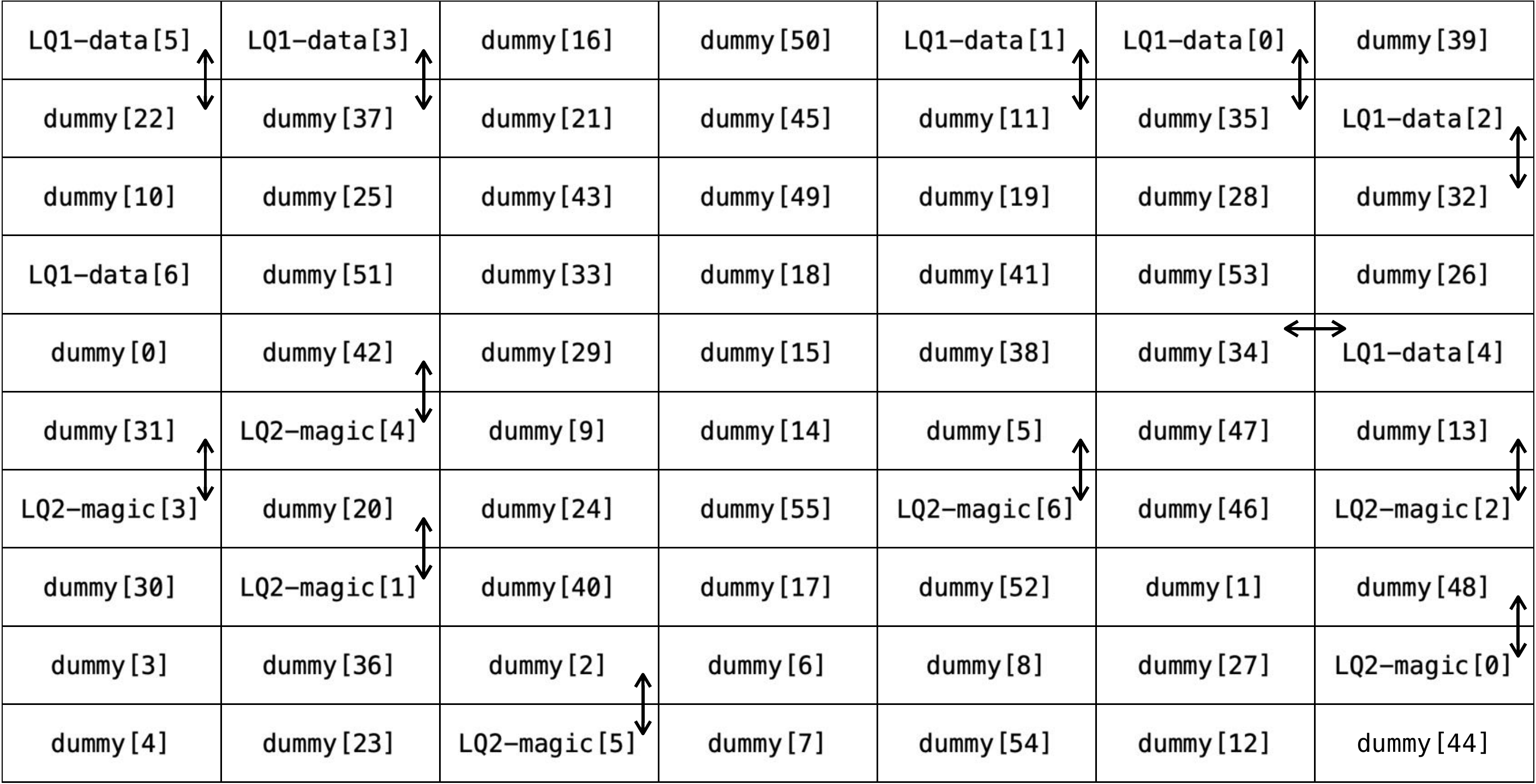}
}
\subfigure[Step 2]{
	\includegraphics[scale=0.35]{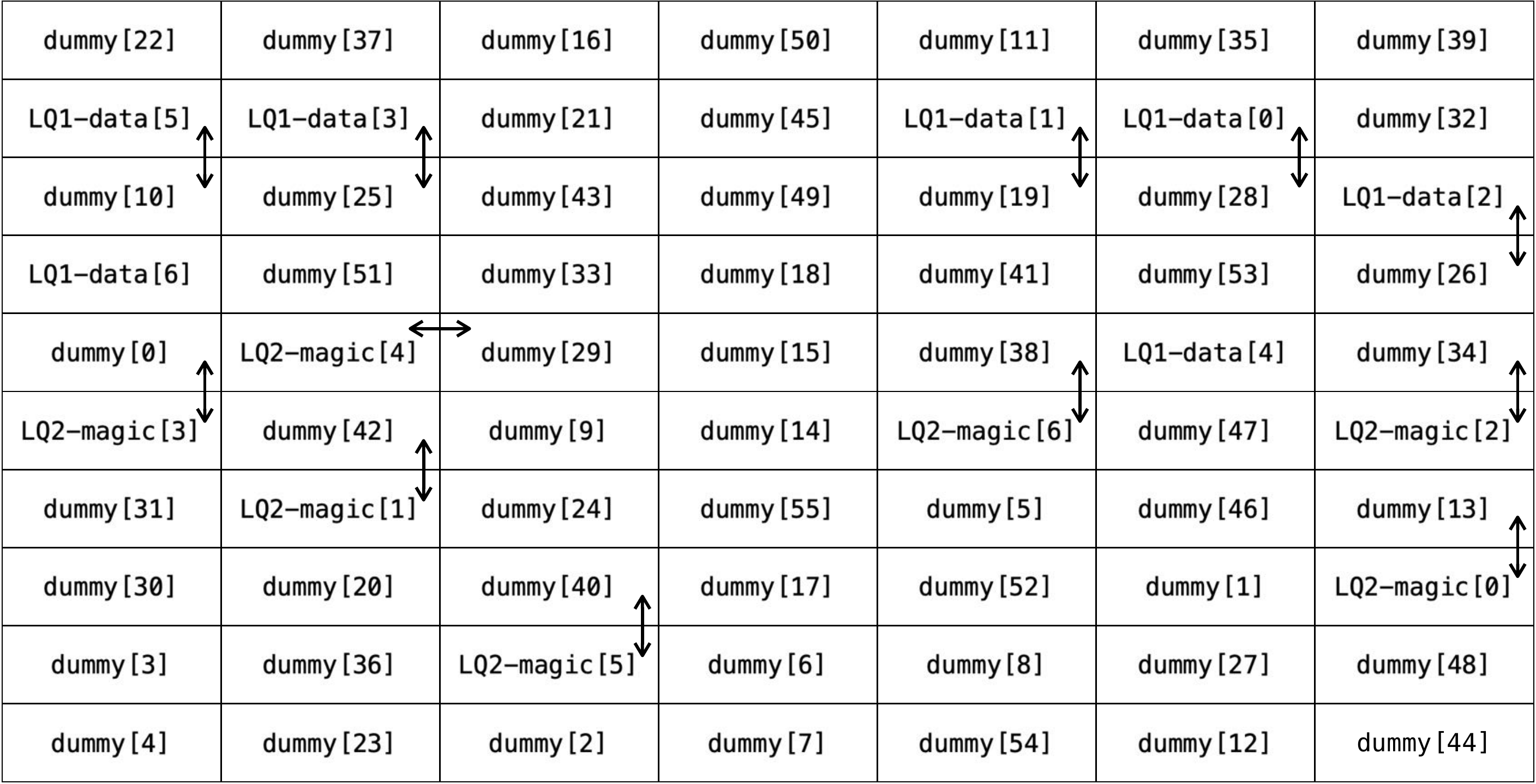}
}
\caption
{
The first part of logical \emph{T} gate operations for qubits arranged vertically: Trasversal \emph{CNOT} between data qubits $data[i]$ and magic qubits $magic[i]$ and \emph{MeasZ} on magic qubits.
(a) The vertically extended qubit layout of logical qubits: data qubits in the north direction and magic qubits in the south direction.
Note that the rectangles, hexagons and bi-directed arrow respectively indicate \emph{S}, \emph{PrepZ}, \emph{MeasZ} and \emph{SWAP} gates.
}
\label{fig:steane_t_vertical_1}
\end{figure*}

\begin{figure*}[t]
\centering
\subfigure[Step 3]{
	\includegraphics[scale=0.35]{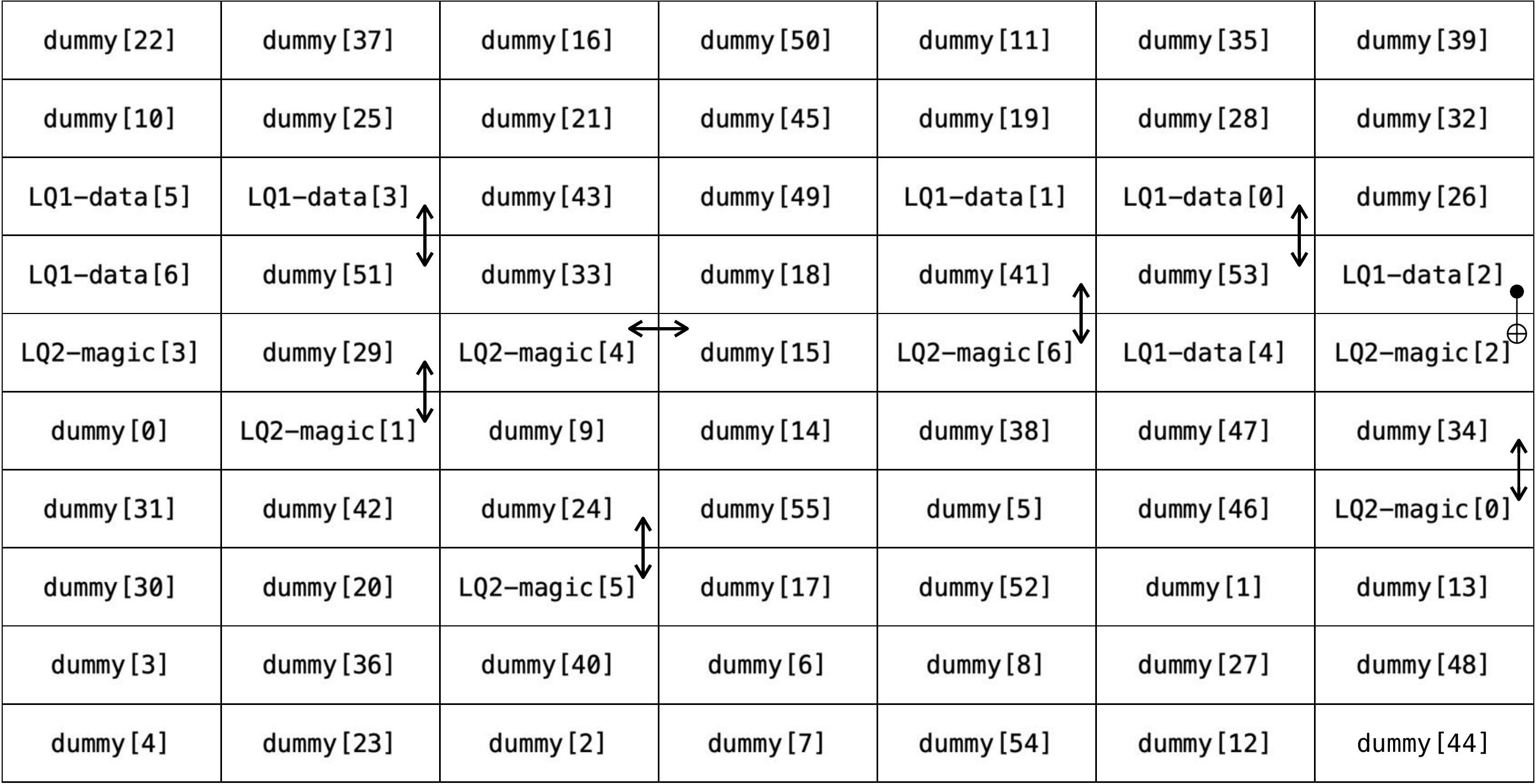}
}
\subfigure[Step 4]{
	\includegraphics[scale=0.35]{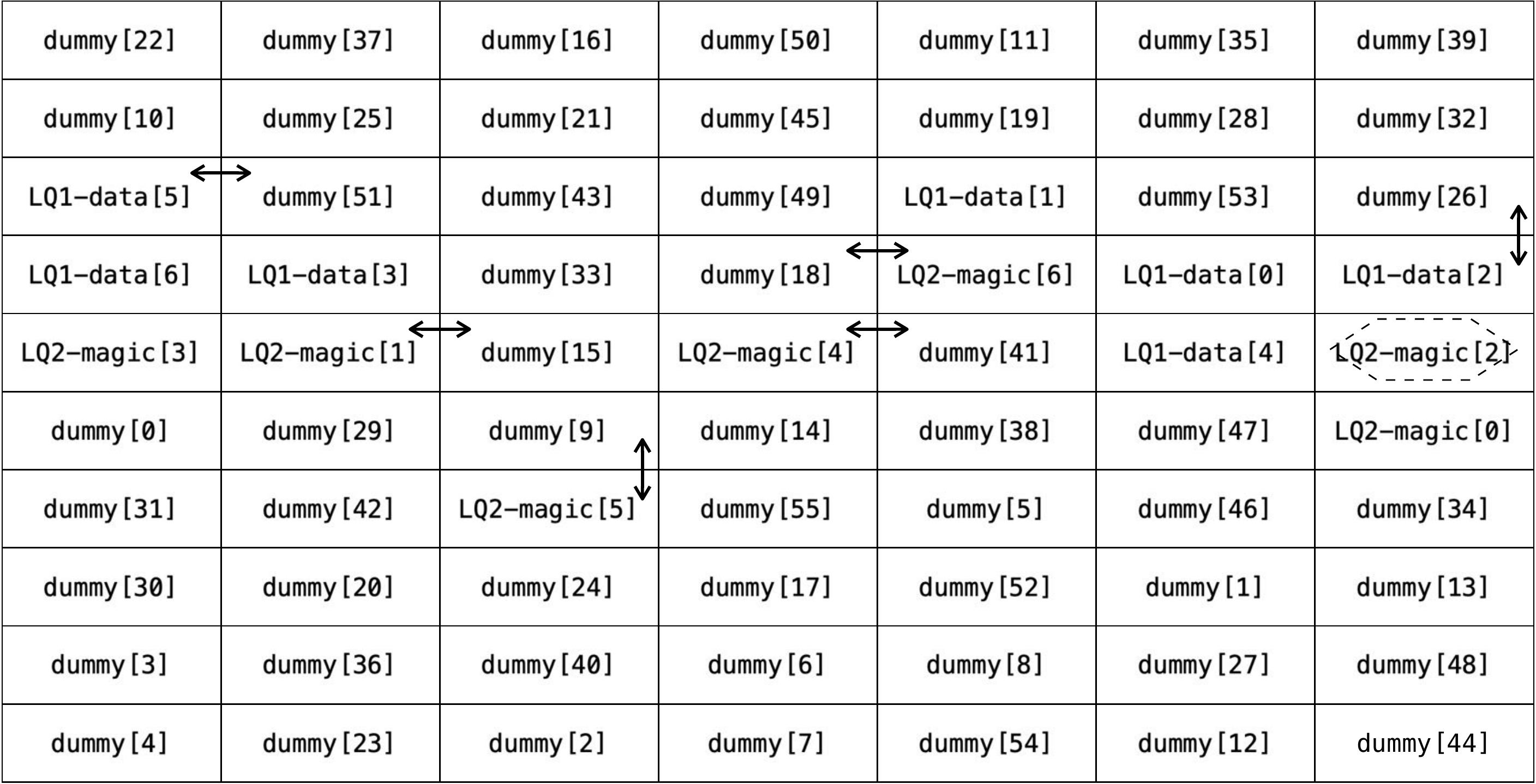}
}
\subfigure[Step 5]{
	\includegraphics[scale=0.35]{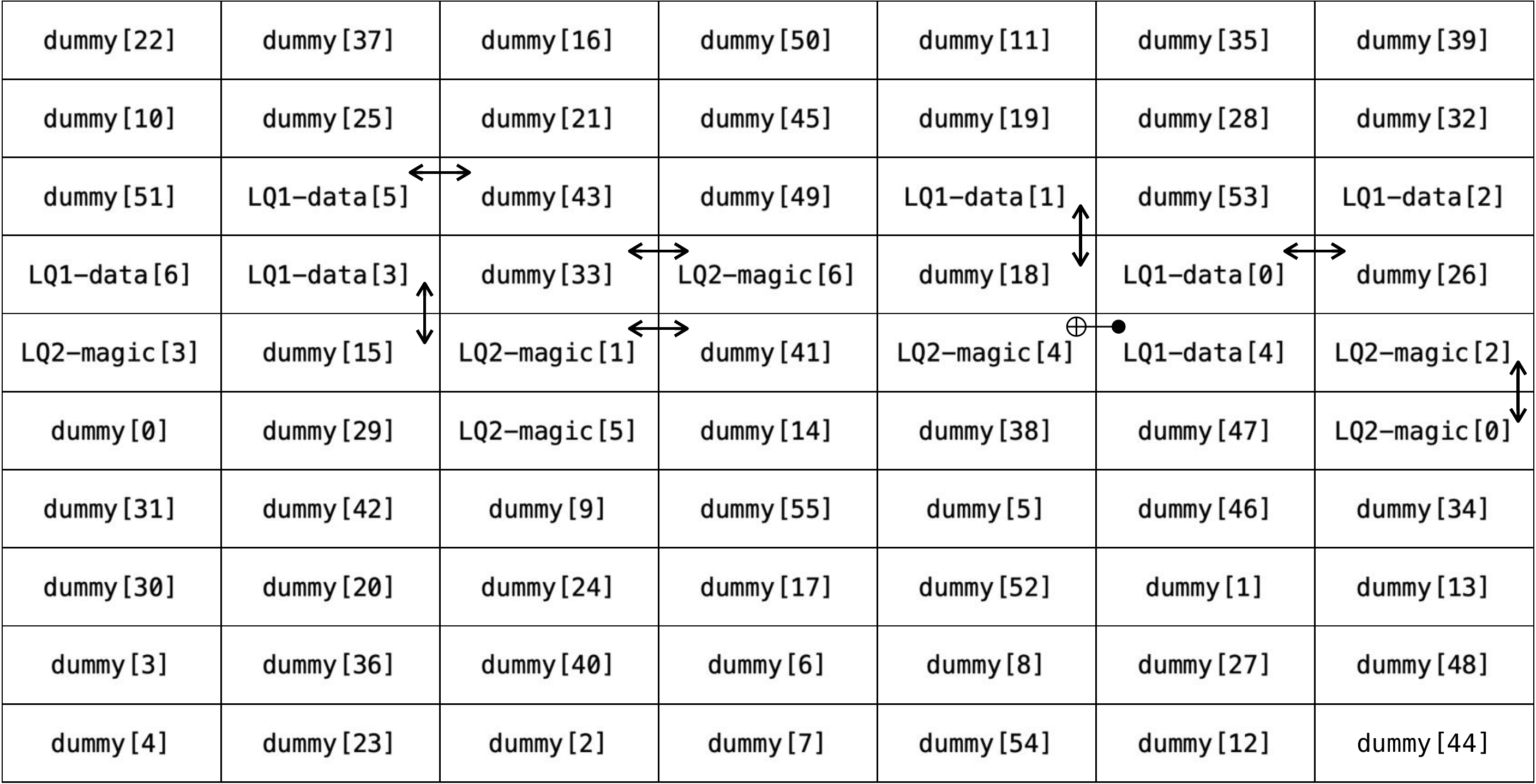}
}
\caption
{
(Continued from Figure~\ref{fig:steane_t_vertical_1}) The first part of logical \emph{T} gate operations for qubits arranged vertically: Trasversal \emph{CNOT} between data qubits $data[i]$ and magic qubits $magic[i]$ and \emph{MeasZ} on magic qubits.
Note that the rectangles, hexagons and bi-directed arrow respectively indicate \emph{S}, \emph{PrepZ}, \emph{MeasZ} and \emph{SWAP} gates.
}
\label{fig:steane_t_vertical_2}
\end{figure*}

\begin{figure*}[t]
\centering
\subfigure[Step 6]{
	\includegraphics[scale=0.35]{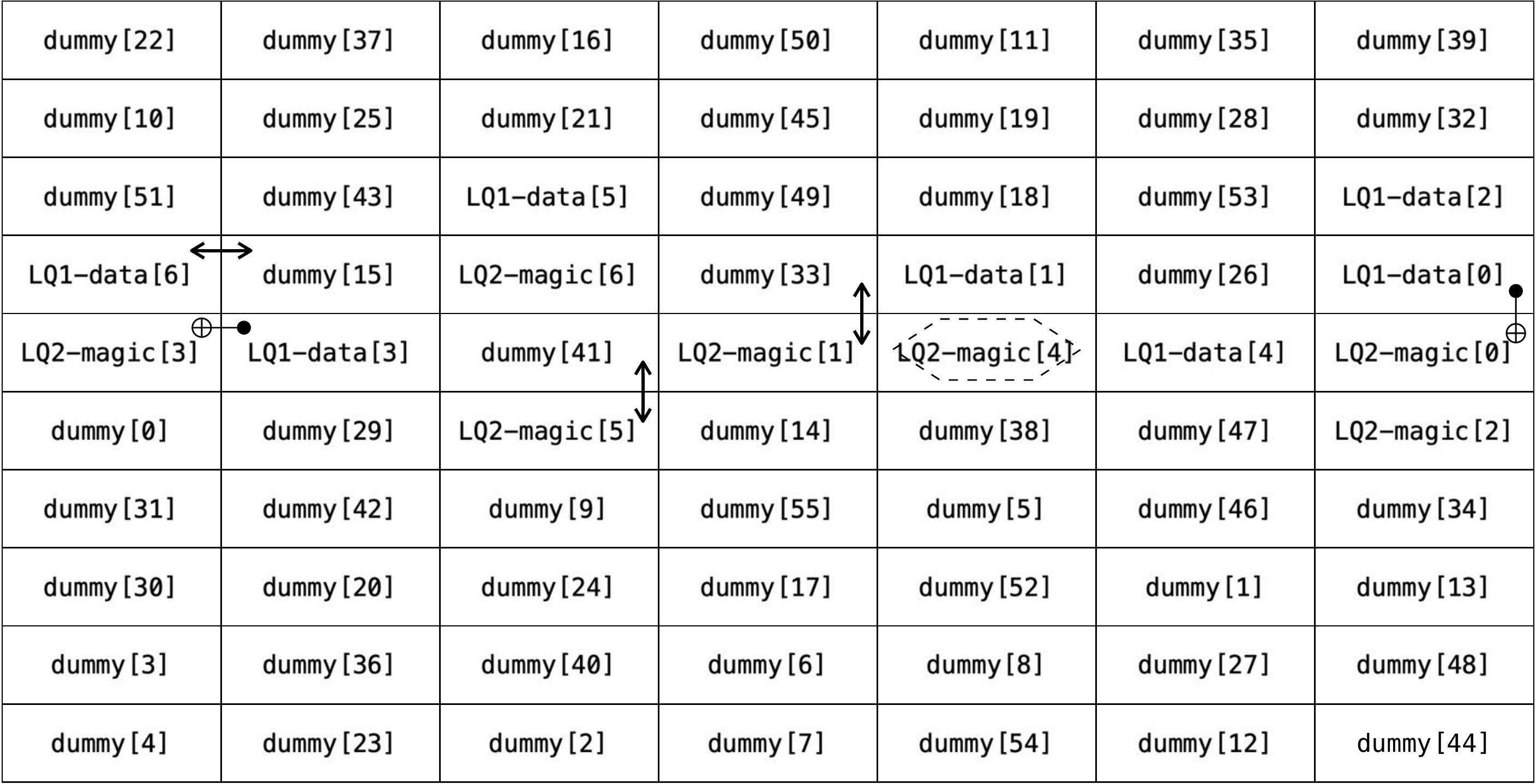}
}
\subfigure[Step 7]{
	\includegraphics[scale=0.35]{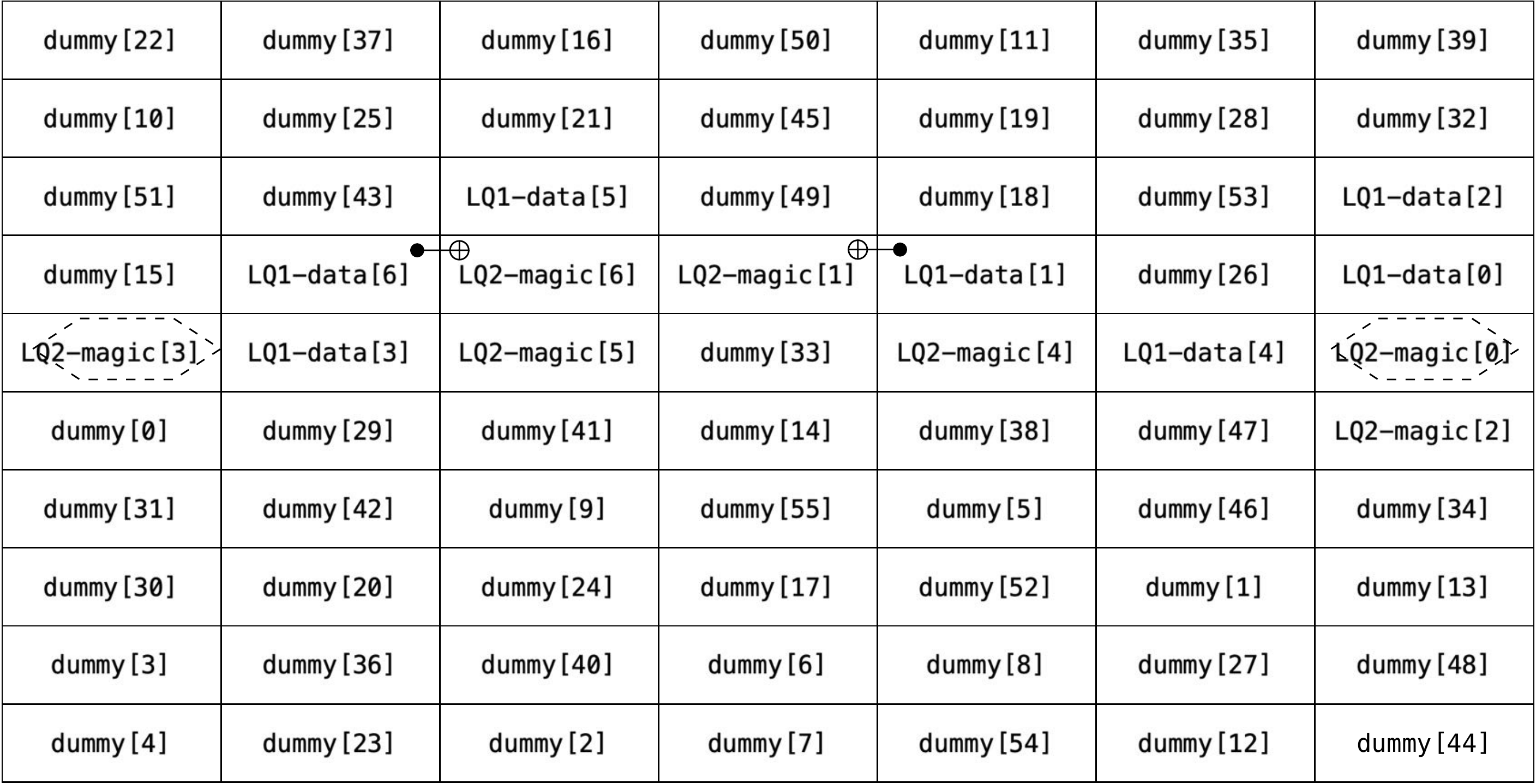}
}
\subfigure[Step 8]{
	\includegraphics[scale=0.35]{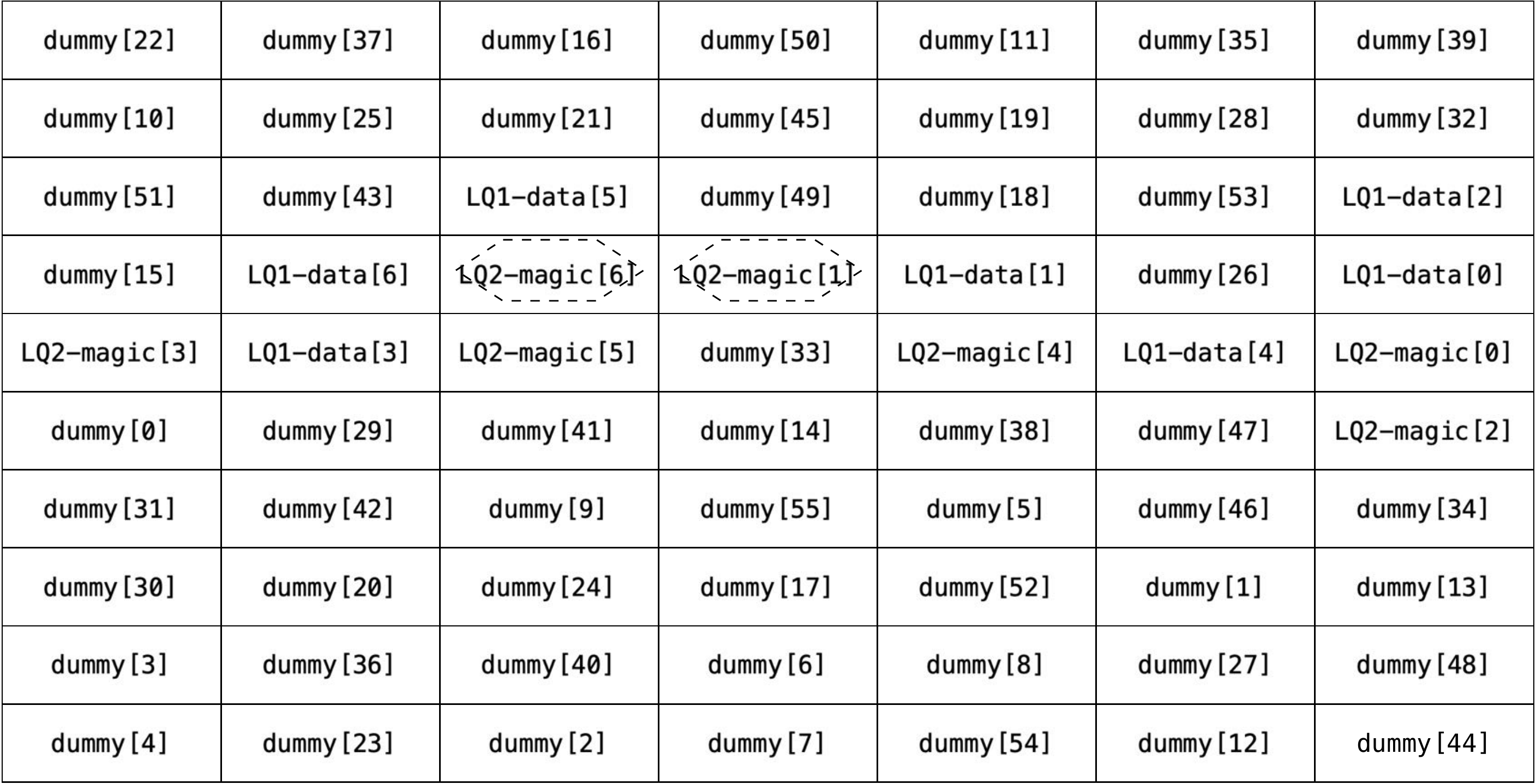}
}
\caption
{
(Continued from Figure~\ref{fig:steane_t_vertical_2}) The first part of logical \emph{T} gate operations for qubits arranged vertically: Trasversal \emph{CNOT} between data qubits $data[i]$ and magic qubits $magic[i]$ and \emph{MeasZ} on magic qubits.
Note that the rectangles, hexagons and bi-directed arrow respectively indicate \emph{S}, \emph{PrepZ}, \emph{MeasZ} and \emph{SWAP} gates.
}
\label{fig:steane_t_vertical_3}
\end{figure*}

\begin{figure*}[t]
\centering
\subfigure[Step 9]{
	\includegraphics[scale=0.35]{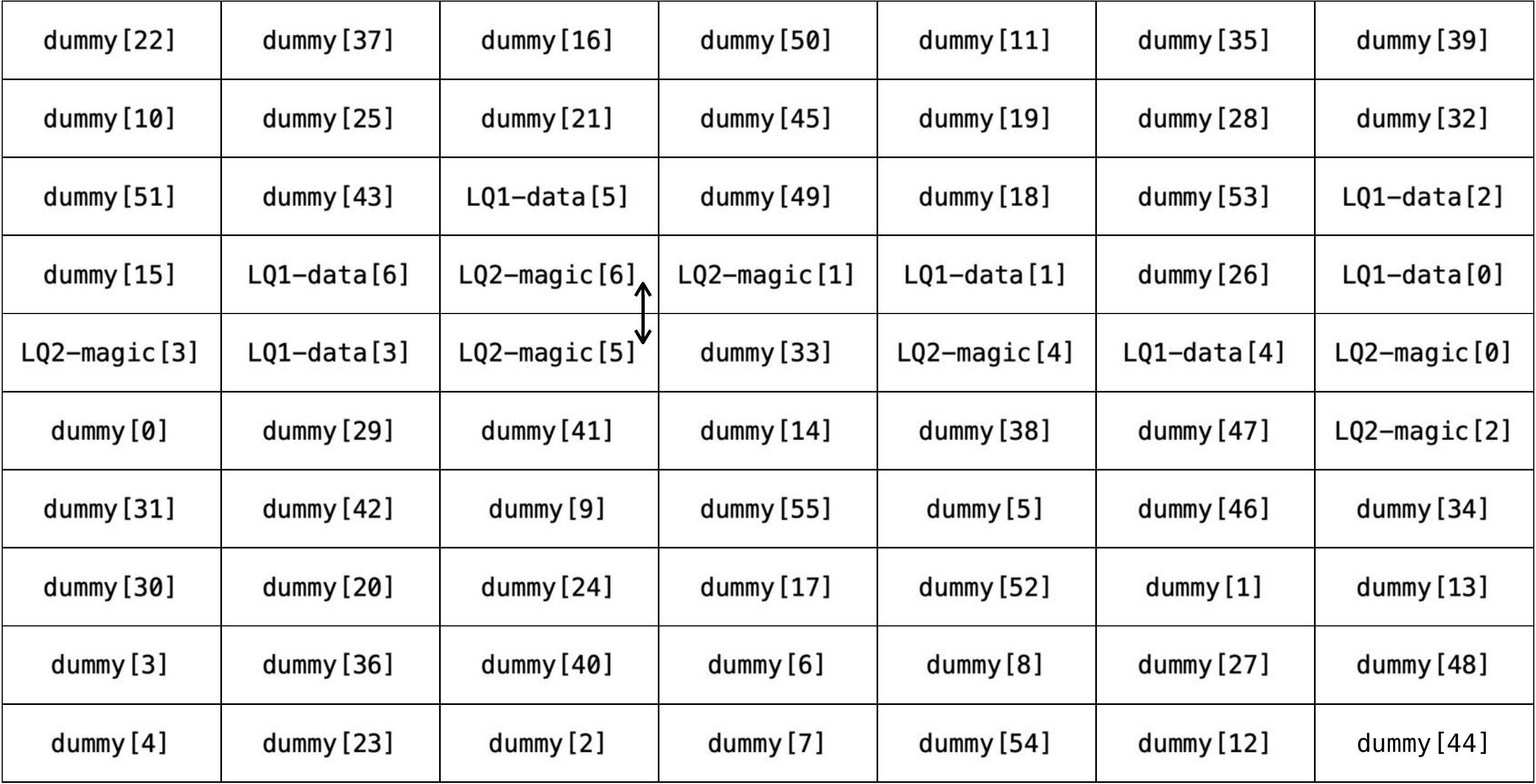}
}
\subfigure[Step 10]{
	\includegraphics[scale=0.35]{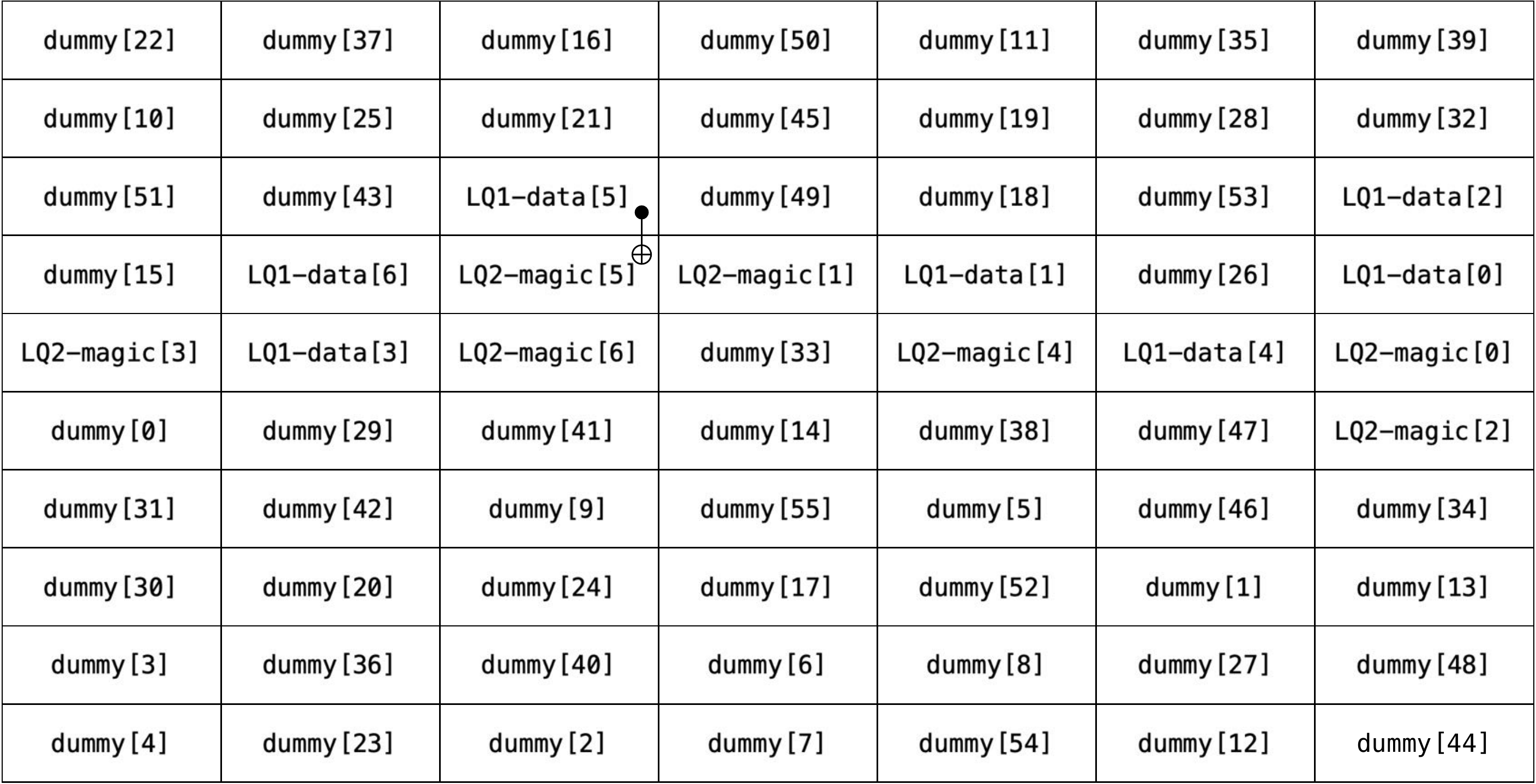}
}
\subfigure[Step 11 (Barrier)]{
	\includegraphics[scale=0.35]{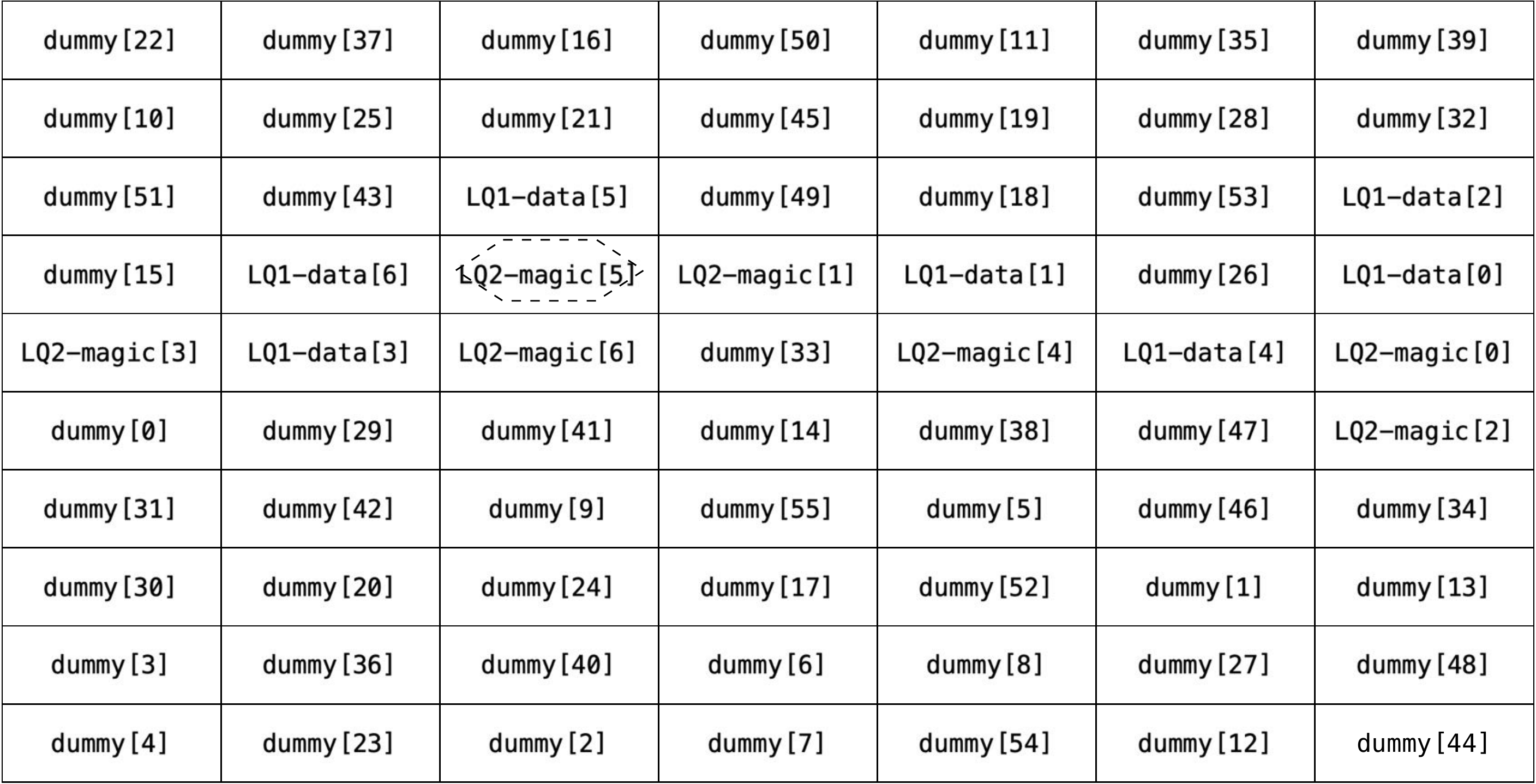}
}
\caption
{
(Continued from Figure~\ref{fig:steane_t_vertical_3}) The first part of logical \emph{T} gate operations for qubits arranged vertically: Trasversal \emph{CNOT} between data qubits $data[i]$ and magic qubits $magic[i]$ and \emph{MeasZ} on magic qubits.
Note that the rectangles, hexagons and bi-directed arrow respectively indicate \emph{S}, \emph{PrepZ}, \emph{MeasZ} and \emph{SWAP} gates.
}
\label{fig:steane_t_vertical_4}
\end{figure*}

\begin{figure*}[t]
\centering
\subfigure[Step 12 (S-correction)]{
	\includegraphics[scale=0.35]{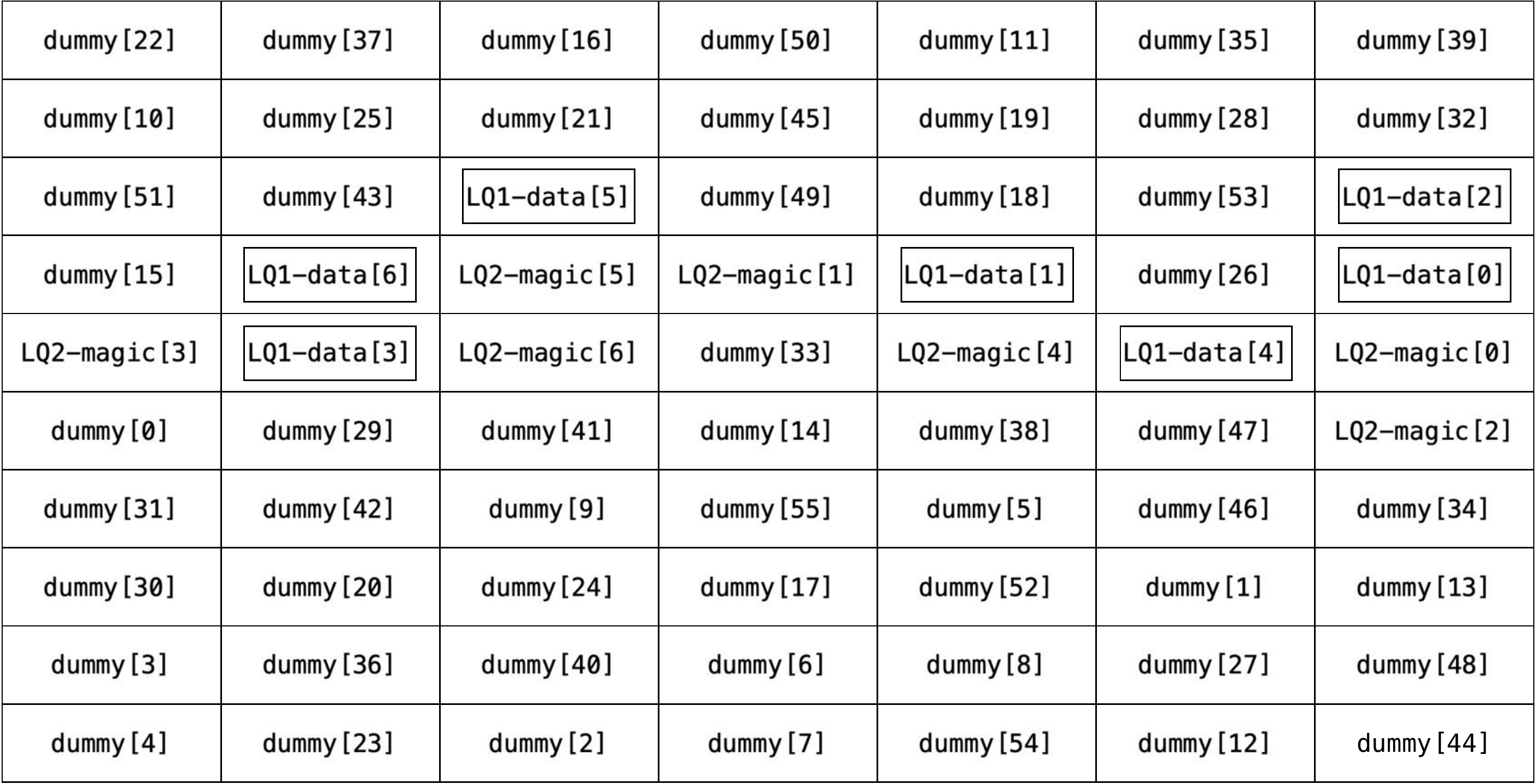}
}
\subfigure[Step 13]{
	\includegraphics[scale=0.35]{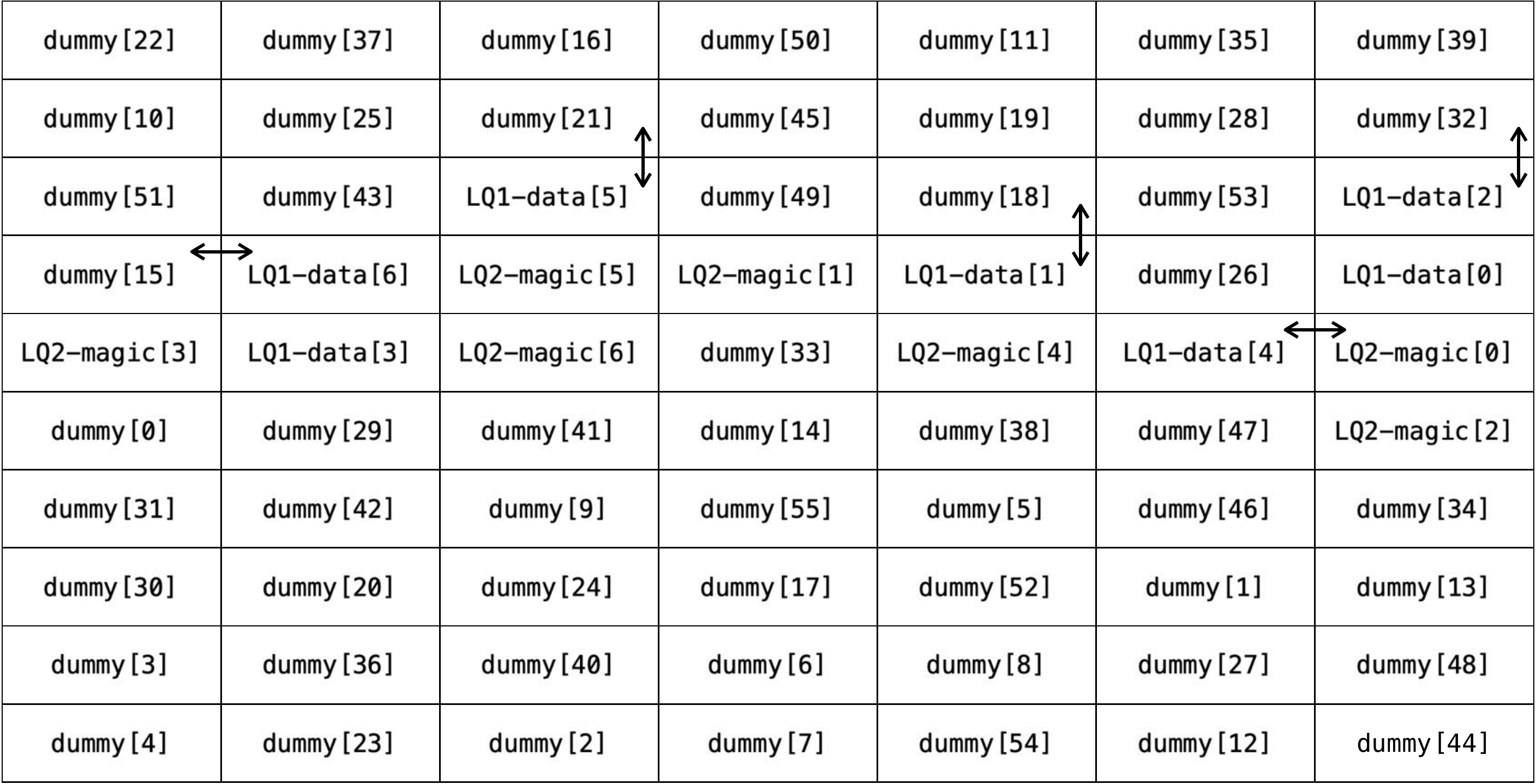}
}
\subfigure[Step 14]{
	\includegraphics[scale=0.35]{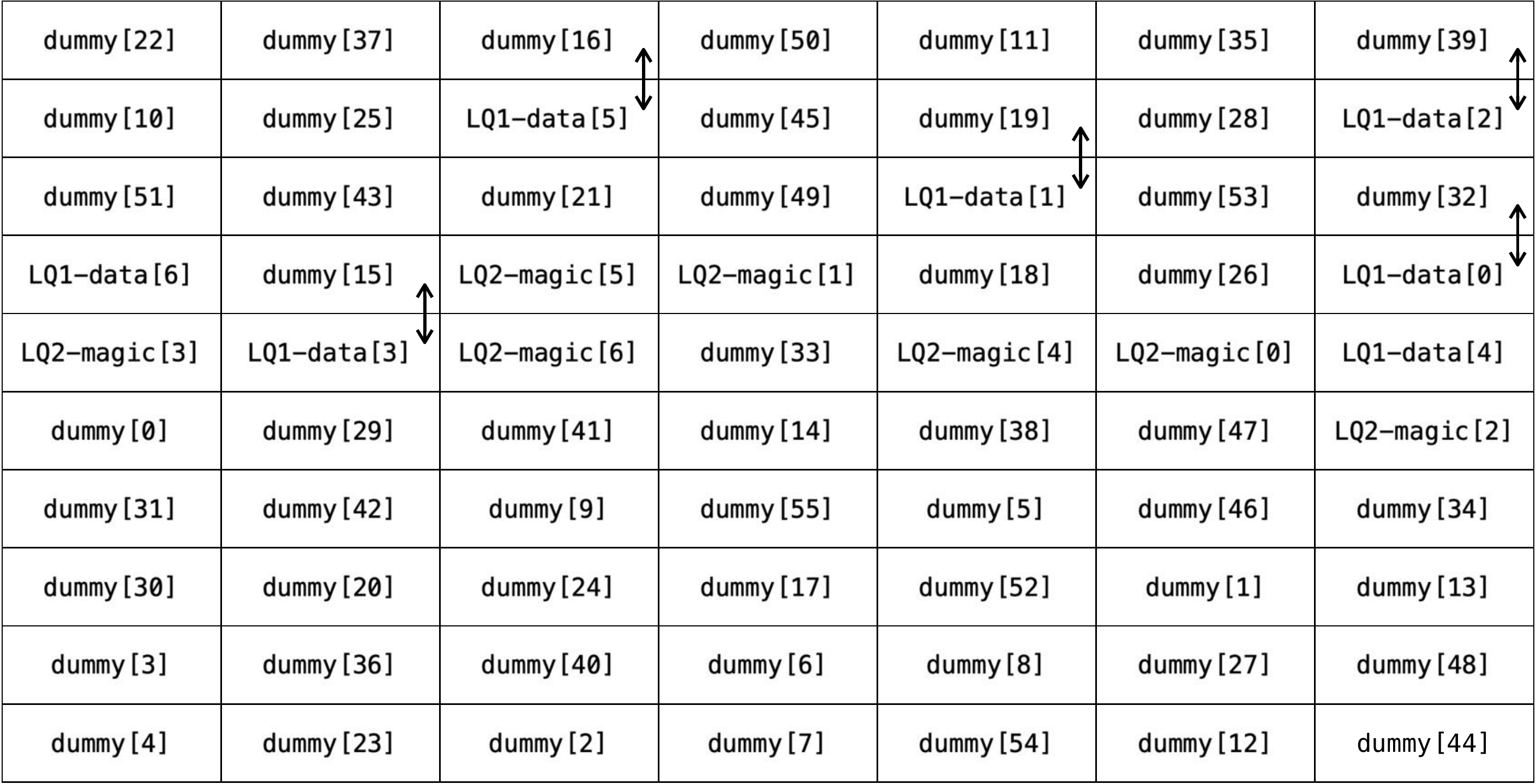}
}
\caption
{
The second part logical \emph{T} gate operations for qubits arranged vertically: Logical \emph{S} gate correction and \emph{Move-Back}.
Please compare the final mapping with the initial mapping (Figure~\ref{fig:steane_t_vertical_1} (a)).
The positions of the data qubits are the same in both mappings, but the magic qubits are not.
Note that the rectangles, hexagons and bi-directed arrow respectively indicate \emph{S}, \emph{PrepZ}, \emph{MeasZ} and \emph{SWAP} gates.
}
\label{fig:steane_t_vertical_5}
\end{figure*}

\begin{figure*}[t]
\centering
\subfigure[Step 15]{
	\includegraphics[scale=0.35]{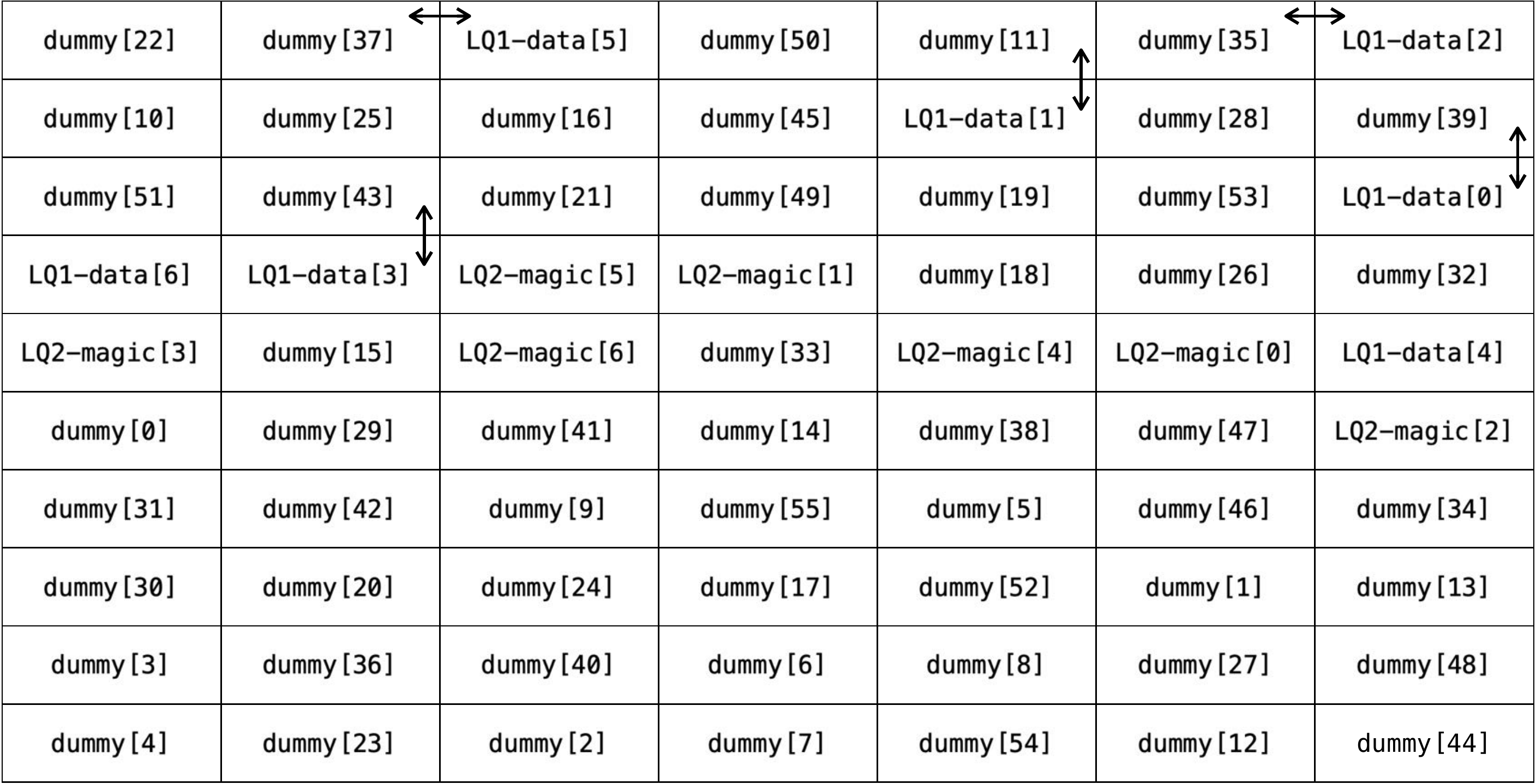}
}
\subfigure[Step 16]{
	\includegraphics[scale=0.35]{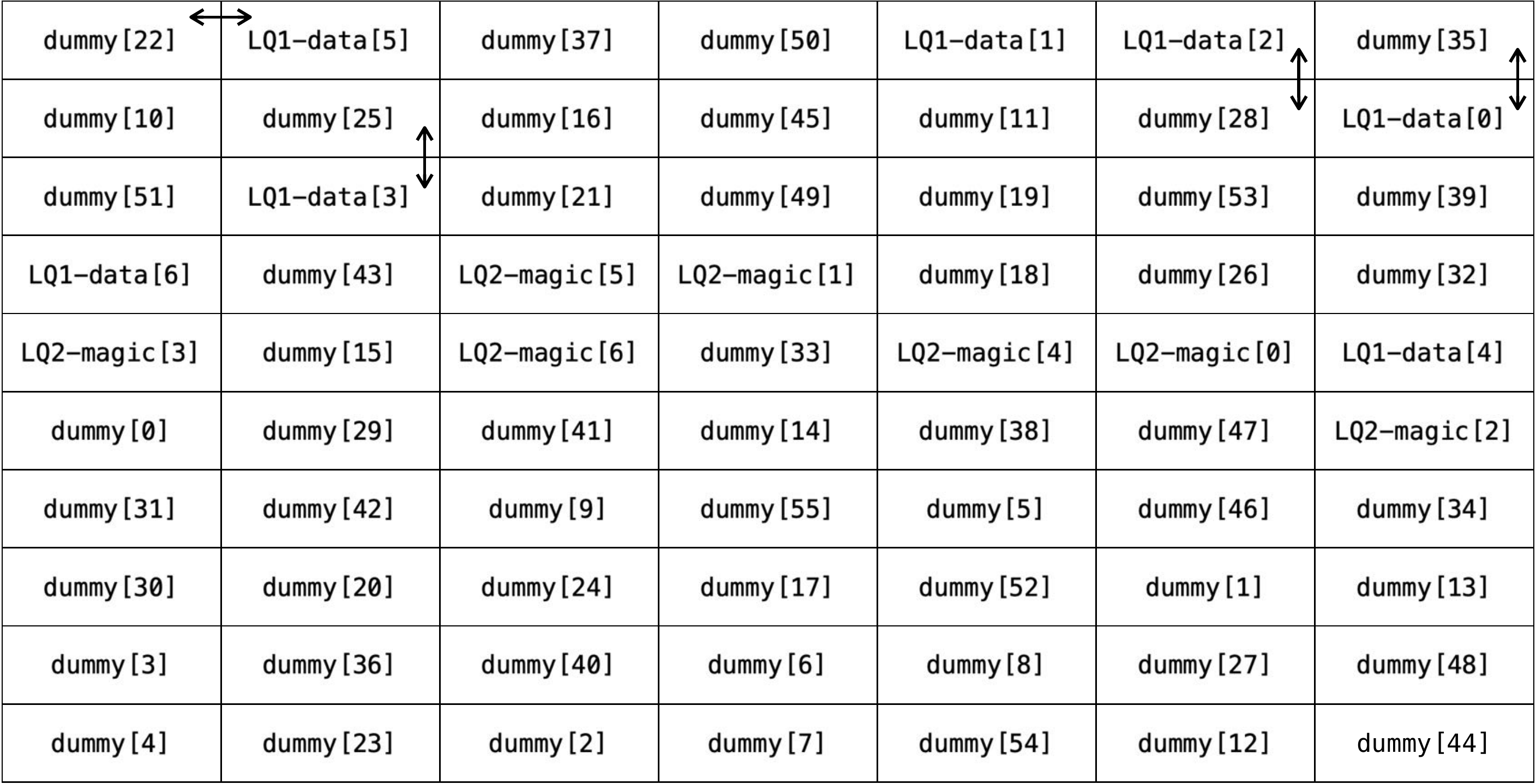}
}
\subfigure[Step 17]{
	\includegraphics[scale=0.35]{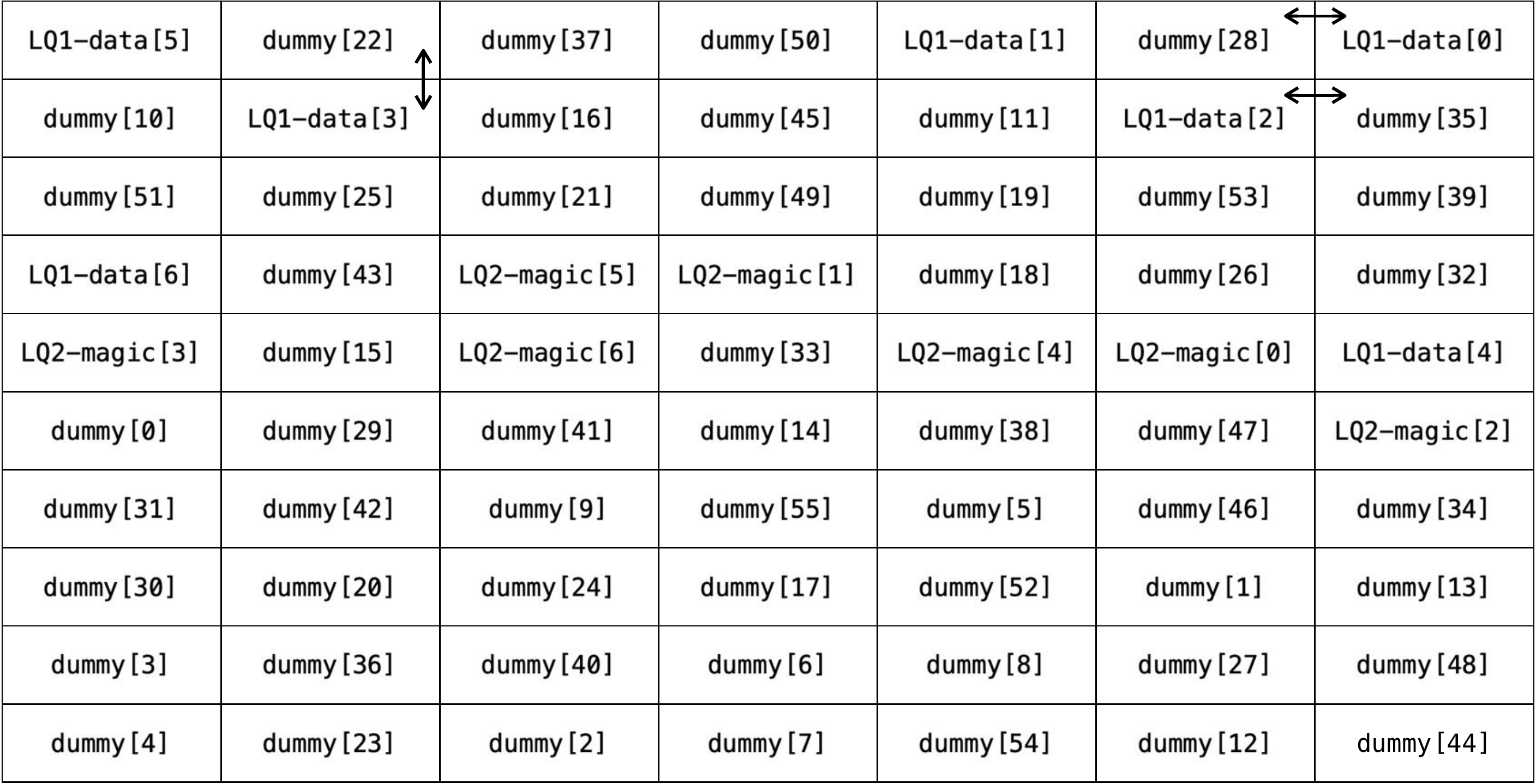}
}
\caption
{
(Continued from Figure~\ref{fig:steane_t_vertical_5}) The second part logical \emph{T} gate operations for qubits arranged vertically: Logical \emph{S} gate correction and \emph{Move-Back}.
Note that the rectangles, hexagons and bi-directed arrow respectively indicate \emph{S}, \emph{PrepZ}, \emph{MeasZ} and \emph{SWAP} gates.
}
\label{fig:steane_t_vertical_6}
\end{figure*}

\begin{figure*}[t]
\centering
\subfigure[Final Mapping]{
	\includegraphics[scale=0.35]{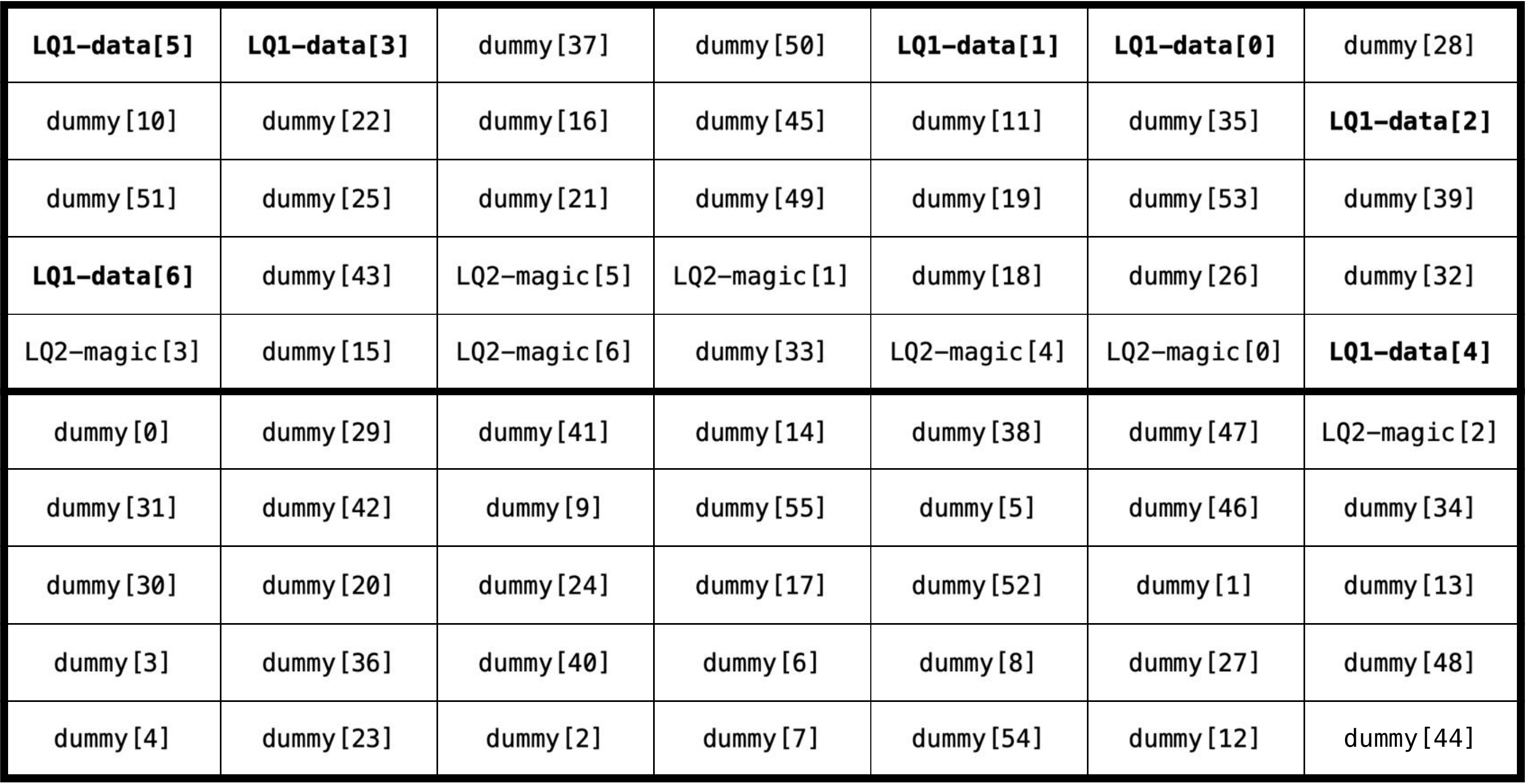}
}
\caption
{
(Continued from Figure~\ref{fig:steane_t_vertical_6}) The second part of logical \emph{T} gate operations for qubits arranged vertically: Logical \emph{S} gate correction and \emph{Move-Back}.
Please compare the final mapping with the initial mapping (Figure~\ref{fig:steane_t_vertical_1} (a)).
The positions of the data qubits are the same in both mappings, but the magic qubits are not.
Note that the rectangles, hexagons and bi-directed arrow respectively indicate \emph{S}, \emph{PrepZ}, \emph{MeasZ} and \emph{SWAP} gates.
}
\label{fig:steane_t_vertical_7}
\end{figure*}

\end{document}